\documentclass{aa} 
\usepackage[varg]{txfonts}
\usepackage{amsmath,mathtools}
\usepackage{graphicx}
\usepackage[T1]{fontenc}
\usepackage{comment}
\usepackage{orcidlink}
\usepackage{enumitem}
\bibpunct{(}{)}{;}{a}{}{,} 

\hypersetup{breaklinks=true,colorlinks=true, citecolor=blue, linkcolor=blue,urlcolor=blue}

\newcommand{\Msun}{\mathrm{M}_{\odot}}
\newcommand{\hi}{\ifmmode{\rm HI}\else{H\/{\scshape i}}\fi}
\newcommand{\msunyr}{{\rm M}_\odot \, {\rm yr^{-1}}} 
 
\newcommand{\kms} {\,{\rm km\,s}^{-1}}
\newcommand{\vlsr}{\ifmmode{v_\mathrm{LSR}}\else{$v_\mathrm{LSR}$}\fi}

\title{Simulating winds in the Galactic centre}
\subtitle{II. Observational biases on \hi\ clouds and a multi-wavelength comparison}

\titlerunning{Galactic centre outflows: observational biases}

\author{Nicolas~Peschken\inst{1,2}\thanks{Contact e-mail: \href{mailto:nicolas.peschken@unifi.it}{nicolas.peschken@unifi.it}}\orcidlink{0000-0002-9925-2974}
 \and  Enrico~M.~Di Teodoro\inst{1,2}\orcidlink{0000-0003-4019-0673}
 \and  Michał~Hanasz\inst{3}\orcidlink{0000-0002-2370-5631}
 \and  Lucia~Armillotta\inst{1,2}\orcidlink{0000-0002-5708-1927} \and \\
 Dominik~W\'{o}lta\'{n}ski\inst{3}
 \and  Artur~Gawryszczak\inst{4}\orcidlink{0000-0002-4435-4594}
 }
\institute{Department of Physics and Astronomy, University of Florence, Largo E. Fermi 2, I-50125 Firenze, Italy
\and INAF - Osservatorio Astrofisico di Arcetri, Largo E. Fermi 5, I-50125 Firenze, Italy
\and Institute of Astronomy, Nicolaus Copernicus University, ul. Grudziadzka 5/7, PL-87-100 Toruń, Poland
\and Nicolaus Copernicus Astronomical Centre, Bartycka 18, 00-716 Warsaw, Poland
}

\date{Received date /  Accepted date }

\begin{document}

\abstract{
High-velocity neutral hydrogen (\hi) clouds found up to several kiloparsecs above the Galactic centre (GC) provide some of the most direct evidence for a nuclear wind in the Milky Way, yet the properties inferred for these clouds are influenced by how they are observed.
Starting from a magnetohydrodynamical simulation in which supernova feedback drives a multiphase nuclear outflow, we generate mock \hi\ observations matched to the angular resolution and sensitivity of Green Bank Telescope (GBT) surveys of the GC and analyse them in the same way as the real data.
We find that observational effects substantially reshape the cloud population: beam smearing makes clouds appear larger, rounder and more diffuse, while noise erases faint clouds and strips low-density gas from the others. 
These effects combined, together with confusion with foreground disk emission, prevent the detection of more than 80\% of the clouds (up to 95\% during more quiescent phases), while those detected appear less massive, less turbulent and less elongated, with column densities underestimated by a factor of two; only their velocities are reliably recovered. 
Once these effects are included, the simulated clouds broadly reproduce the properties and sky distribution of the GBT population, but lack the fastest ($\gtrsim250$ $\kms$) and highest-latitude ($b\gtrsim 10^\circ$) clouds.
Mock UV absorption and soft X-ray observations support this picture, with warm-ionized gas column densities at the low end of the observed range and X-ray emission measures too low at $b\gtrsim10^\circ$.
Our results imply that the true GC cloud population is more numerous, massive, turbulent and elongated than observed, and that fully reproducing observations may require a stronger recent starburst or additional feedback.
}


\maketitle

\section{Introduction}

Over the past decade, radio observations of the inner regions of the Milky Way (MW) have revealed a conspicuous population of atomic hydrogen (\hi) clouds apparently outflowing from the Galactic centre (GC) at high velocities \citep{McClure-Griffiths+13,DiT+18,Bordoloi+2025,Lockman+2026}. These clouds are found up to at least $\sim$2 kpc above the Galactic mid-plane, have velocities of a few hundred $\kms$, and some of them even contain molecular cores \citep{DiTeodoro+2020,DiTeodoro+2026,Cashman+2021,Heyer+2025}. This evidence suggests the presence of strong winds generated at GC, powerful enough to accelerate dense, cold gas to such velocities. This scenario is further supported by multi-wavelength observations of the central outflowing region of our Galaxy: high-velocity warm ionised gas detected in ultraviolet (UV) absorption lines \citep{Fox+15,Bordoloi+17,2017ApJS..232...25S,Cashman+2023}, the eROSITA bubbles seen in soft X-rays \citep{Predehl+2020}, and the Fermi bubbles seen in $\gamma$-rays and extending up to $\sim$10 kpc from the GC \citep{Su+10}, all point towards the presence of high-velocity multiphase gas above and below the MW centre.

It remains unclear, however, which mechanisms drive these outflows and whether the different gas phases observed on different scales originate from the same physical process. Some studies have suggested that active galactic nucleus (AGN) feedback could be the main driver of the Fermi bubbles \citep[e.g.,][]{Zubovas+2012,Bland-Hawthorn+2019}, invoking past activity \citep[e.g.,][]{2012ApJ...756..181G,Bland-Hawthorn+2019,2020ApJ...894..117Z} of the now quiescent central supermassive black hole, Sgr A$^*$. Alternatively, other works have emphasized the role of supernova (SN) feedback in the central region, showing that SN-driven winds can reproduce several of the observed features \citep[e.g.,][]{Crocker+2015,Sarkar+2015,2025ApJ...981..154Z}. 
We explored the latter scenario in our previous paper, \cite[][hereafter Paper~I]{2026A&A...712A.148P}, where we presented a new magneto-hydrodynamical (MHD) simulation of the MW, with a particular focus on the GC and its outflows. In our simulation, the Central Molecular Zone \citep[CMZ, e.g.,][]{Morris1996,Henshaw2023}, a dense, star-forming, ring-like gaseous structure with a radius of $\sim$100--300 pc created by the accumulation of gas driven towards the centre by bar torques \citep[e.g.,][]{1991MNRAS.252..210B,1992MNRAS.259..345A,Sormani+15}, is modelled fully self-consistently and undergoes recurrent cycles of starbursts and quiescent periods. The resulting SN feedback creates powerful hot winds capable of entraining dense \hi\ clouds away from the mid-plane and accelerating them to velocities of hundreds of $\kms$, similarly to the observed \hi\ clouds above the GC. 

In Paper~I, we compared the properties of our simulated clouds with observed \hi\ clouds from \cite{DiT+18} and found generally good agreement, supporting the possibility that these clouds could be accelerated by SN feedback alone.
However, that comparison relied on the intrinsic properties of the simulated clouds, measured directly in three-dimensional space, whereas real clouds can only be characterised from their emission projected on the sky, as recorded by a given instrument. A direct comparison between simulations and observations is therefore not always straightforward, as instrumental effects can cause the measured properties of observed objects to differ, sometimes significantly, from the intrinsic ones. In particular, for emission line observations of dense gaseous clouds such as our simulated \hi\ clouds, previous studies have highlighted biases introduced mainly by two main instrumental effects: the angular resolution of the telescope and its sensitivity.

Finite angular resolution smooths the emission, diluting its observed surface brightness, with the magnitude of the effect depending on the source size relative to the telescope point-spread function (PSF), or beam \citep{2018ApJ...858...16G,2021ApJ...910..109Y}. This ``beam smearing'' can alter the apparent spatial extent and morphology of sources \citep[e.g.,][]{2009A&A...504...15P}, while unresolved velocity gradients can also broaden the observed line profiles and increase the measured velocity dispersion \citep[e.g.,][]{2011ApJ...741...69D,2018MNRAS.474.5076J,2019MNRAS.486.4463Y}.
This effect can be particularly important for small or poorly resolved sources, like molecular clouds or the \hi\ clouds in the nuclear outflow.
In addition, all observations are affected by instrumental noise, whose level depends on the telescope characteristics and integration time and ultimately determines the sensitivity of a given observation. Noise can also significantly affect the properties of observed objects. For example, noise preferentially removes faint emission, such as emission in the line wings, thereby narrowing the recovered line profiles and lowering the measured velocity dispersion \citep{2020MNRAS.498.2440Y}. It can also reduce the recovered spatial extent of sources \citep{2006PASP..118..590R} and cause part of the intrinsic flux to fall below the detection threshold \citep{2021ApJ...910..109Y,2022MNRAS.512.4765S}. 
The combined effect of angular resolution and sensitivity can therefore strongly affect the completeness of observed emission, with recovered fluxes of only $\sim$80\% reported for local molecular clouds \citep{2020ApJ...898...80Y}, and potentially lower completeness expected at larger distances.

The high-velocity clouds found above the MW centre \citep[e.g,][]{DiT+18,Lockman+2026} are likely to be affected by similar observational biases, suggesting that their intrinsic properties may differ from those inferred from the observations. In the context of models aiming to understand the mechanisms responsible for accelerating these clouds, it is therefore essential to take these biases into account when making comparisons between observations and simulations. In this paper, we investigate the impact of observational biases on the simulated \hi\ clouds presented in Paper~I, with particular emphasis on beam smearing and noise. We aim to predict the intrinsic properties of the high-velocity cloud population observed above the GC, while enabling a more direct comparison between our simulation and observations.

The remainder of this paper is structured as follows.
After summarizing our MHD simulation of the GC in Section~\ref{sec:summary_PaperI}, we present our methods to build mock \hi\ observations in Section~\ref{sec:mockdata}. We investigate the impact of different observational effects, including projection, beam smearing and noise, on \hi\ cloud properties in Section~\ref{sec:impact_bias}, then compare our simulation with \hi\ observations in Section~\ref{sec:comp_obs_hi}, accounting for these observational effects. Finally, we further compare our model to observational data in UV absorption and soft X-rays in Section~\ref{sec:comp_obs_multi}, discuss our results in Section~\ref{sec:discussion} and present our conclusions in Section~\ref{sec:conclusion}.


\section{Simulation of nuclear outflows in the MW}
 \label{sec:summary_PaperI}

In Paper I, we presented a self-consistent model of the MW run with the grid-based multi-fluid MHD code {\scshape Piernik} (\citealt{2010EAS....42..275H, 2010EAS....42..281H, 2012EAS....56..363H, 2012EAS....56..367H}), and focusing on the GC using adaptive mesh refinement (AMR). We summarize below the technical details of this simulation, as well as the main results, and refer the reader to Paper I for more details.

\subsection{Presentation of the model}
\label{sec:simu}

Our {\scshape Piernik} model consists of a gaseous disc embedded in an external rotating gravitational potential representing the dark matter and stellar components. We adopt the MW gravitational potential from \cite{2024A&A...692A.216H}, which provides a state-of-the-art, observationally constrained description of the GC and includes a rotating stellar bar that is gradually introduced over one full bar rotation ($\sim$150~Myr). The resulting bar torques drive gas towards the centre, where it naturally forms a dense, highly star-forming ring with a radius of $\sim$200~pc \citep{2015MNRAS.449.2421S,2019MNRAS.488.4663S}, broadly reproducing the morphology and observed properties of the CMZ. 
Thanks to the AMR, we achieve a spatial resolution of 3~pc in a central cylinder of 400~pc radius and 200~pc height, allowing us to capture properly the gas dynamics, star formation and feedback processes in the CMZ. To resolve the outflows and the dense clouds they entrain, we additionally refine dense gas (number densities $n>0.2$~cm$^{-3}$) to 3~pc in a larger cylinder of 800~pc radius and 2~kpc height. The simulation is evolved for a total of 400~Myr.

Our model includes radiative cooling, treated with an exact integration scheme \citep{2009ApJS..181..391T}, using a temperature-dependent cooling coefficient $\Lambda(T)$ based on \citet{Koyama+02} for $T<10^4$~K and on \citet{1993ApJS...88..253S} for $T>10^4$~K. We also include photoelectric heating from far-ultraviolet radiation emitted by massive young stars, as well as heating from ionizing cosmic rays, both treated using an explicit scheme.
The magnetic field is initially introduced in the Galactic disc with a strength of $\sim$0.1~$\mu$G in the CMZ. It is rapidly amplified by several orders of magnitude through turbulence, shear, and the resulting dynamo processes in the central region \citep[e.g.,][]{2022ApJ...924...26S}.

Star formation and stellar feedback are also included and provide the mechanism responsible for launching the outflows. Star formation is modelled using sink N-body particles that form in dense gas and gradually accrete material until reaching 100~$\Msun$. A gas cell is considered star-forming when it has a temperature below $10^4$~K, a negative velocity divergence, a mass exceeding the Jeans mass, and a dynamical time longer than its cooling time. When these conditions are met, the cell transfers a fraction of its mass to the nearest accreting sink particle, following the Schmidt law \citep{1959ApJ...129..243S} with a star formation efficiency per free-fall time of 10\%. Once a sink particle reaches 100~$\Msun$, it becomes locked and can no longer accrete. After 6.5~Myr, it releases $10^{51}$~erg of energy at its current position, corresponding to a single SN event. Following the prescription of \citet{2017ApJ...846..133K}, the energy is injected as a combination of thermal and kinetic energy, or purely as kinetic energy, depending on whether the Sedov-Taylor phase of the remnant is resolved. This SN feedback drives the gas outflows, whose rates closely follow the star formation activity in the model.

The simulated CMZ therefore evolves self-consistently through cycles of gas accretion, star formation, and feedback. Gas is continuously supplied through the dust lanes \citep{1992MNRAS.259..345A}, feeding the CMZ and increasing its gas density and star formation rate (SFR). The SFR then varies in cycles of a few tens of Myr, reaching $\sim$1~$\msunyr$ during starbursts and declining to $\sim$0.1~$\msunyr$ during quieter periods, close to the observed value of $0.07^{+0.08}_{-0.02}~\msunyr$ of today's CMZ \citep{Henshaw2023}. These cycles are self-regulated by stellar feedback: the strong SN activity following a starburst suppresses gas accretion through the dust lanes, while outflows and star formation deplete the CMZ gas reservoir. As the SFR declines, SN activity weakens, allowing gas accretion to resume and initiating a new cycle.

\subsection{Cool nuclear outflows}
\label{sec:hi_clouds}

\begin{figure}
\includegraphics[width=0.5\textwidth]{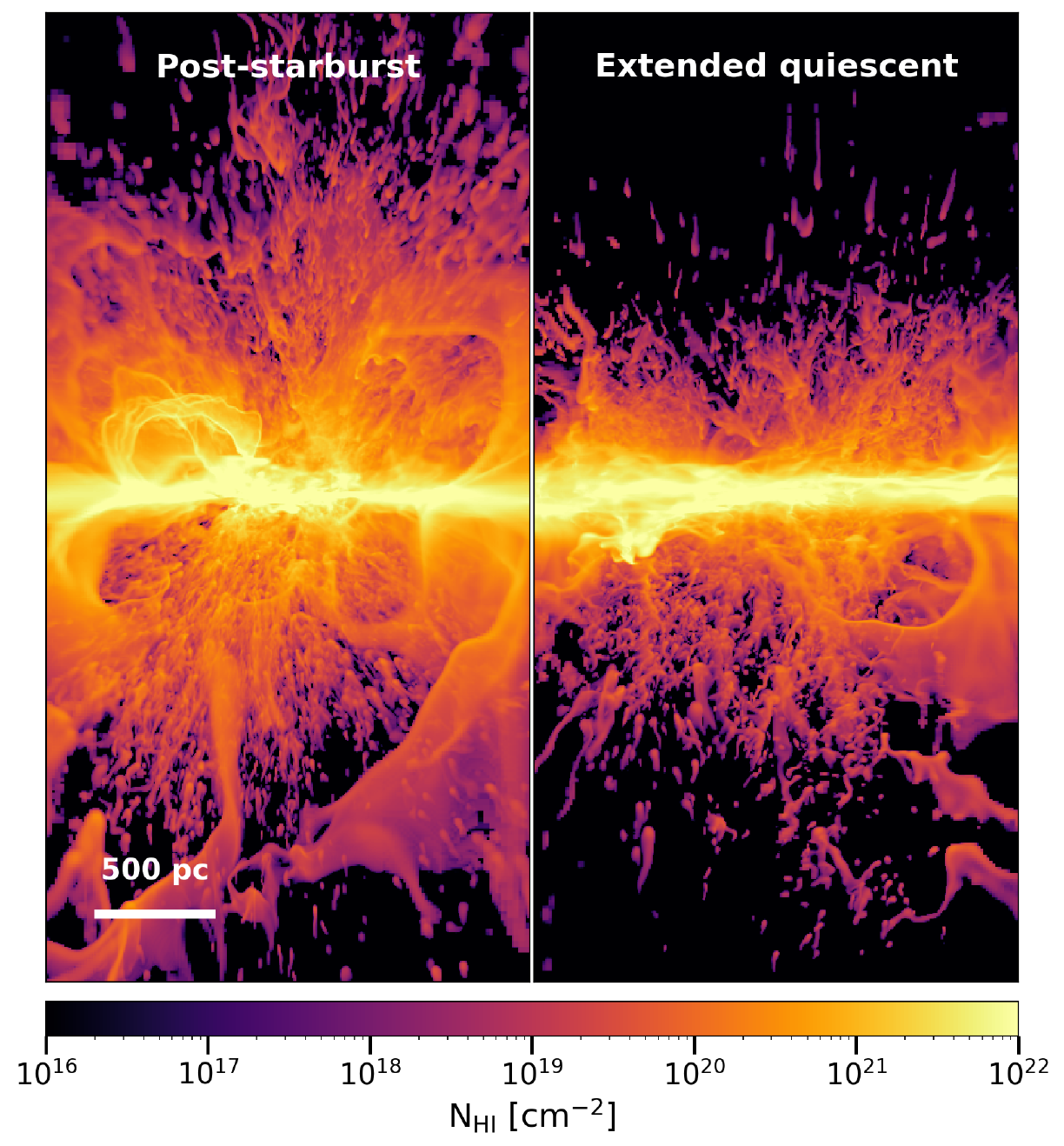}
\caption{\hi\ column density of our {\scshape Piernik} simulated MW centre seen edge-on, for the two specific periods analysed in this paper: the post-starburst time, immediately following a strong starburst (left), and the extended quiescent time, following a long period of low SF  (right). Both epochs have a $\mathrm{SFR}\simeq0.1~\msunyr$, similar to the SFR of today's CMZ.}
\label{fig:hiproj_comp}
\end{figure}

The outflows produced by SN feedback in our {\scshape Piernik} model are multiphase, with two dominant components: hot ($T>10^6$~K), diffuse gas accelerated to velocities of $\sim $1000~$\kms$, and cooler ($T<2\times10^4$~K), denser gas expelled from the disc at velocities of order $100~\kms$. The latter consists of numerous cool, dense clouds lifted from the disc through entrainment in the hot outflow, reaching heights of 1–2~kpc above the midplane. These two phases can therefore be understood as a highly-ionized, energetically-dominant hot wind that accelerates a cool neutral phase in the form of \hi\ clouds and filaments, carrying most of the outflowing mass.

For our analysis, we considered only resolved clouds, defined as those with more than four cells per cloud radius \citep{Leary+2026}. These clouds are elongated along the wind direction, with their elongation increasing with distance from the disc. Although they are typically under-pressured relative to the surrounding hot gas, internal turbulence provides additional support against the external pressure, while the magnetic field plays only a negligible role. Nevertheless, few clouds reach heights greater than 2~kpc, and only $\sim$20\% eventually fall back in a fountain flow, indicating that most clouds are destroyed during their ascent through the hot nuclear wind.
The cloud population varies substantially with the level of star formation. Starburst phases produce stronger outflows that accelerate more numerous \hi\ clouds, which are also more massive, denser, faster, and more turbulent. In contrast, quiescent periods produce weaker outflows and fewer, less massive and slower \hi\ clouds.

In the present paper, we focus on two epochs at which the CMZ SFR matches the observed value of $\sim$0.07~$\msunyr$ \citep{Henshaw2023}: the post-starburst phase ($t \sim 191$~Myr) and an extended quiescent period ($t \sim 210$~Myr, referred to as the first quiet time in Paper I), shown in Fig.~\ref{fig:hiproj_comp} using \hi\ column density maps integrated over 2~kpc across the GC. The post-starburst phase follows an intense burst of star formation that occurred $\sim$10~Myr earlier. Although the SFR has already declined to $\sim$0.1~$\msunyr$, the supernova energy rate remains high ($\sim5\times10^{48}$~erg~yr$^{-1}$) because of the delay between star formation and supernova explosions (see Section~\ref{sec:simu}). Consequently, the outflows remain powerful, producing a large population of outflowing \hi\ clouds (739, excluding unresolved clouds) that reach heights of up to $\sim$2~kpc above the GC with typical velocities of $\sim$200~$\kms$, as visible in the left panel of Fig.~\ref{fig:hiproj_comp}. By contrast, the quiescent (or quiet) phase corresponds to a prolonged period of low star formation, with a similar SFR ($\sim$0.1~$\msunyr$) but a substantially lower supernova energy rate ($\sim1.5\times10^{48}$~erg~yr$^{-1}$). The resulting outflows are therefore weaker, producing fewer \hi\ clouds (543), with typical velocities of $\sim$100~$\kms$ and generally reaching lower heights (see right panel of Fig.~\ref{fig:hiproj_comp}).

We refer the reader to Paper I for a detailed discussion of the temporal evolution of cloud properties in our simulation. The two epochs considered here are particularly well suited for comparison with GC observations because they both reproduce the observed SFR while representing significantly different outflow states. 

\subsection{Identification and characterization of \hi\ clouds}
\label{sec:bbaroloPPP}

\begin{figure*}
\includegraphics[width=1.0\textwidth, trim = 4.1cm 0cm 5.8cm 2.5cm,clip]{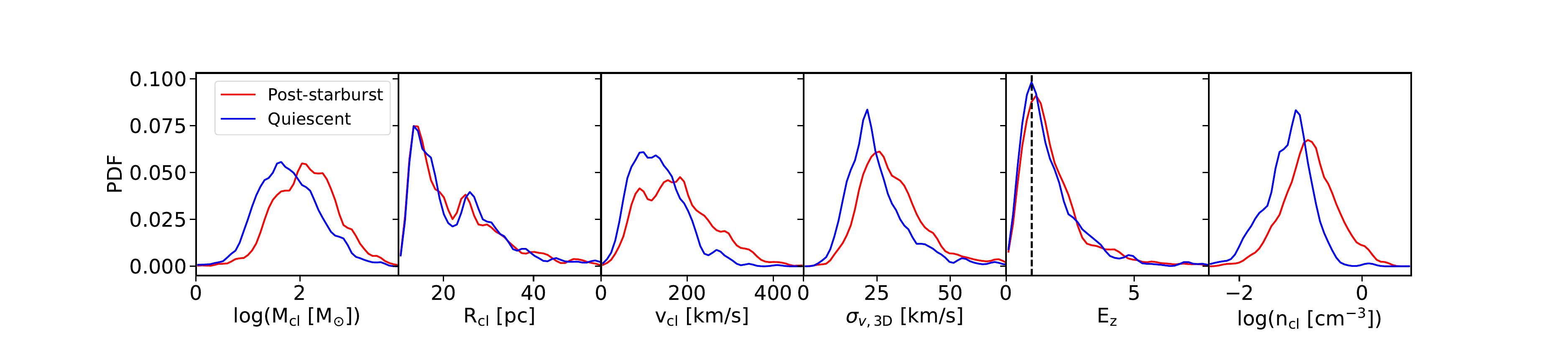}
\caption{Probability density functions (PDFs) of intrinsic cloud properties, smoothed with a Gaussian-weighted moving average over 9 bins. We compare the post-starburst time (red lines) to the extended quiescent time (blue lines). From left to right: masses, radii (in the spherical approximation), total velocities, 3D velocity dispersions, vertical elongations, and mean volume densities. The PDFs show the number of counts per bin, with 50 bins in each panel. The dashed vertical line indicates a vertical elongation $E_z=1$, representing roughly spherical clouds; for $E_z>1$. clouds are elongated in the vertical direction, while $E_z<1$ indicates clouds that are preferentially elongated in a different direction. Clouds at the post-starburst time typically have larger masses, velocities, velocity dispersions, densities, and are slightly more elongated.}
\label{fig:comp_times}
\end{figure*}

We approximate the \hi\ gas content in our simulation by assuming that all gas with $T<10^4$~K is neutral hydrogen, while for gas with $T>10^4$~K we derive the \hi\ fraction from the electron-fraction tabulated values provided by \cite{1993ApJS...88..253S} for solar metallicity (see Paper~I for more details).
To identify and isolate the resulting \hi\ clouds in the nuclear outflowing region of our model, we use the 3D source finder of the $^{\rm 3D}$\textsc{Barolo} code \citep{2015MNRAS.451.3021D}, which identifies and reconstructs objects in a three-dimensional position-position-velocity (PPV) dataset. 
For this purpose, we constructed position-position-position (PPP) data cubes of the \hi\ gas in a volume $2 \times 2 \times 4$~kpc around the MW centre and repurposed the third spatial axis (perpendicular to the mid-plane) as a pseudo-velocity axis, allowing the source finder to be directly applied to the simulation data. This approach allowed us to derive various cloud properties, which we hereafter refer to as ``intrinsic'' properties, since they are measured directly from the full three-dimensional morphology and kinematics of the clouds. This contrasts with properties derived from PPV data cubes (see Section~\ref{sec:mockdata}), where the clouds are projected onto a two-dimensional spatial plane representing the sky, with their velocities projected along the line of sight (LOS).

The source finder identified 739 clouds at the post-starburst time and 543 clouds at the quiescent time. These detections constitute the simulated parent samples used throughout the rest of the paper.
Figure~\ref{fig:comp_times} shows the probability density functions (PDF) of the \hi\ cloud mass, radius, total velocity, 3D velocity dispersion, vertical elongation, and volume density for the resolved clouds at the post-starburst time (red lines) and the extended quiescent time (blue lines). The radius is computed assuming spherical symmetry, $R_\mathrm{C} = (\frac{3}{4} V / \pi)^{1/3}$, where $V$ is the cloud volume, while the velocity is the norm of the 3D velocity vector, density-weighted averaged within each cloud. We derive the cloud elongation from the density-weighted covariance matrix of the cell positions relative to the cloud centroid. The diagonal elements give the variances along the Cartesian $x$, $y$ and $z$ directions, $\sigma_x^2$, $\sigma_y^2$ and $\sigma_z^2$, which characterize the cloud's extent along each direction. We define the vertical elongation as $E_{z}=2\sigma_{z}/(\sigma_{x}+\sigma_{y})$, such that a spherical cloud has $E_{z}=1$, while a cloud elongated in the vertical direction has $E_{z}>1$. 

Most clouds have masses between $\sim$10 and $\sim$1000~$\Msun$, with the distribution peaking around 100~$\Msun$, while radii extend up to $\sim$50~pc and peak around 15~pc. Velocities reach up to 400~$\kms$, although most clouds have velocities of 100--200~$\kms$, while their velocity dispersions peak around 25~$\kms$ and extend up to 50~$\kms$. The vertical elongation distribution peaks around $E_z=1$, but most clouds have $E_z>1$, indicating a preferential elongation in the vertical direction, as expected for cold clouds being accelerated by a hot, vertically directed wind. Finally, average cloud densities span $10^{-2}$ to 1~cm$^{-3}$, with the distribution peaking around 0.1~cm$^{-3}$.
As mentioned in Section~\ref{sec:hi_clouds}, the higher supernova activity during the post-starburst phase produces a larger cloud population, with clouds that are typically more massive, faster, more turbulent, more elongated, and denser.

\section{Mock \hi\ observations}
\label{sec:mockdata}

Cool neutral gas in the MW's nuclear outflow has been studied in recent years thanks to the detection of the \hi\ emission line. 
The deepest and most effective surveys of the GC region have used the single-dish Green Bank Telescope (GBT) \citep{DiT+18,Lockman+2020,Bordoloi+2025,Lockman+2026}.
In Paper~I, we showed that the properties of our simulated clouds broadly agree with those of the \hi\ population presented in \citet{DiT+18}, in terms of \hi\ mass, position, velocity, velocity dispersion, and radius. However, the cloud properties compared with observations in Paper~I were intrinsic (as defined in Section~\ref{sec:bbaroloPPP}), meaning that they were derived directly from the simulation in an idealised setting. In contrast, observed \hi\ clouds can only be characterised from the Sun's position and are affected by projection and instrumental effects. To enable a more direct comparison with the GBT observations of the GC, in this section we investigate how observational effects influence the properties of \hi\ clouds in our model by constructing mock radio observations.
 
\subsection{PPV \hi\ cubes}
\label{sec:PPVcubes}

\hi\ emission-line observations are typically stored as PPV data cubes, where the first two axes represent the position on the plane of the sky (e.g., Galactic or equatorial coordinates) and the third axis (channels) represents the spectral dimension, expressed in frequency or LOS velocity.
Each channel map shows the spatial sky distribution of the gas at a given LOS velocity, while each spatial pixel contains an emission-line spectrum along the spectral axis.
To fairly compare our simulated clouds with the available \hi\ data of the GC, we create \hi\ PPV data cubes that mimic the observational setup of the GBT data by \citet{DiT+18} and \citet{Lockman+2026}.

First, we place the observer at a distance of $R_0 =8.2$~kpc from the GC and at an angle of $30^\circ$ with respect to the bar, corresponding to the position of the Sun \citep[e.g.][]{Bland-Hawthorn+2016,2020RAA....20..159S}. The GBT data cubes are defined in Galactic longitude $\ell$ and latitude $b$, with the spectral dimension given by the Local Standard of Rest (LSR) velocity, $\vlsr$. 
We therefore project the simulation's Cartesian $x$--$y$--$z$ spatial grid onto the $\ell$--$b$ coordinate system and transform the three-dimensional velocity field into the LSR frame (see, e.g., equation 9 in Paper~I), assuming a solar circular velocity around the GC of $V_\odot=240~\kms$ \citep[e.g.,][]{Bland-Hawthorn+2016}. To match the pixel and spectral sampling of the GBT data cubes, we grid our simulation onto a PPV grid with spatial pixels of 115$''$ and velocity channels of 3~$\kms$.
For each $\ell$-$b$ pixel, we divide the corresponding LOS into radial bins of 20~pc. Within each bin, we calculate the mass-weighted LSR velocity and the integrated \hi\ column density $N_{\hi}$ from the \hi\ gas present in that bin. We then convert to peak \hi\ brightness temperature under the assumption that the line emission from that bin is Gaussian and optically thin, i.e.\ $T_\mathrm{HI,peak} = N_{\hi}/(1.82\times 10^{18}\,\sigma_\mathrm{v}\sqrt{2\pi})$ in K \citep[e.g.,][]{Roberts+1975}, where $\sigma_\mathrm{v}$ is the velocity dispersion of the \hi\ gas within the bin in $\kms$ and $N_\hi$ is in cm$^{-2}$. We account for thermal broadening by defining the velocity dispersion as $\sigma_\mathrm{v}=({\sigma_\mathrm{v,LOS}^2+c_s^2})^{\frac{1}{2}}$, where $c_s$ is the sound speed and $\sigma_\mathrm{v,LOS}$ is the velocity dispersion along the LOS.
We finally construct the emission-line spectrum $T_\hi(\vlsr)$ along each LOS by adding the emission contributions from all radial bins. For each bin, the emission is represented by a Gaussian centred on its LSR velocity, with an amplitude given by its peak emission and standard deviation given by its velocity dispersion. The resulting sum of all Gaussians is then sampled onto the velocity grid of the PPV cube.

In our mock PPV cubes, we include only the \hi\ gas belonging to the clouds identified in the PPP data cubes, thereby excluding gas that was not identified as part of a cloud. While this is a strong assumption, it avoids contamination from foreground and background gas at low resolution, which typically appears as large, unresolved structures in the $\ell-b$ plane. 
When comparing to real data (Section~\ref{sec:comp_obs_hi}), we nevertheless account for the foreground and background contamination present in the observations by removing clouds from our simulated sample based on their deviation from the expected rotational motion.

\subsection{Observational effects: spatial resolution and sensitivity}
\label{sec:obs_bias}

In this work, we mainly focus on two observational effects that we expect could significantly affect the measured properties of our \hi\ clouds and their physical interpretation: beam smearing and noise. 
Since we compare our simulated cloud population with observations from the GBT, we account for these effects by adopting spatial resolution and noise levels representative of the GBT observations.
    
Beam smearing arises from the finite spatial resolution of a telescope, whereby the emission is convolved with the telescope's point-spread function, or beam. As a result, emission from neighbouring spatial regions is blended, smoothing small-scale structure and potentially altering the observed spatial and kinematic properties of clouds.
To introduce artificial beam smearing in our simulation, after constructing our PPV data cube, we convolve each channel map with a Gaussian kernel with a full width at half maximum (FWHM) of 570$''$, corresponding to the GBT beam size and equivalent to $\sim$22.6~pc at the GC.

Unlike idealised simulations, observational data are inevitably affected by instrumental noise, which limits the sensitivity to faint emission and can cause low-intensity structures to become undetectable, thereby affecting the measured properties of the observed \hi\ clouds.
Because the GBT survey was carried out over several years, with different integration times across different regions, the root-mean-square (rms) noise level varies across the survey area, ranging from 10 to 90~mK per 3~$\kms$ channel \citep[see Fig.~1 in][]{Lockman+2026}.
To account for this effect in our synthetic observations, we add Gaussian random noise with an rms amplitude of $\sigma_\mathrm{rms}=25$~mK, chosen to be representative of the noise level over large portions of the GBT survey.
Although this prescription cannot reproduce the full spatial variation of the GBT observational noise, it provides a typical level of noise for assessing its impact on the measured properties of the simulated \hi\ clouds.
As in real observations, the noise we add is correlated on the spatial scale of the beam: in all cases considered in this work, the noise is generated as Gaussian random noise, convolved with the $570''$ beam, and rescaled to $\sigma_\mathrm{rms}=25$~mK per channel.

\subsection{Source finding in the PPV domain and cloud properties}
\label{sec:bbaroloPPV}

As for the cloud identification in PPP space described in Section~\ref{sec:bbaroloPPP}, we use the source finder of $^{\rm 3D}$\textsc{Barolo} to reconstruct clouds in the PPV space. 
Before directly comparing our simulated cloud population with observations in Section~\ref{sec:comp_obs_hi}, we aim to quantify the impact of projection and observational effects on the measured properties of \hi\ clouds (Section~\ref{sec:impact_bias}). To this end, it is desirable to establish a one-to-one correspondence between each cloud identified in the PPV cube and its parent cloud in PPP space.
However, when all clouds are placed in a single PPV data cube, distinct clouds can overlap in projected sky position and/or velocity and become blended into a single structure. 
Depending on the observational parameters and source finding criteria, the algorithm may identify such a blended structure as a single source, even though it consists of multiple physically distinct objects, or split it into multiple sources that do not correspond to the original objects. The resulting measurements may therefore either combine the intrinsic properties of multiple clouds or assign different properties to fragments that do not represent the original clouds, making meaningful one-to-one comparisons with the intrinsic cloud properties difficult. 

To avoid this LOS blending and allow a proper quantification of the impact of observational effects on cloud properties, in Section~\ref{sec:impact_bias} we isolate each cloud before running the source finder. 
Specifically, we generate a separate PPV data cube for each cloud in both samples (post-starburst and extended quiescent times) and run the source finder individually on each cube.
This allows the algorithm to characterize each cloud without contamination from neighbouring clouds along the LOS.
In contrast, when comparing our simulated clouds with observational data in Section~\ref{sec:comp_obs_hi}, we place the full population in a single PPV cube, thereby also accounting for the effects of LOS blending.

In the source finding procedure, we always use a primary signal-to-noise ratio (S/N) threshold of 4 and a secondary ``growth'' S/N threshold of 2.5. The primary S/N threshold is used to identify the peak emission region of a detection, while the secondary threshold defines the extent to which the source is grown around the peak.
We reject all detections spanning fewer than two velocity channels (i.e. 6~$\kms$) or fewer than fifteen voxels in $\ell-b-\vlsr$ space, as such detections are unlikely to represent genuine \hi\ clouds and are instead more likely to correspond to spikes in the noise distribution.

Once clouds are identified, we measure and analyse six properties that are commonly calculated from observed \hi\ clouds:
\begin{enumerate}
\item The \hi\ mass $M_\hi = 0.236\, D^2 S_\hi~\Msun$, where $D$ is the distance of the cloud in kpc and $S_\hi$ its integrated flux in Jy~$\kms$. 
\item The cloud radius, derived under the spherical approximation as $R_\mathrm{C} = \sqrt{A/\pi}$, where $A$ is the area covered by the cloud on the sky.
\item The average \hi\ column density over the projected radius, $N_\hi = M_\hi/(\pi R_\mathrm{C}^2)$.
\item The centroid LSR velocity, $v_{\mathrm{LSR}}$, defined as the flux-weighted velocity average of the integrated spectrum.
\item The LOS velocity dispersion, $\sigma_\mathrm{v}$, taken as the flux-weighted second moment of the integrated spectrum.
\item The vertical elongation in the latitude direction, $E_{b}$. Similarly to the vertical elongation derived in the 3D space (Section~\ref{sec:bbaroloPPP}), we define the vertical elongation in the $\ell-b$ space by using the corresponding standard deviations $\sigma_{\ell}$ and $\sigma_b$ so that $E_b=\sigma_b/\sigma_\ell$.
\end{enumerate}

As in observational studies, masses and radii are derived assuming that every cloud lies at the distance of the GC, $D \equiv R_0=8.2$~kpc. For a cloud at a true heliocentric distance $d$, the inferred mass and radius therefore differ from their true values by factors $(R_0/d)^2$ and $R_0/d$, respectively. For clouds within $\pm1$~kpc of the GC along the LOS, this amounts to $\lesssim30\%$ in mass and $\lesssim15\%$ in radius (overestimates for clouds in front of the GC, underestimates for those behind it). Column densities and other quantities are unaffected.


 \section{The impact of observational effects on HI clouds}
  \label{sec:impact_bias}

In this section, we investigate the impact of observational effects on the measured properties of our simulated \hi\ clouds. 
Because we are interested in a one-to-one comparison of measured cloud properties, in this section we always generate a separate PPV data cube for each cloud identified in the PPP data cube.
We first define a reference ``ideal'' case with negligible noise and no beam smearing. 
Beam smearing and noise are then introduced separately, comparing each case to the ideal one in order to isolate their individual impact, before considering both effects together. 
We first focus on the clouds detected during the post-starburst period and discuss the corresponding results for the extended quiescent period in Section~\ref{sec:results_quiet}.

\subsection{Ideal case: projection effects}
\label{sec:ref_sample}

\begin{figure}
\includegraphics[width=0.48\textwidth, trim = 0.0cm 0 0 0, clip]{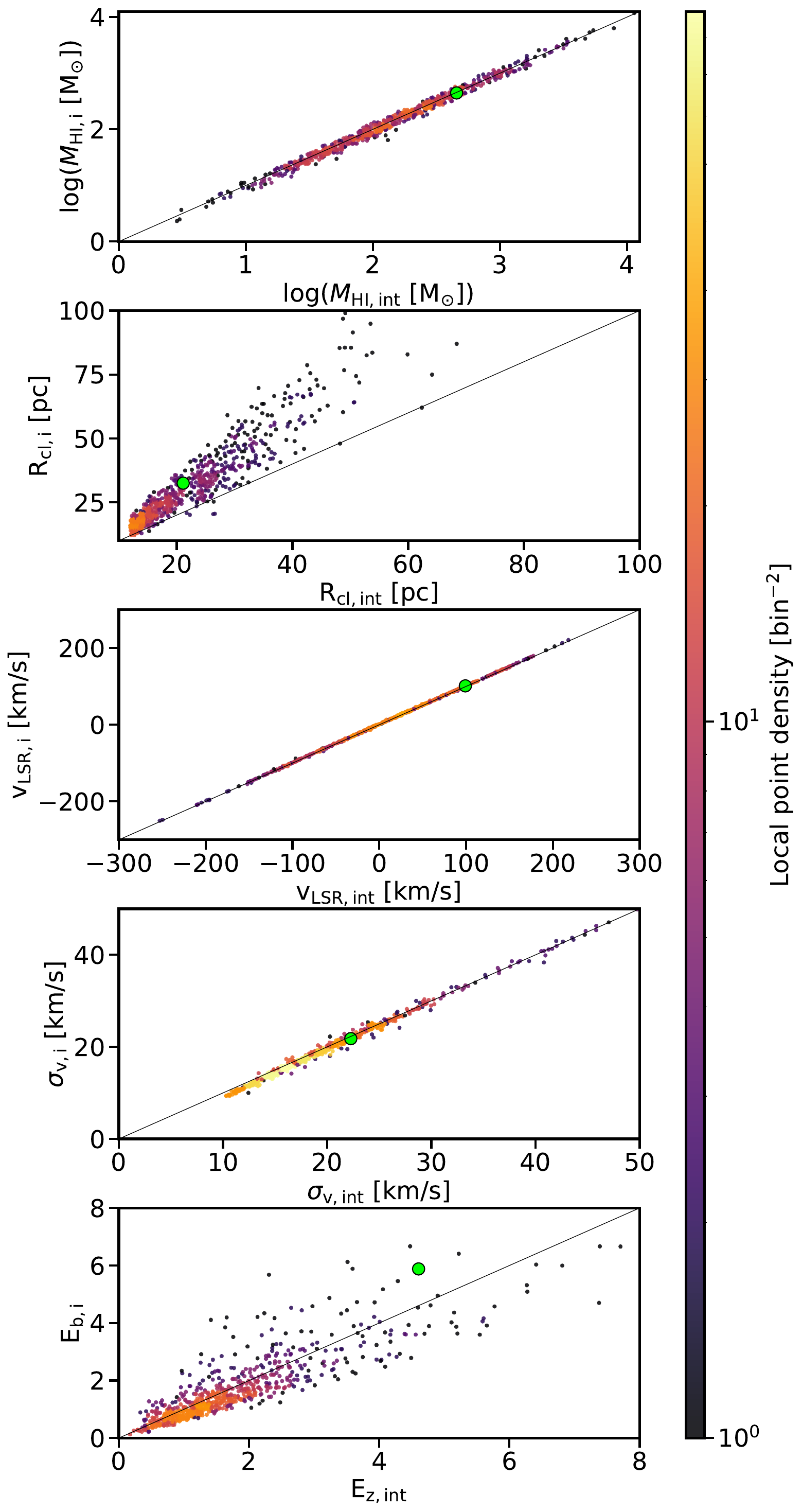}
\caption{Cloud properties at the post-starburst time in ``ideal" observational conditions, i.e. projected onto the sky with neither noise nor beam smearing, as a function of their intrinsic properties. From top to bottom: cloud mass, radius (in the spherical approximation), LSR velocity, LOS velocity dispersion, and vertical elongation. In each panel a thin black line shows the identity function. The colour scale shows the local point density with $50 \times 50$ bins in each panel, and the green dot shows the properties of cloud 398, shown as an illustrative example in Fig.~\ref{fig:1cloud_example}. Projecting clouds onto the sky plane leads to higher measured radii and adds scatter to the vertical elongation.}
\label{fig:corr_3d_ideal}
\end{figure}

\begin{figure*}
\includegraphics[width=1\textwidth, trim = 0.0cm 0cm 0.8cm 0.3cm, clip]{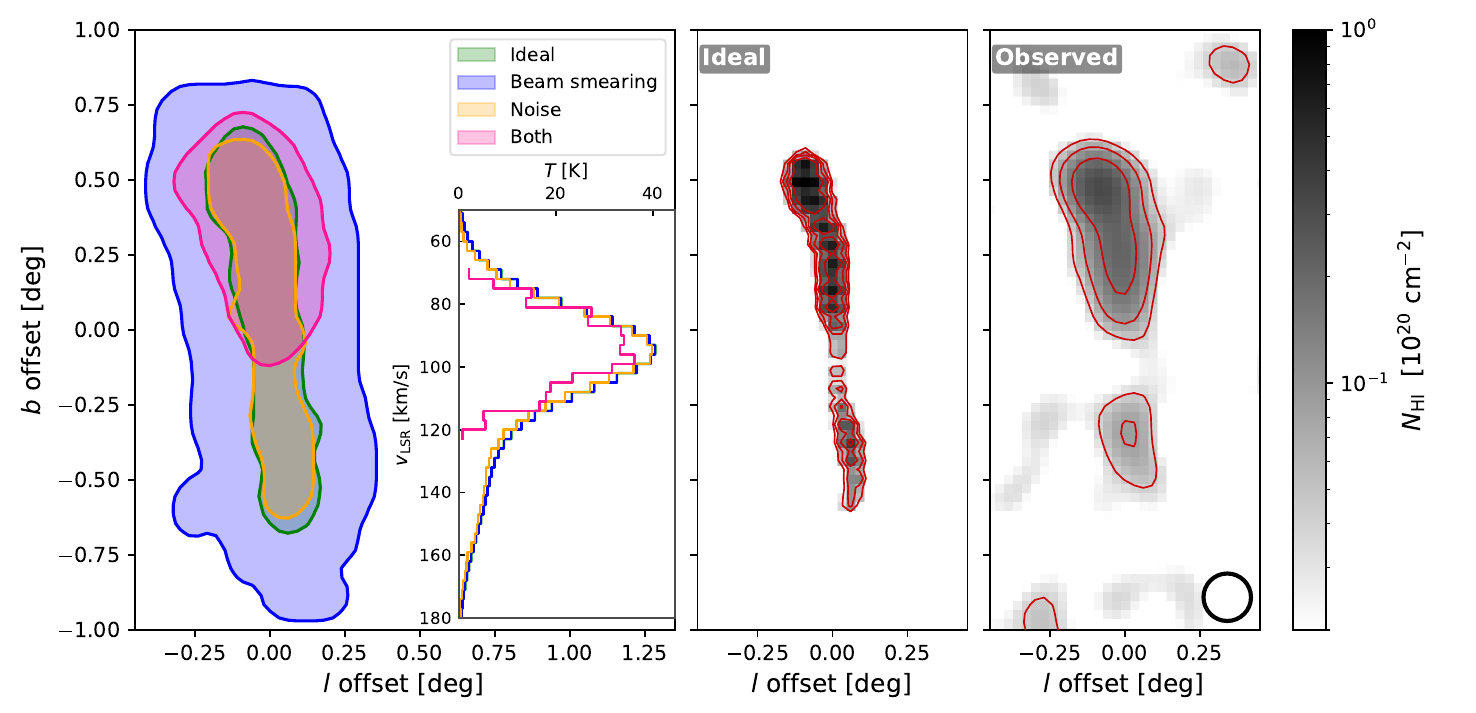}
\caption{Illustrative example of simulated cloud 398 under various observational conditions. The left panel illustrates the change in the morphology of the cloud projected on the sky, showing the ideal case (neither noise nor beam smearing, green), only beam smearing (blue), only noise (yellow), and both combined (pink). 
The inset shows the corresponding integrated spectra (ideal and beam smearing cases show identical spectra). Beam smearing bloats the cloud, while noise removes the lower density parts; combining both leads to a truncated, rounder cloud. 
The middle and right panels show the corresponding \hi\ column density maps around cloud 398 for the ideal case and the ``observed'' case (beam smearing + noise), respectively. Contour lines correspond to $n\times\sigma_{N_{\hi}}$, where $n=2,4,8,16$ and $\sigma_{N_{\hi}} = 2.1 \times 10^{18}$ cm$^{-2}$ is the column density rms noise of the observed map. On the right panel, the black circle shows the FWHM of the Gaussian beam (570$''$).}
\label{fig:1cloud_example}
\end{figure*}

The ``ideal'' case only includes the effect of projecting the clouds from the 3D space to the observational $\ell-b-\vlsr$ space. Therefore, we do not include beam smearing and we adopt a negligible noise level of 0.1~mK. 
Because the source finder identifies emission based on S/N thresholds, the input data cube must contain a non-zero noise level for the algorithm to distinguish emission from noise.
In general, running the source finder on the PPV cube of an individual cloud can lead to three possible outcomes:
\begin{enumerate}
\item a single cloud is identified;
\item multiple spatially and/or spectrally distinct clouds are identified;
\item no cloud is detected.

\end{enumerate}

The second outcome complicates a direct one-to-one comparison between the intrinsic cloud properties and those recovered from the PPV cubes, since a single PPP cloud no longer corresponds to a single detected object. Applying this procedure to the 739 clouds identified during the post-starburst phase, we recover 728 clouds in case~(1), 11 in case~(2), and none in case~(3) because the noise is negligible. 
These 739 clouds define our ideal sample, which serves as the reference population when progressively introducing observational effects in the following sections. When comparing individual cloud properties, we consider only the 728 clouds in case~(1), whereas all 739 clouds are included when comparing distributions.

Beyond providing a reference sample, the ideal population also allows us to quantify an often overlooked observational bias arising purely from projection effects. Because three-dimensional clouds are observed as two-dimensional structures on the plane of the sky, some of their inferred properties may differ from their intrinsic values. Figure~\ref{fig:corr_3d_ideal} compares the properties recovered from the PPV cubes with the intrinsic properties measured directly from the PPP cubes on a cloud-by-cloud basis. The velocities considered in both cases are LSR velocities. To allow for a better comparison with the other cases, we include the contribution of thermal broadening also in the velocity dispersions of the PPP clouds, using $\sigma_\mathrm{v}=(\sigma_\mathrm{v,LOS}^2+c_s^2)^{\frac{1}{2}}$, where $c_s$ is the sound speed based on the average temperature within each cloud. Note that here $\sigma_\mathrm{v,LOS}$ are not the velocity dispersions shown in Fig.~\ref{fig:comp_times}, which are the 3D dispersions and are therefore expected to be larger by a factor of approximately $\sqrt{3}$ for an isotropic gas. 

Overall, the cloud masses (first panel in Fig.~\ref{fig:corr_3d_ideal}), velocities (third panel) and velocity dispersions (fourth panel) are remarkably well recovered after projection.
This is not surprising: the velocities and velocity dispersions are derived from the same LOS in both cases, while projection conserves the total flux and therefore the mass.
The small residual scatter in the masses arises from assuming the same distance for all clouds (Section~\ref{sec:bbaroloPPV}).
Instead, the agreement is poorer for quantities that depend more strongly on the cloud morphology observed along a particular LOS, such as cloud radii (second panel) and elongations (fifth panel). Clouds appear systematically larger in projection than in three dimensions, with this effect becoming stronger with radius. 
This effect arises because the largest clouds tend to be more elongated in the vertical direction, i.e. perpendicular to the LOS. Under the spherical approximation, such geometries lead to smaller intrinsic radii than those inferred from their projected areas. As an example, consider an extreme cuboidal cloud with dimensions $10\times10\times100$~pc$^3$. Its volume of $10^4$~pc$^3$ corresponds to a spherical radius of only 13.4~pc, whereas its projected area in the $x-z$ plane is $10^3$~pc$^2$, corresponding to a projected radius of 17.8~pc. Projection therefore tends to overestimate the sizes of clouds elongated perpendicular to the LOS, as is typically the case for GC outflows.

Cloud elongations are recovered with a considerable scatter. This partly reflects the fact that the Galactic latitude direction does not exactly coincide with our Cartesian $z$ direction, but is likely also due to variations in cloud inclination with respect to the LOS. As shown in Paper~I, although clouds are preferentially elongated along the poloidal direction, their individual inclinations can differ substantially.

\subsection{Beam smearing}
\label{sec:beam_smearing}

\begin{figure*}
\centering
\includegraphics[width=1\textwidth, trim = 3.cm 2cm 1cm 4cm, clip]{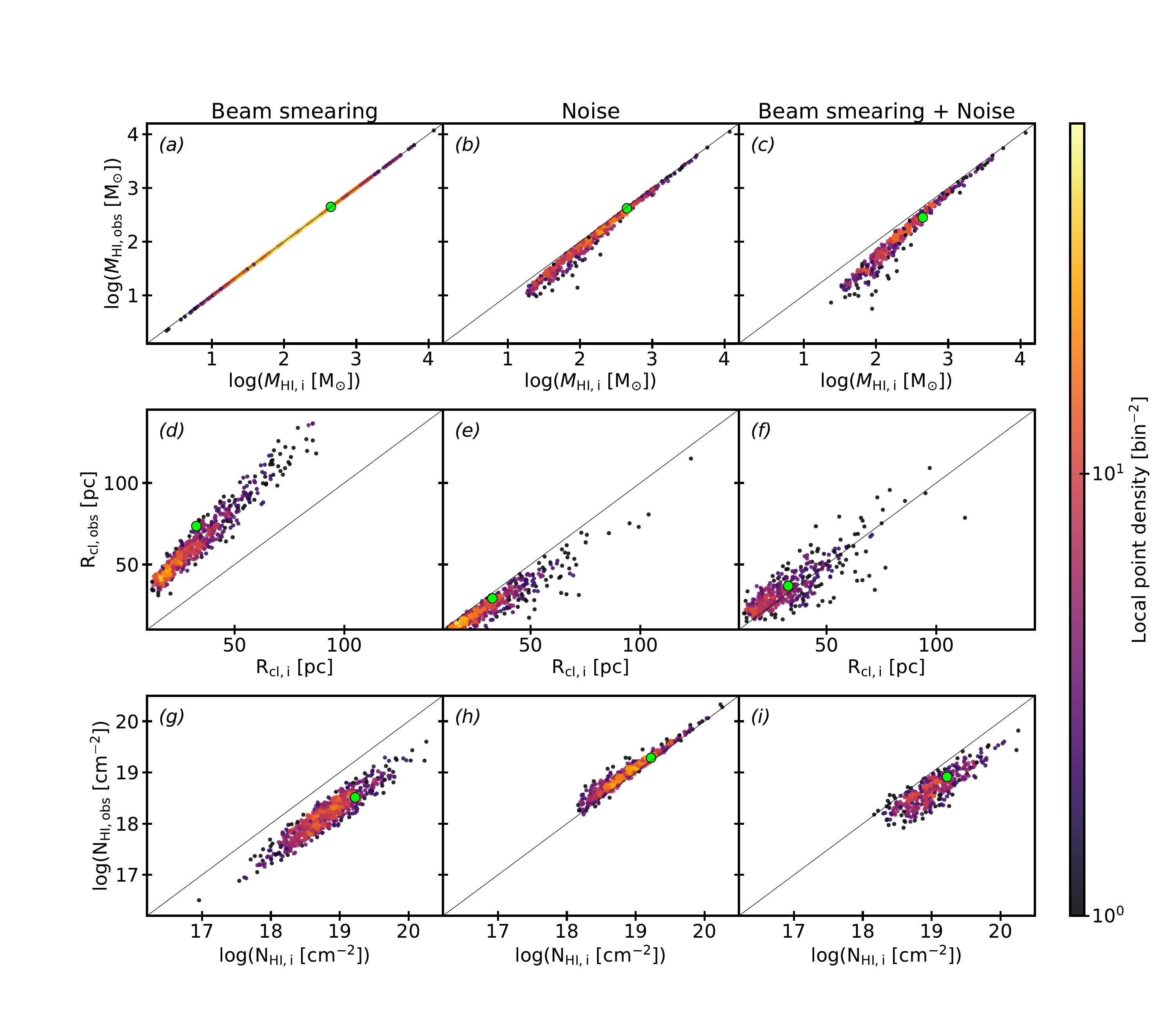}
\caption{Cloud properties at the post-starburst time when adding observational effects. From left to right: beam smearing only, noise only, and both combined, as a function of their properties in the ideal case (projection only, no beam smearing nor noise). From top to bottom: cloud masses, radii (in the spherical approximation), and column densities. Thin black lines show the identity functions. The colour scale shows the local point density with $50 \times 50$ bins in each panel, and the green dot shows the properties of cloud 398, shown as an example in Fig.~\ref{fig:1cloud_example}.}
\label{fig:all_correlations1}
\end{figure*}

\begin{figure*}
\centering
\includegraphics[width=1\textwidth, trim = 2.5cm 2cm 1cm 4cm, clip]{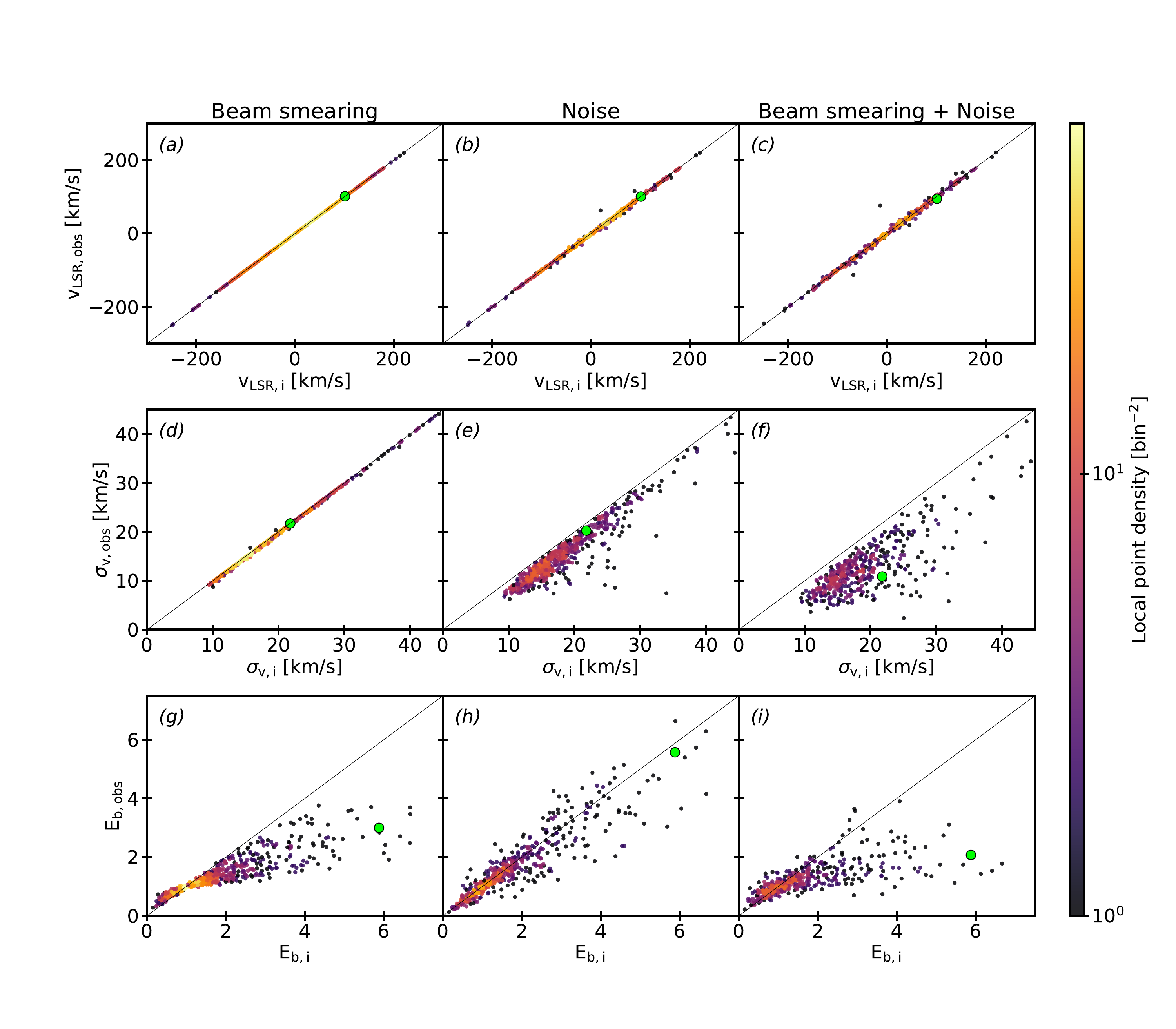}
\caption{Same as Fig.~\ref{fig:all_correlations1}, but showing the LSR velocities, velocity dispersions, and vertical elongations.}
\label{fig:all_correlations2}
\end{figure*}

We now isolate the impact of spatial resolution by applying the same procedure as in the previous section, but adding a Gaussian smoothing of 570$''$, as described in Section~\ref{sec:obs_bias}, while keeping the noise level negligible (0.1~mK). As expected, the smoothing spreads and blurs the cloud emission, causing clouds to appear larger and less dense. This effect is illustrated for a representative example (cloud 398) in the left panel Fig.~\ref{fig:1cloud_example}, which shows the cloud morphology in the $\ell-b$ plane under the different observational conditions explored in this and the following sections, together with the corresponding integrated spectra, shown as an inset. The ideal case is indicated in green, while the smoothed case, which occupies a significantly larger area, is shown in blue.

Beam smearing alone has little impact on the number of clouds detected in the post-starburst population. The source finder recovers 731 single clouds, compared to 728 in the ideal case; the small difference is due to less clouds being detected as multiple detections (case~2 in Section~\ref{sec:ref_sample}) due to the smoothing. The first columns of Fig.~\ref{fig:all_correlations1} and \ref{fig:all_correlations2} compare the cloud properties with and without beam smearing, indicating the illustrative cloud 398 as a green dot. As expected, cloud masses (Fig.~\ref{fig:all_correlations1}$a$), velocities (Fig.~\ref{fig:all_correlations2}$a$), and velocity dispersions (Fig.~\ref{fig:all_correlations2}$d$) are recovered almost unchanged. Although spatial Gaussian convolution spreads the emission over larger areas, it conserves the flux in each velocity channel. As long as the source mask encloses all the smeared emission, the integrated spectrum of the cloud, and hence its total flux, velocity centroid and velocity dispersion, are therefore unchanged, even if the cloud has internal velocity gradients. Beam smearing does broaden the line profiles of individual pixels when the velocity field is not resolved \citep[e.g.,][]{2011ApJ...741...69D,2015MNRAS.451.3021D}, but this does not affect our $\sigma_\mathrm{v}$, which is measured from the integrated spectrum.
This is illustrated by the identical spectra in Fig.~\ref{fig:1cloud_example} (inset plot in the left panel), where the green line of the ideal case is hidden by the blue line of the beam-smeared case. 

In contrast, beam smearing significantly affects the inferred cloud radii, column densities, and elongations.
Cloud radii increase by approximately 80\% relative to the ideal case (Fig.~\ref{fig:all_correlations1}$d$), corresponding to an average increase in the projected cloud area by a factor of $\sim$3.2, consistent with the substantially larger extent of the cloud shown in Fig.~\ref{fig:1cloud_example}. Because the cloud mass remains essentially unchanged, this increase in radius naturally leads to lower inferred column densities (Fig.~\ref{fig:all_correlations1}$g$), by a factor of $\sim$3--5.
Beam smearing also reduces the elongation of vertically extended clouds (Fig.~\ref{fig:all_correlations2}$g$). As illustrated in Fig.~\ref{fig:1cloud_example}, the isotropic broadening introduced by the beam smooths the cloud emission in all directions, causing intrinsically elongated clouds to appear more circular. Because of this rounding effect, most clouds cluster around $E_b \sim 1$.

Figure~\ref{fig:distrib_comp} presents the resulting cumulative distributions (top panels) and PDFs (bottom panels) of cloud properties under the different observational conditions, with the ideal case shown in black and the beam-smearing-only case in blue. 
The intrinsic properties from the PPP space are also shown as dashed lines, as a reference, although, unlike in Fig.~\ref{fig:comp_times}, here we show the LSR velocity and LOS velocity dispersions (including thermal broadening) to allow for more direct comparisons. Unlike the one-to-one comparisons above, these distributions include cases where a single intrinsic cloud is identified as multiple detections. Consistent with the cloud-by-cloud comparison, beam smearing leaves the mass, velocity, and velocity-dispersion distributions essentially unchanged, while shifting the radius distribution towards larger values and the column-density distribution towards lower values. Likewise, the distribution of vertical elongations loses most of its highly elongated clouds and becomes increasingly concentrated around unity, reflecting the tendency of beam smearing to make clouds appear more spherical.
        
\subsection{Noise}
\label{sec:noise}

In this section, we isolate the impact of noise by adding a Gaussian noise with $\sigma_\mathrm{rms} = 25$~mK, as described in Section~\ref{sec:obs_bias}, while neglecting beam smearing and following the same procedure as before. Noise primarily affects the detection of low-density gas, whose emission may fall below the detection threshold of the source finder. As illustrated in Fig.~\ref{fig:1cloud_example} (left panel, yellow), this sometimes removes only the diffuse outer envelope of a cloud, leading to a modest reduction in its apparent size. Consistently, the velocity spectrum shows little change when compared to the ideal case, as shown in the inset of Fig.~\ref{fig:1cloud_example}. In other cases, however, larger portions of the cloud, or even the entire cloud, may fall below the detection threshold. Noise can also fragment individual clouds into multiple detections by removing the low-density gas connecting different cloud components. As a result, the number of multiple detection cases increases from 11 in the ideal sample to 69. Overall, only 581 clouds (79\%) are recovered as single detections, compared to 728 in the ideal sample. 

The second columns of Figs.~\ref{fig:all_correlations1} and \ref{fig:all_correlations2} compare the properties of the recovered clouds with those of their counterparts in the ideal case. Cloud \hi\ masses (Fig.~\ref{fig:all_correlations1}$b$) are generally well recovered for the most massive objects, but become systematically underestimated towards lower masses. This reflects the preferential removal of diffuse gas: in dense and massive clouds, the lost material is largely confined to the outer envelope (as illustrated in Fig.~\ref{fig:1cloud_example}, left panel), whereas in lower-density and less massive clouds, a much larger fraction of the emission can be removed, in some cases reducing the recovered mass by up to 50\%.
Furthermore, the low-mass end of the distribution, corresponding to \hi\ masses $M_\hi\lesssim10\,\Msun$, is entirely lost, as these faint clouds fall below the detection threshold and can no longer be identified (case 3 in Section~\ref{sec:ref_sample}).
The total recovered \hi\ mass in the full sample is $2.26 \times 10^5~\Msun$ (including multiple detections), compared to $2.56 \times 10^5~\Msun$ for the ideal case, which represents a mass loss of $\sim$12\%. A similar loss of faint emission due to noise was found by \cite{2022MNRAS.512.4765S} in synthetic \hi\ 21\,cm observations of the SILCC simulations \citep{2015MNRAS.454..238W}. 

The loss of diffuse emission also decreases the recovered cloud radii by approximately 20\% (Fig.~\ref{fig:all_correlations1}$e$), as the projected area above the detection threshold becomes smaller.
The combined effect of smaller radii and slightly smaller masses results in column densities that are, on average, about 30\% higher than in the ideal case (Fig.~\ref{fig:all_correlations1}$h$). This follows naturally from the preferential removal of low-density gas, which increases the mean column density of the remaining cloud.

\begin{figure*}
\includegraphics[width=1.0\textwidth, trim=1cm 1cm 0.5cm 1cm, clip]{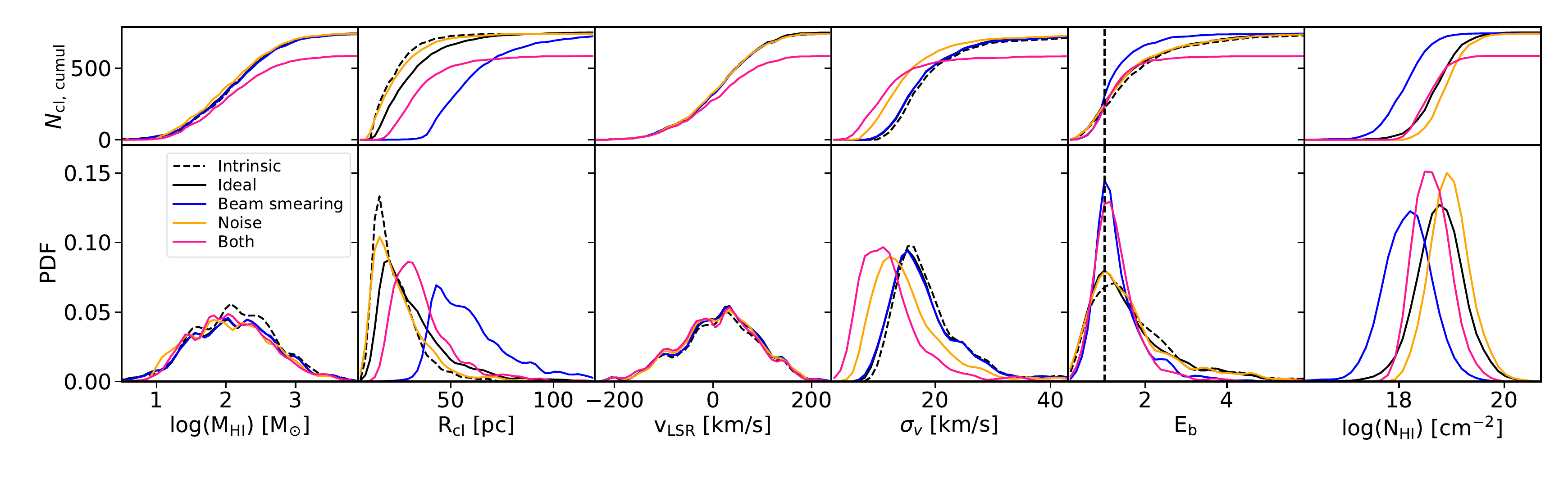}
\caption{Cloud property cumulative distributions (top row) and PDFs (bottom row), smoothed with a Gaussian-weighted moving average over 9 bins, at the post-starburst time compared under various observational conditions: ideal case (no noise nor beam smearing) in black, only beam smearing in blue, only noise in yellow, and both combined in pink. From left to right: cloud masses, radii (in the spherical approximation), LSR velocities, velocity dispersions, vertical elongations ($E_b=1$ is indicated as a dashed vertical line), and column densities. The black dashed lines show the intrinsic properties of the clouds, i.e. derived from the PPP data cube, without projection effects. The PDF represents the fraction of counts per bin, with 50 bins in each panel.}
\label{fig:distrib_comp}
\end{figure*}

As in the previous section, the LSR velocities are recovered remarkably well despite the added noise (Fig.~\ref{fig:all_correlations2}$b$). This reflects the fact that the bulk motion of the clouds is dominated by their densest, highest S/N regions, which remain reliably detected.
Instead, the recovered velocity dispersions are approximately 15\% lower than in the ideal case (Fig.~\ref{fig:all_correlations2}$e$). 
We attribute this reduction to the preferential loss of diffuse gas, which may have slightly different kinematics from the denser gas and is typically warmer, and therefore has larger thermal broadening and possibly stronger turbulent motions. Both effects broaden the line emission, so that removing this gas narrows it.
Finally, the vertical elongations are also recovered reasonably well, albeit with some scatter (Fig.~\ref{fig:all_correlations2}$h$). This scatter is caused by morphological changes induced by the removal of diffuse cloud material, although noise does not appear to introduce a significant systematic bias in the inferred elongations with respect to the ideal case.

The resulting distributions of cloud properties are shown in Fig.~\ref{fig:distrib_comp} as yellow lines. Noise has two distinct effects on the cloud mass distribution: it removes the lowest-mass ($<10~\Msun$), low-density clouds from the sample, while simultaneously shifting the masses of the detected clouds towards lower values, as visible in the cumulated mass distributions (top left panel of Fig.~\ref{fig:distrib_comp}). As a result, the distribution is truncated below $\sim$10~$\Msun$, while being slightly shifted towards lower masses overall. Consistent with the cloud-by-cloud comparison, the recovered population exhibits smaller radii and lower velocity dispersions, while column densities are systematically higher owing both to the loss of diffuse gas within individual clouds and to the preferential detection of denser objects. By contrast, the velocity and elongation distributions remain fully consistent with those of the ideal sample.

\subsection{Combined effect of noise and beam smearing}
\label{sec:both_effects}

Finally, we include both beam smearing and noise simultaneously to reproduce observational conditions similar to those of the GBT data. Their combined effect is particularly interesting, as beam smearing modifies the spatial distribution of the emission and thereby changes how strongly the noise affects the cloud. As discussed in Section~\ref{sec:beam_smearing}, beam smearing spreads the emission over a larger area, lowering the brightness temperature, and thus the apparent column density, of the cloud. When considered alone, this primarily affects the inferred radii, vertical elongations, and column densities. In the presence of noise, however, the consequences are more severe: emission that has been diluted by beam smearing can fall more easily below the noise threshold and consequently be removed from the detected cloud. Thus, the combination of beam smearing and noise can result in substantially more emission being lost than when either effect is considered separately. 

The combined impact of beam smearing and noise is nicely illustrated in Fig.~\ref{fig:1cloud_example} for our representative cloud 398: as before, the left panel highlights the observed sky morphology (pink) compared to the previous cases, while the middle and right panels compare the \hi\ column density maps in the ideal and in the final observed cases, respectively.
In the ideal case, the dense cloud head reaches a peak \hi\ column density of $1.2 \times 10^{20}$ cm$^{-2}$, while the maximum observed value is reduced to $0.3 \times 10^{20}$ cm$^{-2}$.
More importantly, in this specific example, nearly half of the cloud, corresponding to the bottom tail, is no longer recovered. Beam smearing dilutes most of the emission from this region, causing it to fall below the noise level in individual channels and thus preventing its identification as part of the cloud.
Interestingly, the tail still contains some detectable emission at a $\mathrm{S/N}\sim4$, but this signal is limited to a single velocity channel and to a region smaller than the beam size. The source finder therefore does not identify it as genuine, spatially coherent emission and instead rejects it as a noise spike.

The combination of both effects produces the smallest recovered sample, with 442 single detections and 62 multiple detection cases. Overall, about 32\% of the original \hi\ clouds are completely missed, while a further 8\% are fragmented into multiple detections, typically for the largest clouds. Consequently, only about 60\% of the original cloud population is recovered as single detections under realistic observational conditions. The fraction of recovered clouds decreases with increasing latitude, reflecting the corresponding decline in the median cloud density (see Paper I): at $|b| \sim2^{\circ}$, 85\% of the clouds are detected, while this fraction drops to 40\% for $|b| \sim10^{\circ}$. In terms of total \hi\ mass of the cloud population (including multiple detections), only $1.83 \times 10^5~\Msun$ is recovered, corresponding to a loss of $\sim$29\% of the \hi\ gas directly caused by observational biases, both due to missing clouds and to partial removal of cloud regions. 

The third column of Figs.~\ref{fig:all_correlations1} and \ref{fig:all_correlations2} compares the properties of the recovered clouds with those in the ideal sample. As in the noise-only case, the recovered masses (Fig.~\ref{fig:all_correlations1}$c$) are systematically lower, particularly for low-mass clouds, but the effect is stronger because larger fractions of the clouds are lost. The departure from the one-to-one relation can reach an order of magnitude for the lowest-mass clouds, resulting in a combined mass loss of $4.5\times10^4~\Msun$ throughout the sample, i.e. 24\% of the cloud masses. These values quantify how much the masses of detected clouds are underestimated under realistic observational conditions, excluding both missed clouds and clouds fragmented into multiple detections. These findings are in line with previous studies reporting incomplete flux recovery, with losses up to 20\%--40\%, in both observed and synthetic molecular clouds \citep{2006PASP..118..590R,2020ApJ...898...80Y}.

The recovered radii are, on average, slightly larger than in the ideal case, although with substantial scatter (Fig.~\ref{fig:all_correlations1}$f$). This is particularly interesting because beam smearing and noise act in opposite directions when considered separately: beam smearing increases the apparent radius, whereas noise decreases it by removing low-density emission. Despite the larger amount of emission lost when both effects are combined, the recovered radii remain approximately 20\% larger on average. This indicates that the increase caused by beam smearing dominates over the reduction caused by noise.
The recovered column densities are lower than the ideal case values by approximately a factor of two (Fig.~\ref{fig:all_correlations1}$i$). Beam smearing alone strongly decreases the column density (Section~\ref{sec:beam_smearing}), whereas noise alone produces a moderate increase (Section~\ref{sec:noise}). The combined result is therefore dominated again by the effect of beam smearing, indicating that the dilution of the emission has a stronger impact on the inferred column density than the preferential removal of low-density regions.

As in the previous sections, cloud LSR velocities are recovered remarkably well (Fig.~\ref{fig:all_correlations2}$c$), making them the only property considered here that is not significantly affected by the observational biases. As mentioned, the velocity centroid is determined primarily by the densest parts of the clouds, which remain detectable after applying both observational effects. This can also be seen in Fig.~\ref{fig:1cloud_example} (inset of the first panel), where the peak of the velocity spectrum is largely preserved even after beam smearing and noise are included (pink line).
The velocity dispersions, however, are underestimated by approximately 40\% (Fig.~\ref{fig:all_correlations2}$f$). This is the same trend found for noise alone, but stronger here because a larger fraction of the cloud emission is removed. As discussed in Section~\ref{sec:noise}, we attribute this to the preferential loss of low-density, higher-temperature gas, which tends to exhibit stronger turbulent motions and velocities that differ from those of the dense bulk of the cloud. The impact is illustrated in the inset of Fig.~\ref{fig:1cloud_example}, where both the low- and high-velocity tails of the spectrum are significantly suppressed once beam smearing and noise are combined. This result is consistent with previous studies of observational biases in molecular clouds, which have shown that finite sensitivity and angular resolution can significantly affect the measured velocity dispersion \citep{2020MNRAS.498.2440Y}.

As in the beam-smearing case, the vertical elongation of the clouds is strongly underestimated (Fig.~\ref{fig:all_correlations2}$i$), resulting from the combined effects of spatial smoothing and the removal of low-density gas in the extended trailing tails of vertically elongated clouds. Although a weak positive trend remains, indicating that intrinsically elongated clouds tend to remain somewhat elongated after the observational effects are applied, the lower recovered values and large scatter, together with the example shown in Fig.~\ref{fig:1cloud_example}, indicate that much of the intrinsic cloud morphology is lost under realistic observational conditions.

The resulting distributions of cloud properties under both observational effects are shown in Fig.~\ref{fig:distrib_comp} (pink lines). The cumulative distributions in the top panels show that this case has the fewest recovered clouds. Compared to the intrinsic and ideal cases, the mass distribution retains a similar peak around $100~\Msun$, but is truncated below $\sim$10~$\Msun$ because the faintest clouds are not detected, while the bulk of the distribution is somewhat shifted towards lower values. The radius distribution is clearly moved to larger values by a factor of approximately two, due to projection effects (the ideal radii are typically larger than the intrinsic ones), to beam smearing (partly mitigated by noise) and to the preferential loss of small, faint clouds. While the velocity distribution remains broadly unchanged, the velocity-dispersion distribution is shifted towards lower values. The vertical elongation distribution becomes more strongly concentrated around unity and loses most of its high values, indicating that clouds tend to appear more spherical under observational conditions. Finally, the column-density distribution is shifted towards lower values, with a peak around $3\times10^{18}$~cm$^{-2}$ compared to $8\times10^{18}$~cm$^{-2}$ in the ideal sample.

\subsection{Extended quiescent period}
\label{sec:results_quiet}
    
The analysis above was carried out for the post-starburst time, when clouds are numerous and typically more massive, denser, faster, and able to reach greater heights above the disc (see Fig.~\ref{fig:comp_times} and Paper~I). In this section, we investigate how these results change during the extended quiescent period, when the cloud population is smaller and the clouds are typically less massive, more diffuse, slower, and confined to lower heights. Similarly to Fig.~\ref{fig:all_correlations1} and \ref{fig:all_correlations2}, we show in the Appendix~\ref{app:A} the impact of observational effects on the cloud population in the quiet period.

The effects described in Sections~\ref{sec:beam_smearing} and \ref{sec:noise} remain broadly valid for the quiescent time. Beam smearing increases the cloud radii while decreasing their vertical elongations and column densities, whereas noise alone decreases the masses, radii, and velocity dispersions while increasing the column densities. However, when both biases are combined, some important differences emerge. As in the post-starburst case, the recovered masses, velocity dispersions, and vertical elongations are lower than in the ideal case. The radii and column densities, however, behave differently. Despite the large scatter, the recovered radii are on average lower by a factor of $\sim$2 compared to the ideal case. We attribute this to the lower masses and densities of the clouds during the quiet period. Lower-density clouds have larger fractions of their emission falling below the noise level, causing substantial parts of the clouds to be removed and reducing their apparent size. In this case, the effect of noise dominates over the increase in apparent size caused by beam smearing, resulting in a net decrease in the recovered radii. Conversely, the column densities are approximately preserved under observational conditions, although with substantial scatter, rather than showing the decrease found during the post-starburst period. This can be understood in the same way, as the decrease caused by beam smearing is largely compensated by the preferential removal of low-density regions due to noise. However, this balance is probably coincidental for these particular cloud properties coupled with these particular observational conditions.

Interestingly, the differences between the post-starburst and quiet periods occur for the two properties for which beam smearing and noise act in opposite directions when considered separately, namely the cloud radii and column densities. During the post-starburst period, beam smearing dominates over noise, whereas the opposite is true during the quiet period. This difference is driven by the higher density of the post-starburst clouds. More generally, the relative importance of the two observational effects depends on the cloud density: low-density clouds are more strongly affected by noise, leading to a net reduction in their apparent radius, while high-density clouds are more strongly affected by beam smearing, resulting in larger apparent radii. In fact, we find a transition for the correlation between radii recovered under observational conditions and in the ideal case (Fig.~\ref{fig:all_correlations1}$f$ and Fig.~\ref{fig:all_correlations_quiet1}$f$) depending on the clouds column densities: clouds below $\sim$10$^{19}$~cm$^{-2}$ tend to have underestimated radii because noise removes part of their emission, while denser clouds tend to have overestimated radii as beam smearing dominates (see Appendix~\ref{app:B}).

Another important difference between the post-starburst and quiet periods is the number of clouds recovered under observational conditions. During the post-starburst period, 32\% of the clouds are lost when both observational biases are applied (Section~\ref{sec:both_effects}). At the quiescent time, out of the 543 resolved clouds present in the simulation, only 103 are recovered as single detections, with a further 20 producing multiple detections. Thus, approximately 77\% of the clouds are completely missed under observational conditions. This large difference is again a consequence of the lower cloud densities during the quiet period, which cause a much larger fraction of the cloud emission to fall below the noise level. As a result, the total cloud mass recovered under realistic observational conditions is only $3.77 \times 10^4 \, \Msun$ out of the $9.74 \times 10^4 \, \Msun$ present in the ideal case, representing a mass loss of 61\% due both to non-detections and mass underestimates.

Overall, the impact of observational effects depends strongly on the physical state of the cloud population. The lower densities characteristic of the quiescent period lead to substantially fewer detected clouds and cause noise to dominate over beam smearing, resulting in smaller recovered radii. At the same time, the balance between the two effects leads to a surprisingly good recovery of the cloud column densities, although this agreement may be coincidental.

\section{Comparison to observed GC high-velocity clouds}
\label{sec:comp_obs_hi}

\begin{figure}
    \includegraphics[width=0.5\textwidth]{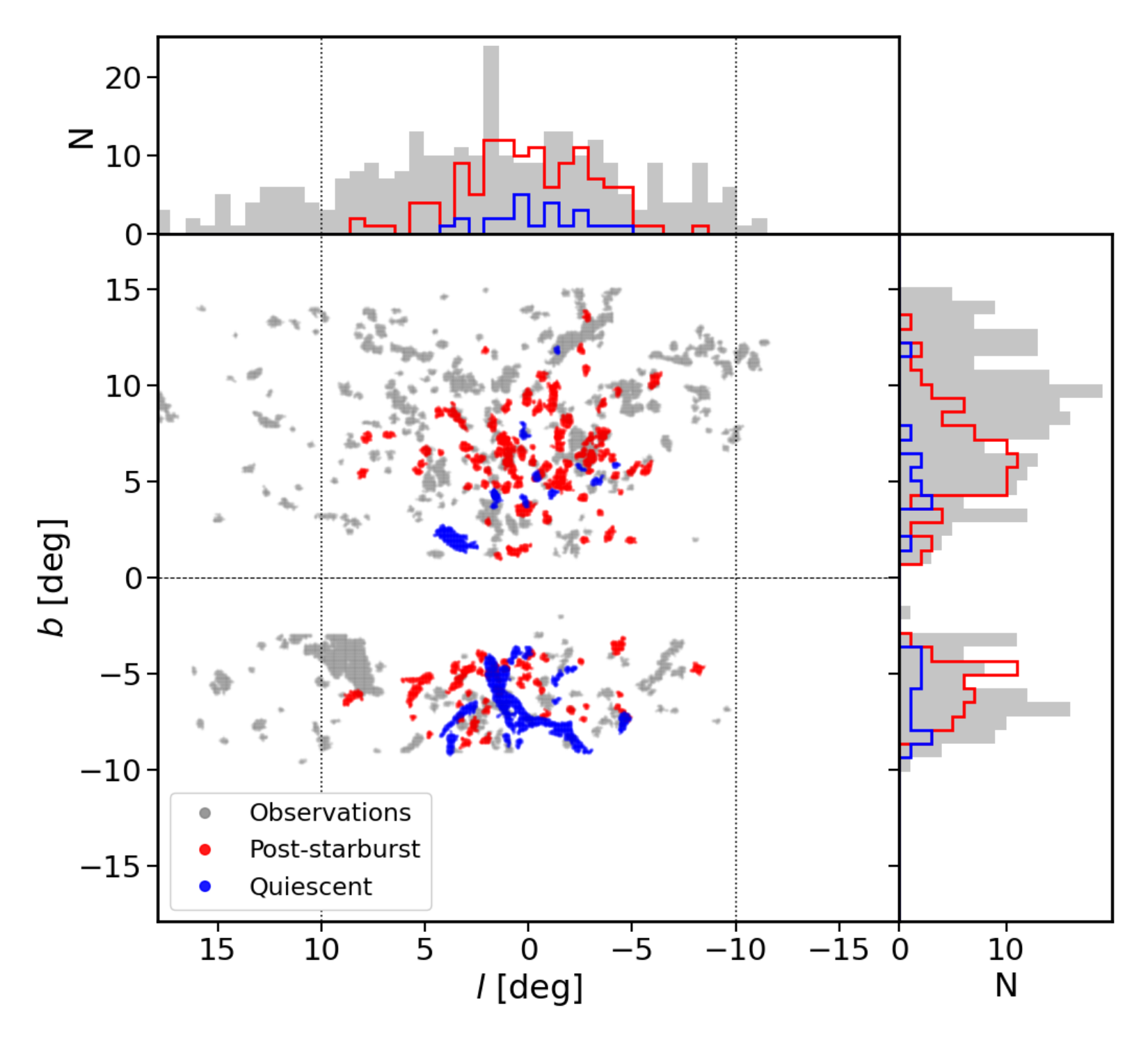}
    \caption{Longitude-latitude positions of our simulated \hi\ clouds observed under GBT-like observational conditions (noise + beam smearing) for the post-starburst (red) and quiet (blue) times, together with the clouds observed with the GBT (grey). We further removed simulated clouds with LSR velocities too close to the rotating gaseous disc on the same LOS, as these low-velocity clouds would likely be obscured by foreground emission in observations. The dotted vertical lines delimit the region $|\ell|<10^\circ$ used for our comparison, while the dotted horizontal line indicates the Galactic mid-plane.}
    \label{fig:clouds_display}
\end{figure}

\begin{figure*}
\includegraphics[width=1.\textwidth, trim = 5.5cm 1.5cm 7cm 2cm, clip]{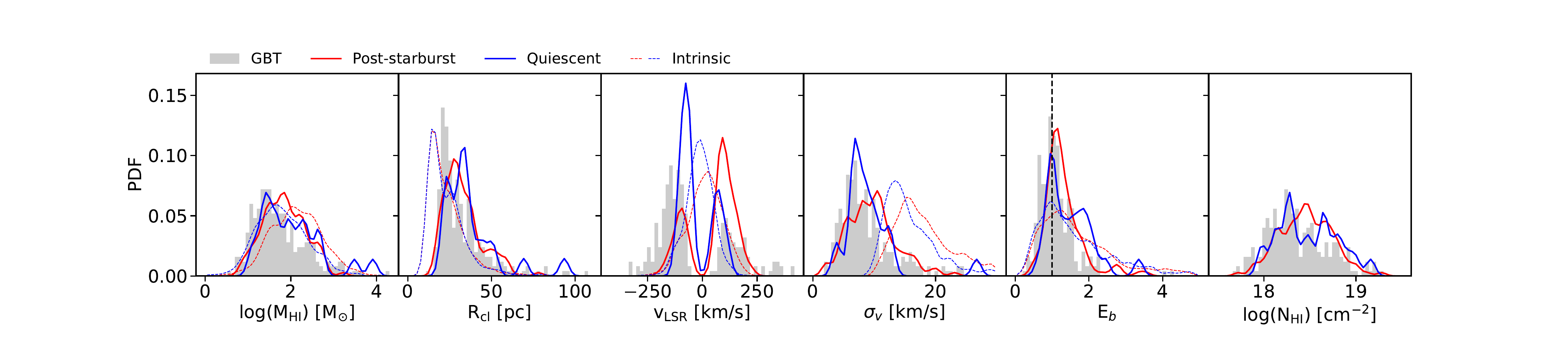}
\caption{Distributions of simulated cloud properties (smoothed with a Gaussian-weighted moving average over 9 bins) under observational conditions (noise + beam smearing + removal of low velocity clouds) at the post-starburst (120 clouds) and quiescent periods (24 clouds), together with observations from the GBT. From left to right: cloud masses, radii, LSR velocities, velocity dispersions, elongation in the latitude direction (the black dashed vertical line indicating $E_b=1$), and column densities. The thin dashed lines represent the intrinsic properties for the post-starburst and quiescent periods.}
\label{fig:comp_obs_clouds}
\end{figure*}

In this section, we compare our simulated clouds with the \hi\ population observed by the GBT in the GC region, accounting for all the observational effects discussed in Section~\ref{sec:impact_bias} to enable a more direct and accurate comparison.
Unlike in Section~\ref{sec:impact_bias}, where clouds were analysed individually, here we place all simulated \hi\ clouds at a given snapshot in a single data cube. 
This allows us to capture projection effects that can cause multiple clouds to overlap along the same LOS. 
As before, to reproduce the angular resolution and sensitivity of the observations, we smooth the mock cubes to the GBT beam size ($570''$) and add Gaussian noise with $\sigma_\mathrm{rms} = 25$ mK. 
We compare the resulting cubes to the most recent GBT survey data presented in \citet{Lockman+2026}, which cover approximately 500 square degrees over the longitude range $\ell \simeq [-12^\circ, 16^\circ]$ for latitudes $b \simeq [-9^\circ, 15^\circ]$.

We apply the same source-finding procedure to both the observational and simulated cubes, applying a primary S/N threshold of 4 and a secondary S/N threshold of 2.5, as in Section~\ref{sec:impact_bias}.
To limit spurious detections, particularly in observational data, we adopt slightly more conservative rejection criteria than in the previous section. 
A reconstructed source is considered genuine only if it spans at least 10 pixels, 15 voxels, and 2 velocity channels.
Furthermore, we reject sources exceeding an angular extent of $3^\circ$ on the sky or spanning more than 50 velocity channels (150~$\kms$).
In the observational data, these upper size limits are particularly important for removing emission from the local interstellar medium, which dominates the signal at low LSR velocities across the entire sky. 
Low-velocity outflowing clouds are drowned out by in this strong foreground emission from the MW disk and therefore cannot be generally recovered in observations.
Because this effect is not naturally present in the simulation, we account for the resulting observational incompleteness by excluding detected simulated clouds whose LSR velocities lie within 50~$\kms$ of the expected rotational velocity of the disc gas projected along the corresponding line of sight, assuming a gaseous disc height of 500~pc. 
Finally, we remove all simulated clouds that are found outside the GBT survey footprint.

\subsection{Cloud properties}
\label{sec:comp_obs_hi_prop}

The source finding procedure results in 290 clouds detected in the observational data, while from the simulated cubes we recover 120 clouds during the post-starburst period and 24 during the quiet period (after removing 116 and 128 low-velocity clouds, respectively, as well as 32 and 51 clouds lying outside the GBT survey footprint). The number of recovered simulated clouds is about six times lower than the 739 cloud starting population during the post-starburst phase, and 23 times lower for the quiescent phase. The detections are also significantly less than the corresponding 586 and 148 clouds (including multiple detections) found in Section~\ref{sec:impact_bias} under the same realistic observational conditions.
The difference arises from the superposition of multiple clouds in a single data cube, the stricter constraints on the spatial and velocity extent of the clouds, and the restrictions on the low-velocity clouds and on the region covered by the GBT.

Figure~\ref{fig:clouds_display} shows the observed (grey) and simulated (red and blue) populations in the $\ell-b$ plane.
The spatial distribution of the simulated clouds is broadly in agreement with the observations, with clouds extending up to $\sim$15$^\circ$ in latitude and tracing a bi-conical outflow originating from the GC. However, in our model the number of clouds decreases with latitude, with very few objects found at $b>10^\circ$, while GBT clouds are numerous at least up to $b\simeq15^\circ$.
Simulated clouds also typically do not extend in longitude as much as the observed population, partly because our PPV data cube is restricted to a $2 \times 2$~kpc region centred on the GC, corresponding to a maximum longitude of $|\ell_{\mathrm{max}}|\sim7$--$10^\circ$, depending on the simulation time. We therefore exclude observed clouds with $|\ell|>10^\circ$ from our comparison, as indicated by the dotted vertical lines in Fig.~\ref{fig:clouds_display}, leaving a sample of 250 observed clouds.
The agreement is better for the post-starburst population than for the extended quiescent one, where most clouds remain closer to the mid-plane.

In Fig.~\ref{fig:comp_obs_clouds}, we further compare the cloud property distributions of the simulated (red and blue lines) and observed (grey histograms) cloud populations. 
As a reference, we also plot the corresponding intrinsic properties as thin dashed lines. 
The simulated mass distributions (first panel) show good agreement with observations, with most clouds having masses between $\sim$10 and $1000~\Msun$.
However, the mass distribution of the GBT population peaks at around $50~\Msun$, compared to $\sim$100~$\Msun$ for our simulated clouds.
In addition, the observed population contains a smaller fraction of high-mass clouds ($M_\hi > 100~\Msun$) than the simulated populations.
The simulated cloud radii are also consistent with the observations, spanning approximately 10 to 50~pc (second panel). However, the simulated distribution peaks at around 30~pc, while the observed clouds are concentrated at smaller radii, with a sharp peak around 20~pc. 
This concentration is mostly a resolution effect: at the distance of the GC, the GBT beam ($\mathrm{FWHM}\simeq23$~pc) sets a minimum apparent cloud size, so that unresolved clouds pile up at radii of $\sim$20~pc.
Together with the differences in mass distributions, the offset in the radius distribution suggests that our simulation produces a somewhat smaller fraction of small, low-mass clouds than are observed by the GBT. 
This discrepancy may partly be driven by the limited spatial resolution of the simulation, which can hinder the formation and survival of small clouds, as well as by our exclusion of clouds with radii smaller than four resolution elements.

The LSR velocity distributions (third panel) of the simulated clouds broadly cover the observed GBT range. However, the simulation contains fewer high-velocity clouds and fails to reproduce the highest observed velocities, above $250~\kms$. The simulated distribution peaks around $|v_\mathrm{LSR}|\sim100~\kms$, compared to $\sim$150~$\kms$ in the observations. The impact of confusion with the foreground MW disc is also evident: excluding clouds close to the disk rotation velocity progressively suppresses the distributions towards lower $|v_\mathrm{LSR}|$, relative to the intrinsic distributions, which peak at $v_\mathrm{LSR}\sim0$.
The velocity dispersions (fourth panel) of the simulated clouds span from a few to $\sim$25~$\kms$, also consistent with the range observed by the GBT, although the distributions peak at different values: $\sim$10~$\kms$ for the simulated clouds and $\sim$5~$\kms$ for the observed population.
The \hi\ column densities (last panel) of the simulated clouds are likewise broadly consistent with the observations, ranging from approximately $10^{18}$ to $10^{19}$~cm$^{-2}$.

The distributions of cloud elongations (fourth panel) along the latitude direction are fairly similar between our model and observations, both peaking around unity. Thus, as in our simulated population after accounting for observational effects, the majority of observed clouds appear relatively round, with no strong preference for elongation along either Galactic longitude or latitude. However, as shown in the previous section, the combination of beam smearing and noise can substantially obscure the intrinsic cloud morphology. Given the overall agreement between the simulated and observed elongation distributions, it is plausible that the real clouds in the GC are also preferentially elongated in the vertical direction, even if this is not apparent in the observed distributions.

Overall, the comparison reveals good agreement and general consistency between the simulated and observed cloud populations, alongside a few systematic differences.
The relative scarcity of low-mass and small clouds in the simulation may partly result from the loss of under-resolved clouds and, more generally, from our limited spatial resolution. By contrast, the deficit of high-velocity clouds, as well of high-latitude clouds, are harder to explain through observational or numerical effects. Instead, it suggests that the simulated CMZ does not accelerate clouds to the highest velocities and latitudes observed by the GBT.
Additional or stronger feedback, whether through more energetic SN activity, AGN feedback, or cosmic rays, could potentially be required to reproduce the full range of observed \hi\ cloud velocities and spatial distribution.

The cloud populations during the post-starburst and quiescent periods remain substantially different. The quiet period produces fewer clouds with lower velocities, making the discrepancy with the observations more pronounced than during the post-starburst period. This further suggests that strong feedback is required to reproduce the observed cloud population. Although the SN activity during the post-starburst period still fails to produce enough high-velocity clouds to fully match the observations, it provides a closer agreement than the quiescent period. This supports the possibility that, if SN feedback is the only driving mechanism, a recent episode of enhanced star formation may be required to explain the population of outflowing clouds in the GC. 
We note, however, that the observed population is closer to the simulated population at the quiescent time in terms of velocity dispersion and column density (although based on just 24 clouds).

Finally, we emphasize the significant differences between the measured and intrinsic distributions (thin dashed lines).
A direct comparison of the intrinsic cloud properties with the GBT observations would lead to rather different conclusions from those obtained using realistic observational conditions: the intrinsic cloud masses are somewhat higher than observed, their radii are smaller, their velocities are substantially lower (largely because low velocity clouds have not been removed), their velocity dispersions are considerably higher, and their elongations are more pronounced. These differences highlight the importance of accounting for projection effects, beam smearing, and noise when comparing simulated and observed \hi\ clouds.

    \subsection{Benchmarking simple models of a radial bi-conical wind}
     \label{sec:comp_models_hi}

Previous observational studies have modelled the kinematics of the GC \hi\ cloud population using a simplified description of a bi-conical galactic wind originating from the GC and expanding radially within a certain opening angle.
In these models, the wind velocity is assumed to be either constant \citep{McClure-Griffiths+13,DiT+18} or to increase with distance from the GC, up to a given maximum value \citep{Lockman+2020,Lockman+2026}.
By matching the predicted and observed longitude, latitude and LSR velocity of each cloud, these kinematic models can constrain their radial velocity and their three-dimensional positions, including their distances from the GC, and provide estimates of their lifetimes.
Here we use the simulated cloud populations of the post-starburst epoch, described in Section~\ref{sec:impact_bias}, to quantify the uncertainties associated with estimating cloud positions using simplified kinematic models.

The models are defined in a Cartesian reference frame centred on the GC, with the Sun located at $(x,y,z)=(-R_0,0,0)$, such that the $x$-axis points away from the Sun, the $y$-axis lies in the Galactic plane and defines the direction of increasing longitude, and the $z$-axis points towards the North Galactic Pole \citep[see also][]{DiT+18}.
We treat the simulated clouds as if they were observed, using their Galactic coordinates $(\ell,b)$ and LSR velocities $\vlsr$ as input to the kinematic models. 
After testing both constant-velocity and accelerated wind scenarios, we find that the ``observed'' simulated data are adequately described by a radial wind with a constant velocity of $200~\kms$ (similar to the median poloidal velocity of clouds, see Paper I) and a wide opening angle of $160^\circ$. Unlike the GBT observed clouds \citep{Lockman+2020}, the simulated clouds do not show any strong signature of acceleration in the $\ell-b-\vlsr$ space. Clearly, this model provides only a rough approximation of the full 3D velocity field of the simulated wind (see Paper I). Based on this simple model, we estimate the 3D position of clouds by requiring an object observed at $(\ell, b)$ to match the $\vlsr$ of the wind at a given position.

\begin{figure}
\includegraphics[width=0.5\textwidth]{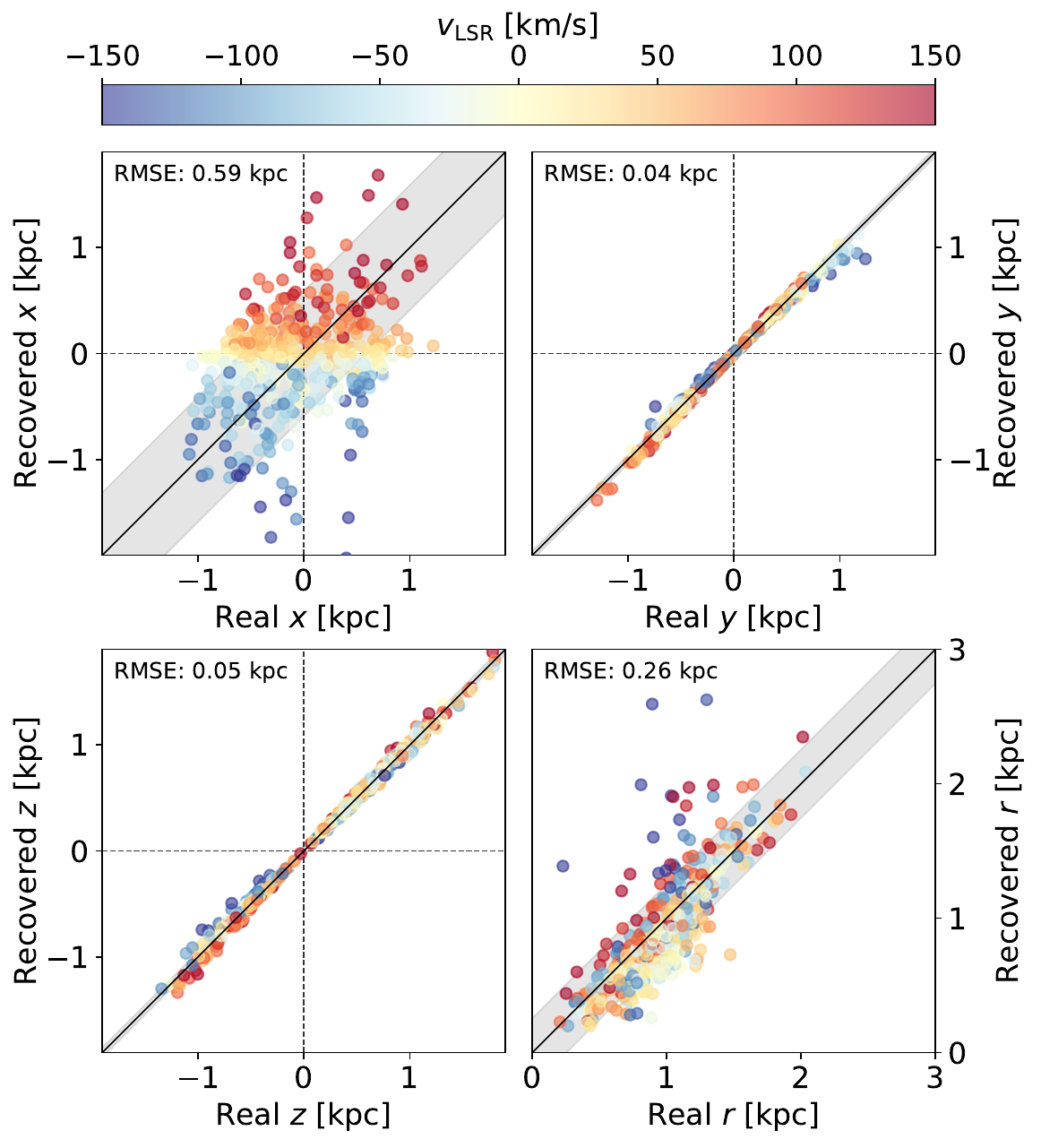}
\caption{Comparison between kinematical-model-derived (recovered) and intrinsic (real) Galactocentric Cartesian coordinates ($x, y, z$) and distance ($r$) for the simulated clouds. Points represent individual simulated clouds, colour-coded by their LSR velocity; the solid black line marks the ideal 1:1 identity line, and the grey shaded band indicates the $\pm 1$ RMSE uncertainty interval around the 1:1 relation in each panel.}\label{fig:sim_kinmodel_comp}
\end{figure}

\begin{figure*}
     \includegraphics[width=\textwidth, trim = 1.5cm 0 0.5cm 1cm, clip]{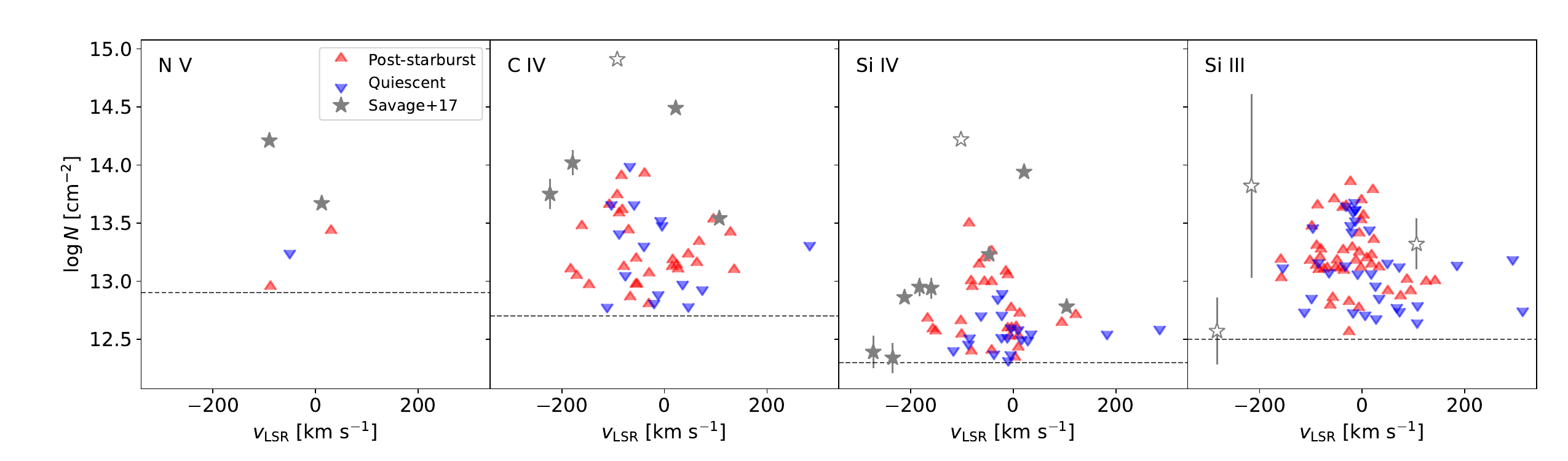}
     \caption{UV absorption components identified mock {\scshape Trident} spectra of our model, shown in the LSR velocity-column density plane, and compared to observations of LS 4825 \citep{2017ApJS..232...25S}. From left to right:  \ion{N}{v}, \ion{C}{iv}, \ion{Si}{iv}, and \ion{Si}{iii}. Red upward triangles show the post-starburst period, blue downward triangles the quiescent period, and grey stars the observations, with empty stars showing saturated components with column densities provided as lower limits. The dashed horizontal lines represent the column density detection floors.}
     \label{fig:mock_UV}
\end{figure*}

Figure~\ref{fig:sim_kinmodel_comp} compares the model-derived Cartesian $x-y-z$ positions and Galactocentric distance $r$ with their intrinsic values for each simulated cloud, colour-coded by their LSR velocity. 
As expected, the $y$ and $z$ coordinates, which are primarily constrained by the observed Galactic longitude and latitude, respectively, are recovered with high accuracy and show little dependence on the adopted kinematic model. The corresponding root mean square errors (RMSEs) are only 0.04 kpc and 0.05 kpc (grey bands in Fig.~\ref{fig:sim_kinmodel_comp}).
In contrast, the recovery of the $x$ coordinate is far more sensitive to the adopted kinematic model, as it depends on the inferred LOS distance and is therefore more strongly affected by the assumptions about the wind velocity field.
The RMSE along the $x$-axis increases to 0.59 kpc, with the recovered values showing a noticeable compression towards $x \approx 0$ kpc relative to their intrinsic range.
The recovered $x$-coordinates show a strict stratification with $\vlsr$, which is a necessary consequence of the projection along the LOS of a radially expanding wind \citep[see][for a detailed discussion]{Lockman+2026}: low-velocity clouds are placed near $x\sim0$, while highly blueshifted and redshifted clouds are placed at negative and positive $x$, respectively.
We note that, as explained in Section~\ref{sec:comp_obs_hi}, most low-$\vlsr$ clouds would be undetectable in real observations, because of confusion with foreground emission from the interstellar gas in the solar neighbourhood. 

As a result of the large scatter along the $x$-axis, the accuracy with which the distance $r$ from the GC is recovered also depends on the adopted kinematic model, reflecting the uncertainty in the inferred LOS positions, although the impact of these errors is mitigated by the small uncertainties in the $y$ and $z$ coordinates. In fact, the RMSE on $r$ is reduced to 0.26 kpc. Moreover, because the model tends to compress $\vert{}x\vert{}$ around 0 kpc, the recovered $r$ is systematically underestimated for clouds at larger radii ($r \gtrsim 0.5$ kpc).
The large error along the LOS also translates in a similar uncertainty of $\sim$0.58~kpc in the estimated distance of a cloud from the Sun. 
Finally, we note that the most significant outliers in the $x$ and $r$ coordinates are all clouds with high velocities, $|\vlsr|\gtrsim100$ $\kms$. This effect is the result of the mismatch between the simulated velocity field and the simplified constant radial flow: clouds with high $\vert{}\vlsr\vert{}$ are assigned to large $\vert{}x\vert{}$, even when their intrinsic LOS offset from the GC is small. In the simulation, however, such large LSR velocities can also arise from non-radial motions or from clouds that move faster than the assumed wind speed. The kinematic model attributes this excess LOS velocity to a geometric displacement, which can place clouds at the wrong LOS distance, or even on the wrong side of the GC, and inflates $r$ in the most extreme cases.

In summary, while simple kinematic models provide a reliable first-order framework for reconstructing the spatial distribution of the clouds, the large uncertainty along the LOS must be kept in mind when placing individual clouds within the Galactic wind and when interpreting the trends of observed cloud properties with distance from the GC.

\section{Comparison to other multi-wavelength data}
 \label{sec:comp_obs_multi}

Beyond the high-velocity \hi\ clouds considered thus far, the multiphase nuclear outflow of the MW has been probed with a range of multi-wavelength tracers. 
The $\gamma$-ray Fermi bubbles \citep{Su+10} trace high-energy non-thermal particles, while their infrared \citep{Bland-Hawthorn+2003}, microwave \citep{Dobler+2008} and polarized radio \citep{Carretti+2013} counterparts trace dust and magnetic fields, respectively. 
Hot, highly ionized gas and warm ionized gas have been observed in X-ray \citep[eROSITA bubbles,][]{Predehl+2020} and through UV absorption lines \citep{Fox+15,Bordoloi+17,2017ApJS..232...25S,Cashman+2023}, respectively. Molecular gas embedded within outflowing \hi\ clouds has also been detected through CO emission \citep{DiTeodoro+2020,DiTeodoro+2026,Heyer+2025}.
Because our current simulation does not include cosmic rays or a chemical network to model dust and molecular gas, we cannot produce realistic mock observations of $\gamma$-ray, infrared and radio continuum, nor of CO emission lines. 
We can however extend the comparison of our model to observations of thermal gas detectable through UV absorption lines and soft X-ray continuum.

\subsection{UV absorption}
\label{sec:UV}

Most high-velocity UV absorption lines are detected at high latitudes ($b>15^\circ$) within the Fermi bubbles \citep[e.g.,][]{Bordoloi+17}.
However, since the resolution of our simulation decreases beyond 2~kpc above the MW centre (corresponding to $b\sim 14^\circ$) and beyond 1~kpc horizontally (corresponding to $\ell\sim 7-10^\circ$), we restrict the UV comparison to the only sightline probing the region within this height and longitude. We therefore compare our results with the UV absorption observations of \cite{2017ApJS..232...25S}, who studied the spectrum of the B1 Ib-II star LS 4825, located 9.9~kpc from the Sun \citep{Cashman+2021} at Galactic coordinates $(\ell,b)=(1.67^\circ,-6.63^\circ)$, using the Space Telescope Imaging Spectrograph (STIS) on the Hubble Space Telescope (HST). They detected strong absorption in high and intermediate ions with multiple velocity components spanning $-283$ to $107~\kms$, which we hereafter refer to as UV clouds. These diffuse absorption components indicate the presence of outflowing warm ionized gas with typical temperatures of $T\simeq10^5$ K, providing an additional test for our model.
We note that UV absorption-line observations with HST/STIS are subject to a different set of observational biases. Unlike emission-line observations, which provide spatially resolved information over an extended field, absorption spectroscopy probes only the gas located along the sightlines to suitable background UV-bright sources.
The main instrumental limitation is the sensitivity to weak absorption features, which is determined by the S/N of the background continuum and, in turn, by the source brightness, exposure time, and instrumental background.

To mimic UV absorption observations, we use {\scshape Trident} \citep{2017ApJ...847...59H}, an open source \textsc{Python} package that computes ionic abundances and generates synthetic absorption line spectra from astrophysical hydrodynamical simulations. We select sightlines from the Sun's position in our model ($R_0=8.2$~kpc, 30$^\circ$ angle to the bar) and compute the corresponding column densities of different ions, including a line spread function (LSF) and Gaussian noise consistent with the observations. To match the data of \cite{2017ApJS..232...25S}, we consider the four high and intermediate ions \ion{N}{v}, \ion{C}{iv}, \ion{Si}{iv}, and \ion{Si}{iii}, with S/N per channel of 12, 35, 20, and 8, respectively. These values approximate the wavelength dependent noise in the STIS E140M observations, estimated from the continuum scatter in Figs. 5 and 8 of \cite{2017ApJS..232...25S}. We further adopt a Gaussian LSF with a FWHM of $6.6~\kms$, corresponding to the HST/STIS E140M spectral resolution. 
Ion number densities are computed by \textsc{Trident} assuming solar metallicity and ionization equilibrium, including collisional ionization and photoionization by the metagalactic UV background of \citet{2012ApJ...746..125H} in the optically thin limit, while neglecting local radiation sources.

For each ion, we fit the resulting velocity spectrum with Gaussian components to identify individual UV clouds and derive their velocities as the centroid of the fitted Gaussian converted to LSR, and column densities by integrating over their absorption profile. 
Since \citet{2017ApJS..232...25S} do not quote formal detection limits for these ions, we estimate $3\sigma$ detection thresholds from the column-density uncertainties of the weakest isolated components in their profile fits, which include both statistical and continuum-placement errors. This yields $\log\, (N\, / \,\mathrm{cm^{-2}}) \approx 12.9$, 12.7, 12.3, and 12.5 for \ion{N}{v}, \ion{C}{iv}, \ion{Si}{iv}, and \ion{Si}{iii}, respectively, where the \ion{Si}{iii} value is derived from the foreground star HD~167402, observed with the same instrumental setup, since all \ion{Si}{iii} absorption towards LS~4825 is saturated.
To obtain a larger sample of UV clouds, we consider several lines of sight. Since the velocity of the outflowing gas depends on its height above the Galactic disc (as shown in Paper I), we keep the same absolute latitude as LS 4825 and vary only the longitude. We therefore consider 10 sightlines, combining five longitudes ($-3^\circ$, $-1.67^\circ$, $0^\circ$, $1.67^\circ$, $3^\circ$) with the two latitudes ($-6.63^\circ$, $6.63^\circ$), and extending to 10~kpc distances from the Sun. As before, we perform these mock observations for both the post-starburst and quiescent periods.

Figure~\ref{fig:mock_UV} shows the absorption components obtained for each ion in the LSR velocity - column density plane, compared with the observations of LS 4825 from \citet{2017ApJS..232...25S}. For \ion{N}{v},
the agreement is rather poor: while the velocities are consistent with observations, the two post-starburst and the single quiescent UV clouds identified in our simulation have column densities lower than observed. 
The agreement is somewhat better for \ion{C}{iv} (second panel), for which we identify 21 UV clouds during the post-starburst period and 12 during the quiescent period. The velocity and column density ranges are broadly consistent with observations, although the observed UV clouds tend to have higher column densities. We find a similar trend for \ion{Si}{iv} (third panel), for which we identify 33 clouds during the post-starburst period and 32 during the quiescent one. Finally, for \ion{Si}{iii} (last panel), the column densities are consistent with the lower limits in the observations (as lines are saturated), although our UV clouds (56 for post-starburst, 40 for quiescent times) tend to have lower velocities. 

Overall, our simulated UV clouds tend to have lower column densities than those observed towards LS~4825, with the discrepancy increasing towards more highly ionized species. In addition, the simulated warm gas seems to have slightly lower velocities. 
Taken at face value, this suggests that the simulated wind is not dense and fast enough compared with the nuclear outflow of the MW.
As for the \hi\ clouds, the post-starburst period produces more numerous UV clouds with higher column densities than the extended quiescent period. 
However, we stress that the comparison of column densities should be interpreted with some caution. \citet{2017ApJS..232...25S} find that the high-ion ratios towards LS~4825 are best reproduced by non-equilibrium radiative cooling and turbulent mixing-layer models, whereas \textsc{Trident} computes ion fractions assuming ionization equilibrium with a uniform UV background. In addition, gas at $T\sim10^5$~K mostly resides in thin interfaces between the cool clouds and the hot wind, which are only marginally resolved in our simulation, and the column densities scale linearly with the assumed metallicity, which might be supersolar in the GC \citep[e.g.,][]{Cunha+2007,Ashley+2022}. The column densities of our UV clouds therefore provide only a qualitative comparison, whereas their velocities are a more robust test of the simulated outflow.

\subsection{Soft X-rays}
\label{sec:xray}

\begin{figure*}
     \includegraphics[width=\textwidth]{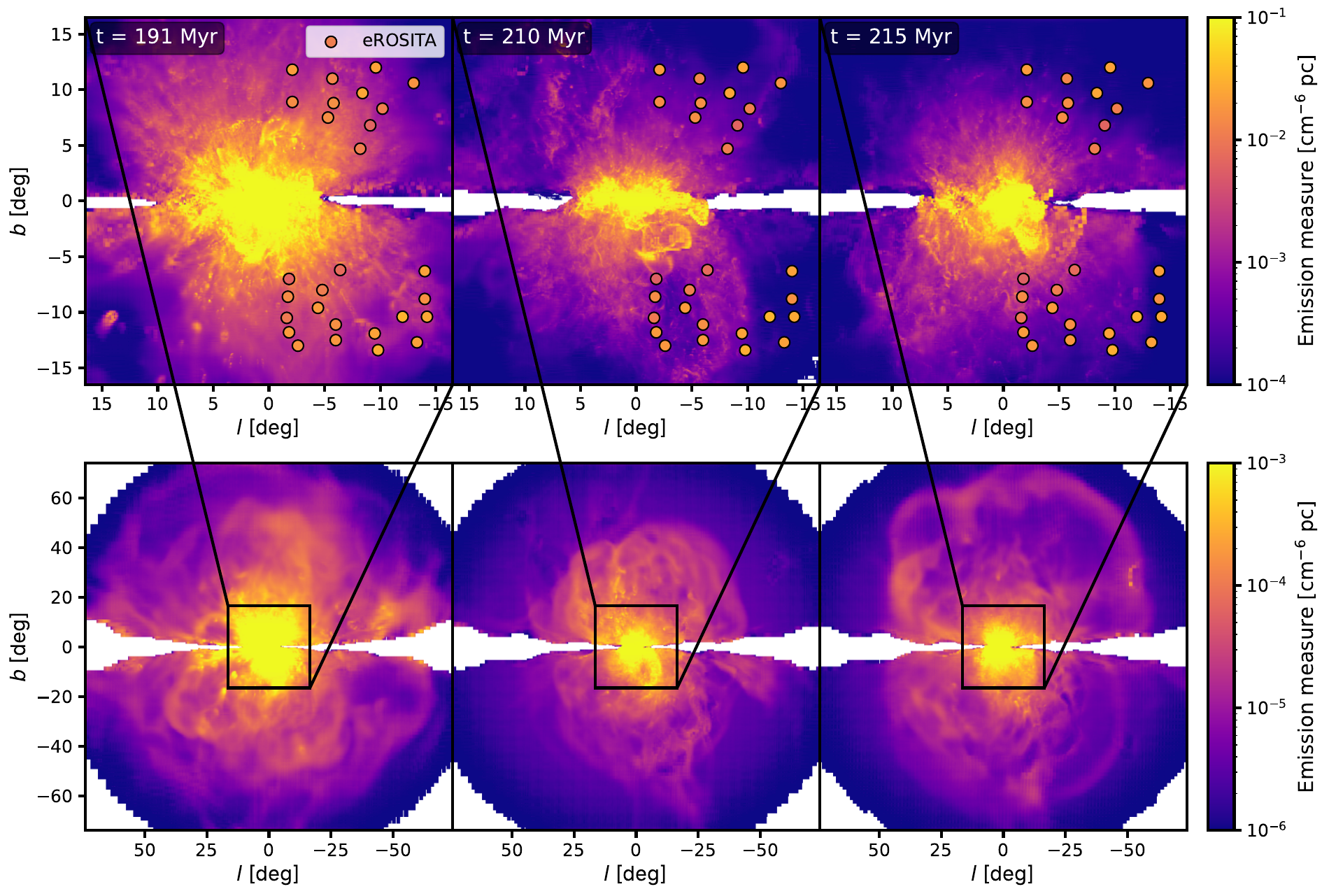}
     \caption{Mock images of X-ray emission measure of our {\scshape Piernik} simulated Galactic centre at the post-starburst time (left column), at the quiescent time (middle column), and at $t=215$~Myr (right column), close to the GC (top row), and at a larger scale (bottom row). In the top row we further compare with the emission measures of the eROSITA bubbles from \citet{2024A&A...690A.399Y}, shown as circles with the same colour scale.}
     \label{fig:mock_Xray}
\end{figure*}

An important piece of evidence for nuclear outflows in the MW is the existence of the eROSITA bubbles: giant, hourglass-shaped X-ray structures extending from the GC to nearly 14~kpc above the Galactic disc. Here we investigate whether similar features are present in our model and compare the simulated X-ray emission measure (EM) with the values estimated for the eROSITA bubbles. Specifically, we adopt the EMs of the eROSITA-bubble component from the best-fit spectral models of \citet{2024A&A...690A.399Y}, in which the bubble emission in each spectral bin is modelled as a single-temperature plasma in collisional ionization equilibrium with solar abundance, on top of the local hot bubble, circumgalactic medium (CGM), Galactic corona and cosmic X-ray background components.
In this comparison, the main source of uncertainty is the modelling and decomposition of the X-ray spectra.
We therefore compare our simulation directly with the fitted EM of the bubbles, without attempting to create realistic X-ray data that account for the complex instrumental response and observational effects of eROSITA.

We compute the EM maps by integrating the product of the electron and hydrogen number densities along each line of sight, EM = $\int n_{\rm e} n_{\rm H}dl$, considering only gas with temperatures above $10^6$~K. We assume this gas to be fully ionised and adopt hydrogen and helium abundances of 90\% and 10\% by number, respectively, while neglecting the contribution of metals. Under these assumptions, each hydrogen atom contributes one free electron and each helium atom contributes two, giving $n_{\rm e}=n_{\rm H}+2n_{\rm He}\simeq1.2n_{\rm H}$. We therefore calculate the EM directly from the simulated hydrogen density as $\mathrm{EM}=\int 1.2n_{\rm H}^2dl$.
As before, we place the observer at the Sun's position looking towards the GC and integrate the emission along the full LOS.
Figure~\ref{fig:mock_Xray} presents the resulting EM maps close to the midplane (top panels) and on larger scales, up to 10 kpc above the Galactic disk (bottom panels).
We show the maps at both the post-starburst and quiescent times (first and second columns, respectively), as well as at $t=215$~Myr, i.e. 5~Myr after the quiescent time (third column).
On the zoomed-in maps, we also overlay the eROSITA bubble EM measures from \citet{2024A&A...690A.399Y} as circles.

Close to the disk (top panels), we find broad agreement with the observed EM in the post-starburst case, with a median of the relative error lying at -0.79 dex for $\ell<10^\circ$ data points in the post-starburst case. However the agreement becomes poorer at higher latitudes ($b\gtrsim10^\circ$), where the EM in our simulation decreases more rapidly than in the eROSITA observations. During the extended quiescent period, the agreement is generally worse, with our simulation often underestimating the EM even close to the disc (relative error median = -1.57 dex for $\ell<10^\circ$). 
At larger heights (bottom panels), bubble-like structures extending up to $b\sim50^\circ-60^\circ$ emerge, qualitatively resembling the eROSITA bubbles in morphology.
These features trace shock waves produced by starburst periods propagating outwards from the CMZ and are therefore transient. 
As seen throughout the different panels of Fig.~\ref{fig:mock_Xray}, they expand from the GC to heights of order 10~kpc within 10-15~Myr, and we observe about ten such bubbles forming and disappearing during the last 250~Myr evolution of our simulation (i.e.\ its high-resolution phase).
However, on these scales, the simulated EMs are about two orders of magnitude lower than observed values. For example, at a latitude of $\sim$50$^\circ$, eROSITA estimates EM of order $10^{-3}$~cm$^{-6}$~pc, against $10^{-5}$~cm$^{-6}$~pc in our simulation. 
We note that a more recent, dedicated analysis of the eROSITA bubbles at $b\gtrsim12^\circ$ by \citet{2026A&A...714A..85Y} finds that their emission is better described by two plasma components ($kT\simeq0.2$ and 0.6~keV) with subsolar abundances ($Z\simeq0.05$--$0.2\,Z_\odot$); since for metal-line dominated emission the fitted EM scales approximately inversely with the assumed abundance, their values are about an order of magnitude higher (e.g. $\sim$0.3--0.4~cm$^{-6}$~pc at $|b|\simeq12$--$18^\circ$), so that the discrepancy with our simulation is likely even larger, possibly also close to the disk.

The increasing discrepancy with height between the simulated and observed EM values may result from a combination of several factors, including the strength of feedback, the decreasing resolution with height, and the absence of a realistic CGM in our simulation.
Stronger feedback would drive a denser hot wind to greater distances from the Galactic mid-plane.
In fact, consistently with our previous conclusions, the post-starburst period provides a closer match to the observables, including the X-ray emission, at least in the region close to the Galactic plane. 
Moreover, the spatial resolution of our simulation degrades with distance from the GC. Within the disc and the inner outflow region (out to $b\sim15^\circ$ and $\ell\sim8^\circ$), dense gas is resolved at 3~pc, while the diffuse hot phase is often resolved only at 6~pc. Beyond this region, the resolution decreases rapidly, reaching 96~pc at latitudes of $b\sim50^\circ$. Simulations of starburst-driven winds have shown that X-ray luminosity tends to increase with spatial resolution \citep{2018ApJ...860..135S,lin2026illuminatingm82simulatingxray}, as resolving the mixing layers between the hot wind and the cooler ambient gas is essential for capturing their contribution to the X-ray emission. 
The coarse resolution at large heights may therefore contribute to the growing discrepancy between our simulated EMs and the eROSITA values at high latitudes.

Finally, because our simulation does not model the CGM around the MW, gas densities beyond the immediate vicinity of the disc are extremely low ($n<10^{-4}$~cm$^{-3}$).
The bubbles therefore expand into a medium that can be more than an order of magnitude less dense than expected in the MW, especially in the inner halo \citep[$n\simeq10^{-2}-10^{-3}$~cm$^{-3}$, e.g.,][]{Miller+2013, Kataoka+2015}.
This difference can significantly affect the evolution of the simulated bubbles and, in particular, reduce their X-ray emission by weakening the emission from shock-heated gas in the surrounding medium.
For example, consider a strong shock that compresses the ambient gas by a factor of four \citep[e.g.,][]{Draine+2011} and produces an emitting shell with a thickness of $l=100$ pc. For an initial CGM density of $n\sim 10^{-3}$~cm$^{-3}$, the post-shock density would be $\sim4\times10^{-3}$~cm$^{-3}$, yielding an EM of approximately $\mathrm{EM}\sim2\times10^{-3}$~cm$^{-6}$~pc, comparable to the observed order of magnitude.
Instead, for a much lower ambient density of $n\sim10^{-4}$~cm$^{-3}$, as in our simulation, the corresponding emission measure drops to $\sim2\times10^{-5}$~cm$^{-6}$~pc, consistent with the low values we obtain.


   
\section{Discussion}
\label{sec:discussion}
In this work, we have used the simulation of Paper~I to quantify how observational effects shape the high-velocity \hi\ cloud population above the GC, and to compare our model with data in the observational domain.
Under GBT-like conditions, beam smearing and noise bias the detected cloud properties towards lower masses, velocity dispersions and column densities and rounder shapes, while leaving their LSR velocities essentially unaffected.
Once these effects and the confusion with foreground Galactic emission are accounted for, our simulated clouds broadly reproduce the observed GBT population, but lack its fastest and highest-latitude members. In the following, we examine the implications of these findings for the interpretation of \hi\ observations and for the origin of the GC outflow.

\subsection{Implications for \hi\ observations of the GC outflow}
\label{sec:implications_obs}

Our results show that the properties of high-velocity \hi\ clouds measured with single-dish telescopes such as the GBT can differ significantly from their intrinsic values. Assuming that the clouds in our post-starburst snapshot provide a reasonable representation of those above the GC, the results of Section~\ref{sec:impact_bias} can be inverted to estimate how the intrinsic cloud population may differ from the observed one. These estimates are model-dependent and should be regarded as indicative. They are, however, likely conservative: clouds in the quiescent phase are more diffuse and therefore, on average, more strongly affected by observational effects (Section~\ref{sec:results_quiet}), and the rms noise of the GBT survey ranges from 10 to 90\,mK across the field (Section~\ref{sec:obs_bias}), so that the biases in the noisier regions should exceed those obtained for our representative value of 25\,mK.

The most direct implication concerns the number and mass of the clouds. Since about one third of the clouds are missed due to beam smearing and noise under GBT-like conditions, the number of high-velocity clouds above the GC should be at least $\sim$50\% larger than observed.
In addition, the masses of the detected clouds are underestimated by $\sim$24\%, implying intrinsic masses $\sim$32\% higher.
Because the missed clouds are predominantly faint, low-mass objects, these two effects do not simply multiply: comparing total masses directly, the \hi\ mass in the GC outflow could be at least $\sim$40\% higher than the detected one.
Confusion with foreground Galactic emission further removes about half of the remaining clouds: in our full mock survey, 116 of the 236 clouds detected after applying beam smearing and noise have LSR velocities too close to those of the disc gas to be recovered (Section~\ref{sec:comp_obs_hi_prop}). Taken together, these effects suggest that the observed sample may contain only about one third of the outflowing \hi\ clouds above the GC.
Assuming that the masses of the clouds lost to foreground confusion are similar to those of the recovered clouds, the total \hi\ mass in the outflow may be almost a factor of three ($\sim$175\%) higher than the observed value.
For example, the \hi\ mass measured in the 290 high-velocity clouds detected in the latest GBT survey is $\sim 7 \times 10^4~\Msun$\footnote{We note that this number is an order of magnitude lower than that reported in \citet{DiT+18} for a smaller area. Their equation (2) contains an error that leads to an overestimation of the \hi\ mass. In Paper~I, we compared to the sample of \citet{DiT+18} after recalculating their masses.}. Based on the above, we may expect a total \hi\ mass in the GC outflow that exceeds $2\times10^5~\Msun$.
Furthermore, not all clouds observed in the outflows are necessarily outflowing, as some may be returning to the disc as part of a galactic fountain. 
In our simulated cloud population, $\sim$90\% of the clouds are outflowing. This suggests that approximately 10\% of the clouds observed by the GBT may instead be inflowing, i.e. falling towards the disc.

Observational effects also alter the internal properties of the clouds. 
Column densities are underestimated by about a factor of two, and velocity dispersions by $\sim$40\%, implying intrinsic dispersions about two thirds higher than observed. Because the combined thermal and turbulent pressure scales as $\sigma_\mathrm{v}^2$, the internal support of the clouds against the pressure of the surrounding hot wind, which plays an important role in their survival (Paper~I), would be underestimated by a factor of $\sim$3 if inferred directly from the observed line widths. 
These line widths, however, include thermal broadening. 
The observed GBT clouds have dispersions peaking at $\sim$5~$\kms$ (Fig.~\ref{fig:comp_obs_clouds}), below the thermal width of warm neutral gas ($\sigma_\mathrm{th}=\sqrt{k_\mathrm{B}T/m_\mathrm{H}}\approx8$--$9\kms$ for $T\approx8000$--$10^4$\,K), and even after correcting for the $\sim$40\% bias their intrinsic dispersions ($\sim$8~$\kms$) remain comparable to this value. Unless their turbulence is negligible, this implies that a substantial fraction of their \hi\ is colder ($T\lesssim3000$\,K) than in our simulated clouds, which are mostly at $T\sim10^4$\,K (Paper~I), possibly because the formation and survival of a cold neutral phase in outflowing clouds is not resolved at our 3\,pc resolution. Since our correction is calibrated on these warm simulated clouds, it should be applied with caution to clouds with narrower intrinsic lines.
Larger intrinsic column densities and colder gas may also help to explain the presence of molecular cores in some of these \hi\ clouds \citep{DiTeodoro+2020,DiTeodoro+2026,Heyer+2025}.

Cloud sizes are affected in a density-dependent way: dense clouds appear larger than they are, with intrinsic radii $\sim$20\% smaller on average, whereas clouds with $N_\hi\lesssim10^{19}$\,cm$^{-2}$ appear smaller because part of their emission falls below the noise (Appendix~\ref{app:B}). Finally, the elongated, head-tail morphologies expected for clouds entrained in a wind are largely erased at the GBT resolution, so the roughly round shapes of the observed clouds do not rule out a strong intrinsic elongation along the outflow direction. Recovering this morphology requires higher angular resolution, for instance from interferometric observations, although our results indicate that a gain in resolution must be accompanied by sufficient surface-brightness sensitivity to avoid losing the diffuse cloud tails.

LSR velocities are the only property robustly recovered, which makes kinematics the most reliable diagnostic of the outflow. 
Simple radial-wind models, constrained by the observed sky positions and LSR velocities of the clouds, return outflow velocities that are consistent with the median poloidal velocity of the underlying cloud population, despite the greater complexity of the intrinsic wind velocity field. 
Reconstructing the three-dimensional distribution of the clouds nevertheless remains quite uncertain: radial-wind models recover positions perpendicular to the LOS to within $\sim$50~pc, but carry uncertainties of $\sim$0.6~kpc along the LOS and systematically underestimate Galactocentric distances beyond $\sim$0.5~kpc (Section~\ref{sec:comp_models_hi}). Moreover, in a nearly radial wind the clouds with low $|v_{\rm LSR}|$ are those moving close to the plane of the sky near $x\approx0$, precisely the clouds lost to foreground confusion, so that this region remains poorly sampled by observations.

Accounting for these biases is essential to interpret the physical and chemical state of the observed clouds. Underestimated column densities and velocity dispersions affect estimates of their volume density, thermal and turbulent support and ability to shield themself from the radiation field and to form and retain molecular gas \citep{Krumholz+2009,Sternberg+2014}, while underestimated \hi\ masses bias molecular-to-atomic gas ratios towards higher values \citep{Noon+2023,Noon+2026,DiTeodoro+2026}.
All these quantities regulate the ability of clouds to survive their journey through the hot wind \citep[e.g.,][]{Scannapieco&Bruggen15,Schneider+17,Gronke+20}.
In addition, the large uncertainties on LOS positions propagate into the inferred Galactocentric distances and dynamical ages of the clouds, and hence into any trend of their physical or chemical properties with distance from the GC \citep{Lockman+2026}.
Finally, the total \hi\ mass of the cloud population may be underestimated by up to a factor of $\sim$3, with important consequences for the inferred properties of the nuclear wind. Since cloud velocities are reasonably well recovered, the mass outflow rate and kinetic energy of the cold phase would be underestimated by a similar factor, and so would the mass loading of the wind. \hi\ masses and mass outflow rates derived from the observed cloud population \citep[e.g.][]{DiT+18} should therefore be regarded as lower limits.

\subsection{Implications for the nuclear outflow}
\label{sec:implications}

In Paper~I, by comparing the intrinsic properties of our simulated clouds with the observed \hi\ population, we concluded that SN feedback from the star formation cycles of the CMZ could be sufficient to explain the high-velocity clouds above the GC. The observationally consistent, multi-wavelength comparison presented here provides a more nuanced picture. Once observational effects are accounted for, our model reproduces the masses, sizes, column densities and overall sky distribution of the observed clouds, but falls systematically short in all the tracers considered: it lacks \hi\ clouds with $|\vlsr|\gtrsim250\,\kms$ and at $|b|\gtrsim10^\circ$, its UV absorption from warm ionized gas returns column densities typically lower than observed, increasingly so for the most highly ionized species, and its X-ray EM from hot ionized gas a few kpc away from the GC is at least an order of magnitude below the eROSITA values (although possibly mostly due to the absence of a CGM in our simulation).
These discrepancies consistently point to a wind that is not sufficiently energetic, fast or dense at large heights above the plane.
As described in Section~\ref{sec:comp_obs_hi} and \ref{sec:comp_obs_multi}, stronger feedback could accelerate \hi\ clouds to higher velocities and latitudes, while also producing higher column densities in UV absorption and larger EMs in soft X-rays.

If SN feedback is the main driver of the observed outflow, our results favour a recent episode of enhanced star formation, possibly stronger than the starbursts in our simulation, which peak at $\sim$1~$\msunyr$. This is in line with \citet{2025ApJ...981..154Z}, whose simulated clouds in a MW-like simulation reach velocities and heights comparable to the observed ones during a starburst with a SFR of $\sim$2~$\msunyr$. 
However, the star formation history of the GC does not seem support such an intense starburst: the SFR of the inner regions of the MW has been $\sim$0.1~$\msunyr$ over the last $\sim$5~Myr \citep{Barnes+17}, and $\sim$0.2--0.8~$\msunyr$ over the last $\sim$30~Myr \citep{2020NatAs...4..377N}, consistent with the moderate starbursts of our simulation but not with a burst of $\sim$2~$\msunyr$.
The weaker outflow in our model may also partly stem from our treatment of stellar feedback. In the dense gas of the CMZ, the Sedov-Taylor phase of individual SN remnants is often unresolved, in which case the SN is injected as terminal momentum rather than thermal energy \citep{2017ApJ...846..133K}. Such momentum-based schemes efficiently regulate star formation but produce little hot gas and drive weaker, less mass-loaded winds than resolved energy injection \citep{Smith+2018,Hu+2019}, while the hot, volume-filling phase that powers galactic winds is mostly generated by spatially and temporally clustered SNe in superbubbles \citep{Gentry+2017,Li+2017,Fielding+2017}. 
In addition, our model does not include pre-SN feedback from stellar winds and radiation. Pre-SN feedback can lower the ambient density around young massive stars before they explode, but it also reduces SN clustering, and recent simulations find that the latter effect can suppress outflow mass and energy loadings by up to an order of magnitude at a similar SFR \citep{Rathjen+2021,Smith+2021}. Its net effect on the nuclear wind is therefore uncertain, whereas a better-resolved SN energy injection would likely strengthen the simulated wind at a fixed SFR.

Alternatively, an additional source of feedback may contribute to the outflow. Although Sgr~A$^*$ is currently quiescent, there is evidence that it was considerably more active in the recent past, such as a Seyfert-like flare $\sim$3.5\,Myr ago \citep{Bland-Hawthorn+2019}, and AGN activity has been proposed to explain both the Fermi bubbles \citep[e.g.][]{Zubovas+2012,2020ApJ...894..117Z} and the eROSITA bubbles \citep{2022NatAs...6..584Y,2025arXiv250713665Z}. An AGN-driven outflow could supply the additional energy and momentum needed to accelerate \hi\ clouds to the highest observed velocities and latitudes, and increase the density of the warm and hot phases probed by UV absorption and X-ray emission. Since SN feedback is an unavoidable consequence of star formation in the CMZ, the observed outflow may well result from the combined action of both mechanisms.

Finally, cosmic rays (CRs), which are not included in our simulation as a dynamical component (only as a heating term), have also been shown to alter SN-driven outflows. Accelerated in SN remnants, CRs propagate through the Galaxy via advection and diffusion and can escape the disk, building up a pressure gradient that pushes gas away from the mid-plane through momentum transfer to the thermal gas \citep[e.g.][]{1993A&A...269...54B,2013ApJ...777L..38H}. CR-driven outflows are typically smoother and cooler, and carry more cold gas than purely thermally driven winds \citep[e.g.,][]{2018MNRAS.479.3042G,2023MNRAS.522.5529P}, making them particularly relevant for the outflowing \hi\ clouds studied here. Moreover, CRs could potentially increase the X-ray emission by enhancing the production of warm-hot ionized gas through heating by the dissipation of Alfvén waves \citep{2024ApJ...964...99A,Kim+2026}.
We will investigate the impact of CRs on the nuclear outflows of the MW in a forthcoming paper.


 \section{Summary and conclusions}
  \label{sec:conclusion}

In this paper, we analysed from an observational perspective the simulation of the Milky Way's Galactic centre introduced in \citet{2026A&A...712A.148P}, in which recurrent cycles of starburst and quiescent periods generate nuclear outflows capable of accelerating \hi\ clouds to heights of a few kpc and velocities of a few hundred $\kms$. We investigated how instrumental effects, particularly the spatial resolution and the sensitivity of a single-dish telescope like the GBT, influence the detection and characterization of the \hi\ cloud population. 
This allowed a more direct and consistent comparison between our simulation and observations of outflowing \hi\ clouds in the Galactic centre. 
In addition, we extended this comparison to other tracers of the multi-phase nuclear wind: UV absorption lines, which trace the warm ionized gas, and soft X-ray emission, which probes the hot, highly ionized phase.

We focus our analysis on two snapshots with star formation rates comparable to that of the present-day Galactic centre: a post-starburst time, $\sim$10~Myr after an intense burst of star formation, when delayed supernovae from the burst still drive a strong outflow, and a quiescent time, following a prolonged period of low star formation, when the outflow is weaker.
The main findings of our work can be summarized as follows:

\begin{itemize}[label=$\bullet$, itemsep=2pt]

    \item Finite angular resolution (beam smearing) primarily affects the morphology and surface-brightness distribution of the clouds. It inflates them, increasing their measured radius and decreasing their observed column density, while making them appear rounder and less elongated. Their masses, velocities and velocity dispersions are essentially unaffected.
    
    \item Limited sensitivity determines the faintest emission that can be detected above the instrumental noise level. 
    Noise therefore prevents the detection of low-density, low-mass clouds, and removes the diffuse outer envelopes  of denser ones. As a result, detected clouds tend to appear less massive, smaller, denser, and have lower velocity dispersions than their intrinsic values.

    \item The combined effect of beam smearing and noise at levels typical of current GBT surveys (beam FWHM of 570$''$ and rms noise of 25 mK) has a significant impact on detection of \hi\ clouds: in our post-starburst snapshot, about one third of the cloud population and of the mass can be missed. Furthermore, the individual recovered clouds have typically 24\% lower masses, 40\% lower velocity dispersions, and 50\% lower column densities than their real counterparts, and tend to lose their intrinsic head-tail morphology. Cloud radii also change, but the direction of this change depends on the cloud column density: dense clouds tend to have observed radii larger than their intrinsic radii, while the opposite is true for low density clouds. The velocities of \hi\ clouds are the only observable quantity that does not change significantly as a result of these observational effects.

    \item Under realistic, GBT-like observational conditions, and after excluding clouds whose velocities are too close to those of the Galactic disc gas to be observable, only about one sixth of the parent cloud population is detected at the post-starburst time.
    We find generally good agreement between the simulated and observed \hi\ cloud populations in terms of their spatial distribution and overall properties, particularly at the post-starburst time. However, although the simulated properties broadly span the observed ranges, the simulation systematically lacks clouds at the highest velocities ($|\vlsr|\gtrsim250~\kms$) and latitudes ($b\gtrsim10^\circ$).
    
    \item Simple radial-wind models, commonly used to deproject the line-of-sight velocities and sky positions of observed \hi\ clouds, provide a first-order approximation of the median outflow velocity, They can also accurately recover their intrinsic positions perpendicular to the line of sight (within $\sim$50\,pc), but carry uncertainties of $\sim$0.6\,kpc along it.
    
    \item Mock observations of UV absorption lines and soft X-ray emission suggest that our simulation underproduces the warm and hot gas phases of the outflows. The simulated UV absorption column densities lie at the lower end of the observed values. The X-ray emission measures are broadly consistent with eROSITA bubble estimates close to the Galactic centre, but become increasingly underestimated at higher latitudes ($b\gtrsim10^\circ$), although this discrepancy may partly reflect limitations of our current simulation.

\end{itemize}

\noindent These findings highlight the importance of accounting for observational effects when analysing data and when comparing simulations with observations. 
For outflowing \hi\ clouds in the Galactic centre, our results suggest that observations with a telescope such as the GBT may only probe a small fraction of the total \hi\ population in the outflow.
The measured properties of individual clouds will also be affected: masses, velocity dispersions, and column densities will be underestimated with respect to the underlying population, and their morphologies will appear rounder and less elongated than they are intrinsically.
These biases must be taken into account when interpreting the physical and chemical state of the observed clouds and when inferring the properties of the wind, like mass loadings and energetics.

Finally, the lack of high-velocity atomic gas clouds at large heights above the Galactic plane, together with the underproduction of extraplanar ionized gas, suggests that supernova feedback in our model is insufficient to fully reproduce the data.
If star formation is the main driver of the outflows in the Galactic centre, this would require a significantly higher level of star formation activity over the past $\sim$10--20~$\mathrm{Myr}$ than in our current simulation or than estimated observationally.
Alternatively, or in addition, other feedback mechanisms such as AGN activity or cosmic rays may be needed to fully account for the observational evidence of the multiphase nuclear outflow of the Milky Way.

\vspace{0.5cm}
\begin{acknowledgements}
Calculations were carried out at the center of Informatics - Tricity Academic Supercomputer \& networK (TASK) and on the HYDRA cluster at the Institute of Astronomy of Nicolaus Copernicus University in Toruń.  
We acknowledge ISCRA for awarding this project access to the LEONARDO supercomputer, owned by the EuroHPC Joint Undertaking, hosted by CINECA (Italy).
EDT and NP were supported by the European Research Council (ERC) under grant agreement no.\ 101040751.
\end{acknowledgements}

\bibliographystyle{aa} 
\bibliography{biblio}

@ARTICLE{Lockman+2026,
       author = {{Lockman}, Felix J. and {Di Teodoro}, Enrico M. and {Cary}, Savannah and {Fox}, Andrew J.},
        title = "{High-velocity Neutral Gas in the Fermi Bubbles: New Kinematic Limits and Spatial Structure}",
      journal = {ApJ},
         year = 2026,
        month = oct,
       volume = {1009},
       number = {2},
          eid = {147},
        pages = {147},
          doi = {10.3847/1538-4357/ae99db},
       adsurl = {https://ui.adsabs.harvard.edu/abs/2026ApJ..1009..147L}
}

@ARTICLE{2024ApJ...964...99A,
       author = {{Armillotta}, Lucia and {Ostriker}, Eve C. and {Kim}, Chang-Goo and {Jiang}, Yan-Fei},
        title = "{Cosmic-Ray Acceleration of Galactic Outflows in Multiphase Gas}",
      journal = {\apj},
         year = 2024,
        month = mar,
       volume = {964},
       number = {1},
          eid = {99},
        pages = {99},
          doi = {10.3847/1538-4357/ad1e5c},
archivePrefix = {arXiv},
       eprint = {2401.04169},
 primaryClass = {astro-ph.GA},
       adsurl = {https://ui.adsabs.harvard.edu/abs/2024ApJ...964...99A}
}

@ARTICLE{Kim+2026,
       author = {{Kim}, Chang-Goo and {Armillotta}, Lucia and {Ostriker}, Eve C. and {Moon}, Sanghyuk and {Lancaster}, Lachlan and {Kim}, Jeong-Gyu and {Linzer}, Nora B. and {Hix}, Ronan N.},
        title = "{Cosmic-Ray Feedback in Galactic Disks: Star Formation, Cosmic-Ray Transport, and Multiphase Outflows in TIGRESS++ Simulations}",
      journal = {ApJ},
         year = 2026,
        month = aug,
       volume = {1006},
       number = {2},
          eid = {123},
        pages = {123},
          doi = {10.3847/1538-4357/ae76d6},
archivePrefix = {arXiv},
       eprint = {2605.26238},
 primaryClass = {astro-ph.GA},
       adsurl = {https://ui.adsabs.harvard.edu/abs/2026ApJ..1006..123K}
}

@ARTICLE{Sternberg+2014,
       author = {{Sternberg}, Amiel and {Le Petit}, Franck and {Roueff}, Evelyne and {Le Bourlot}, Jacques},
        title = "{H I-to-H$_{2}$ Transitions and H I Column Densities in Galaxy Star-forming Regions}",
      journal = {ApJ},
         year = 2014,
        month = jul,
       volume = {790},
       number = {1},
          eid = {10},
        pages = {10},
          doi = {10.1088/0004-637X/790/1/10},
archivePrefix = {arXiv},
       eprint = {1404.5042},
 primaryClass = {astro-ph.GA},
       adsurl = {https://ui.adsabs.harvard.edu/abs/2014ApJ...790...10S}
}

@ARTICLE{Noon+2026,
       author = {{Noon}, Karlie A. and {Krumholz}, Mark R. and {McClure-Griffiths}, Naomi M. and {Di Teodoro}, Enrico M. and {Armillotta}, Lucia},
        title = "{Modelling the non-equilibrium chemistry of the Milky Way's cold nuclear wind}",
      journal = {MNRAS},
         year = 2026,
        month = may,
       volume = {548},
       number = {3},
          eid = {stag715},
        pages = {stag715},
          doi = {10.1093/mnras/stag715},
archivePrefix = {arXiv},
       eprint = {2602.13580},
 primaryClass = {astro-ph.GA},
       adsurl = {https://ui.adsabs.harvard.edu/abs/2026MNRAS.548ag715N}
}

@ARTICLE{Krumholz+2009,
       author = {{Krumholz}, Mark R. and {McKee}, Christopher F. and {Tumlinson}, Jason},
        title = "{The Atomic-to-Molecular Transition in Galaxies. II: H I and H$_{2}$ Column Densities}",
      journal = {ApJ},
         year = 2009,
        month = mar,
       volume = {693},
       number = {1},
        pages = {216-235},
          doi = {10.1088/0004-637X/693/1/216},
archivePrefix = {arXiv},
       eprint = {0811.0004},
 primaryClass = {astro-ph},
       adsurl = {https://ui.adsabs.harvard.edu/abs/2009ApJ...693..216K}
}

@ARTICLE{Barnes+17,
       author = {{Barnes}, A.~T. and {Longmore}, S.~N. and {Battersby}, C. and {Bally}, J. and {Kruijssen}, J.~M.~D. and {Henshaw}, J.~D. and {Walker}, D.~L.},
        title = "{Star formation rates and efficiencies in the Galactic Centre}",
      journal = {\mnras},
         year = 2017,
        month = aug,
       volume = {469},
       number = {2},
        pages = {2263-2285},
          doi = {10.1093/mnras/stx941},
archivePrefix = {arXiv},
       eprint = {1704.03572},
 primaryClass = {astro-ph.GA},
       adsurl = {https://ui.adsabs.harvard.edu/abs/2017MNRAS.469.2263B}
}

@ARTICLE{2025arXiv250713665Z,
       author = {{Zhang}, Ruiyu and {Guo}, Fulai and {Xie}, Shaokun and {Zhang}, Ruofei and {Wang}, Shumin and {Mou}, Guobin and {Duan}, Xiaodong},
        title = "{The Double-Episode Jet Genesis of the eROSITA and Fermi Bubbles}",
      journal = {arXiv e-prints},
         year = 2025,
        month = jul,
          eid = {arXiv:2507.13665},
        pages = {arXiv:2507.13665},
          doi = {10.48550/arXiv.2507.13665},
archivePrefix = {arXiv},
       eprint = {2507.13665},
 primaryClass = {astro-ph.HE},
       adsurl = {https://ui.adsabs.harvard.edu/abs/2025arXiv250713665Z}
}

@ARTICLE{2015MNRAS.454..238W,
       author = {{Walch}, S. and {Girichidis}, P. and {Naab}, T. and {Gatto}, A. and {Glover}, S.~C.~O. and {W{\"u}nsch}, R. and {Klessen}, R.~S. and {Clark}, P.~C. and {Peters}, T. and {Derigs}, D. and {Baczynski}, C.},
        title = "{The SILCC (SImulating the LifeCycle of molecular Clouds) project - I. Chemical evolution of the supernova-driven ISM}",
      journal = {\mnras},
         year = 2015,
        month = nov,
       volume = {454},
       number = {1},
        pages = {238-268},
          doi = {10.1093/mnras/stv1975},
archivePrefix = {arXiv},
       eprint = {1412.2749},
 primaryClass = {astro-ph.GA},
       adsurl = {https://ui.adsabs.harvard.edu/abs/2015MNRAS.454..238W}
}

@ARTICLE{Bordoloi+2025,
       author = {{Bordoloi}, Rongmon and {Fox}, Andrew J. and {Lockman}, Felix J.},
        title = "{A New High-latitude H I Cloud Complex Entrained in the Northern Fermi Bubble}",
      journal = {ApJL},
         year = 2025,
        month = jul,
       volume = {987},
       number = {2},
          eid = {L32},
        pages = {L32},
          doi = {10.3847/2041-8213/addd16},
archivePrefix = {arXiv},
       eprint = {2504.21091},
 primaryClass = {astro-ph.GA},
       adsurl = {https://ui.adsabs.harvard.edu/abs/2025ApJ...987L..32B}
}

@ARTICLE{2023MNRAS.522.5529P,
author = {{Peschken}, N. and {Hanasz}, M. and {Naab}, T. and {W{\'o}lta{\'n}ski}, D. and {Gawryszczak}, A.},
        title = "{The phase structure of cosmic ray driven outflows in stream fed disc galaxies}",
      journal = {\mnras},
         year = 2023,
        month = jul,
       volume = {522},
       number = {4},
        pages = {5529-5545},
          doi = {10.1093/mnras/stad1358},
archivePrefix = {arXiv},
       eprint = {2210.17328},
 primaryClass = {astro-ph.GA},
       adsurl = {https://ui.adsabs.harvard.edu/abs/2023MNRAS.522.5529P}
}

@ARTICLE{2013ApJ...777L..38H,
       author = {{Hanasz}, M. and {Lesch}, H. and {Naab}, T. and {Gawryszczak}, A. and
         {Kowalik}, K. and {W{\'o}lta{\'n}ski}, D.},
        title = "{Cosmic Rays Can Drive Strong Outflows from Gas-rich High-redshift Disk Galaxies}",
      journal = {\apjl},
         year = 2013,
        month = nov,
       volume = {777},
       number = {2},
          eid = {L38},
        pages = {L38},
          doi = {10.1088/2041-8205/777/2/L38},
archivePrefix = {arXiv},
       eprint = {1310.3273},
 primaryClass = {astro-ph.GA},
       adsurl = {https://ui.adsabs.harvard.edu/abs/2013ApJ...777L..38H}
}

@INPROCEEDINGS{2010EAS....42..275H,
       author = {{Hanasz}, M. and {Kowalik}, K. and {W{\'o}lta{\'n}ski}, D. and
         {Paw{\l}aszek}, R.},
        title = "{The PIERNIK MHD code - a multi-fluid, non-ideal extension of the relaxing-TVD scheme (I)}",
    booktitle = {EAS Publications Series},
         year = 2010,
       editor = {{Go{\.z}dziewski}, K. and {Niedzielski}, A. and {Schneider}, J.},
       series = {EAS Publications Series},
       volume = {42},
        month = apr,
        pages = {275-280},
          doi = {10.1051/eas/1042029},
archivePrefix = {arXiv},
       eprint = {0812.2161},
 primaryClass = {astro-ph},
       adsurl = {https://ui.adsabs.harvard.edu/abs/2010EAS....42..275H}
}

@INPROCEEDINGS{2010EAS....42..281H,
       author = {{Hanasz}, M. and {Kowalik}, K. and {W{\'o}lta{\'n}ski}, D. and
         {Paw{\l}aszek}, R. and {Kornet}, K.},
        title = "{The PIERNIK MHD code - a multi-fluid, non-ideal extension of the relaxing-TVD scheme (II)}",
    booktitle = {EAS Publications Series},
         year = 2010,
       editor = {{Go{\.z}dziewski}, K. and {Niedzielski}, A. and {Schneider}, J.},
       series = {EAS Publications Series},
       volume = {42},
        month = apr,
        pages = {281-285},
          doi = {10.1051/eas/1042030},
archivePrefix = {arXiv},
       eprint = {0812.2799},
 primaryClass = {astro-ph},
       adsurl = {https://ui.adsabs.harvard.edu/abs/2010EAS....42..281H}
}

@INPROCEEDINGS{2012EAS....56..363H,
       author = {{Hanasz}, M. and {Kowalik}, K. and {W{\'o}lta{\'n}ski}, D. and
         {Paw{\l}aszek}, R.},
        title = "{PIERNIK MHD code - a multi-fluid, non-ideal extension of the relaxing-TVD scheme (III)}",
    booktitle = {EAS Publications Series},
         year = 2012,
       editor = {{de Avillez}, Miguel A.},
       series = {EAS Publications Series},
       volume = {56},
        month = sep,
        pages = {363-366},
          doi = {10.1051/eas/1256059},
archivePrefix = {arXiv},
       eprint = {0812.4839},
 primaryClass = {astro-ph},
       adsurl = {https://ui.adsabs.harvard.edu/abs/2012EAS....56..363H}
}

@INPROCEEDINGS{2012EAS....56..367H,
       author = {{Hanasz}, M. and {Kowalik}, K. and {W{\'o}lta{\'n}ski}, D. and
         {Paw{\l}aszek}, R.},
        title = "{PIERNIK MHD code - a multi-fluid, non-ideal extension of the relaxing-TVD scheme (IV)}",
    booktitle = {EAS Publications Series},
         year = 2012,
       editor = {{de Avillez}, Miguel A.},
       series = {EAS Publications Series},
       volume = {56},
        month = sep,
        pages = {367-370},
          doi = {10.1051/eas/1256060},
archivePrefix = {arXiv},
       eprint = {0901.0104},
 primaryClass = {astro-ph.GA},
       adsurl = {https://ui.adsabs.harvard.edu/abs/2012EAS....56..367H}
}

@ARTICLE{1993ApJS...88..253S,
       author = {{Sutherland}, Ralph S. and {Dopita}, M.~A.},
        title = "{Cooling Functions for Low-Density Astrophysical Plasmas}",
      journal = {\apjs},
         year = 1993,
        month = sep,
       volume = {88},
        pages = {253},
          doi = {10.1086/191823},
       adsurl = {https://ui.adsabs.harvard.edu/abs/1993ApJS...88..253S}
}

@ARTICLE{2017ApJ...846..133K,
       author = {{Kim}, Chang-Goo and {Ostriker}, Eve C.},
        title = "{Three-phase Interstellar Medium in Galaxies Resolving Evolution with Star Formation and Supernova Feedback (TIGRESS): Algorithms, Fiducial Model, and Convergence}",
      journal = {\apj},
         year = 2017,
        month = sep,
       volume = {846},
       number = {2},
          eid = {133},
        pages = {133},
          doi = {10.3847/1538-4357/aa8599},
archivePrefix = {arXiv},
       eprint = {1612.03918},
 primaryClass = {astro-ph.GA},
       adsurl = {https://ui.adsabs.harvard.edu/abs/2017ApJ...846..133K}
}

@ARTICLE{2024A&A...692A.216H,
       author = {{Hunter}, Glen H. and {Sormani}, Mattia C. and {Beckmann}, Jan P. and {Vasiliev}, Eugene and {Glover}, Simon C.~O. and {Klessen}, Ralf S. and {Soler}, Juan D. and {Brucy}, No{\'e} and {Girichidis}, Philipp and {G{\"o}ller}, Junia and {Ohlin}, Loke and {Tress}, Robin and {Molinari}, Sergio and {Gerhard}, Ortwin and {Benedettini}, Milena and {Smith}, Rowan and {Hennebelle}, Patrick and {Testi}, Leonardo},
        title = "{Testing kinematic distances under a realistic Galactic potential: Investigating systematic errors in the kinematic distance method arising from a non-axisymmetric potential}",
      journal = {\aap},
         year = 2024,
        month = dec,
       volume = {692},
          eid = {A216},
        pages = {A216},
          doi = {10.1051/0004-6361/202450000},
archivePrefix = {arXiv},
       eprint = {2403.18000},
 primaryClass = {astro-ph.GA},
       adsurl = {https://ui.adsabs.harvard.edu/abs/2024A&A...692A.216H}
}

@ARTICLE{2009ApJS..181..391T,
       author = {{Townsend}, R.~H.~D.},
        title = "{An Exact Integration Scheme for Radiative Cooling in Hydrodynamical Simulations}",
      journal = {\apjs},
         year = 2009,
        month = apr,
       volume = {181},
       number = {2},
        pages = {391-397},
          doi = {10.1088/0067-0049/181/2/391},
archivePrefix = {arXiv},
       eprint = {0901.3146},
 primaryClass = {astro-ph.SR},
       adsurl = {https://ui.adsabs.harvard.edu/abs/2009ApJS..181..391T}
}

@ARTICLE{1992MNRAS.259..345A,
       author = {{Athanassoula}, E.},
        title = "{The existence and shapes of dust lanes in galactic bars.}",
      journal = {\mnras},
         year = 1992,
        month = nov,
       volume = {259},
        pages = {345-364},
          doi = {10.1093/mnras/259.2.345},
       adsurl = {https://ui.adsabs.harvard.edu/abs/1992MNRAS.259..345A}
}

@ARTICLE{2024A&A...690A.399Y,
       author = {{Yeung}, Michael C.~H. and {Ponti}, Gabriele and {Freyberg}, Michael J. and {Dennerl}, Konrad and {Liu}, Teng and {Locatelli}, Nicola and {Mayer}, Martin G.~F. and {Sanders}, Jeremy S. and {Sasaki}, Manami and {Strong}, Andy and {Zhang}, Yi and {Zheng}, Xueying and {Gatuzz}, Efrain},
        title = "{The SRG/eROSITA diffuse soft X-ray background: I. The local hot bubble in the western Galactic hemisphere}",
      journal = {\aap},
         year = 2024,
        month = oct,
       volume = {690},
          eid = {A399},
        pages = {A399},
          doi = {10.1051/0004-6361/202451045},
archivePrefix = {arXiv},
       eprint = {2410.23345},
 primaryClass = {astro-ph.GA},
       adsurl = {https://ui.adsabs.harvard.edu/abs/2024A&A...690A.399Y}
}

@ARTICLE{McClure-Griffiths+13,
   author = {{McClure-Griffiths}, N.~M. and {Green}, J.~A. and {Hill}, A.~S. and 
	{Lockman}, F.~J. and {Dickey}, J.~M. and {Gaensler}, B.~M. and 
	{Green}, A.~J.},
    title = "{Atomic Hydrogen in a Galactic Center Outflow}",
  journal = {\apjl},
archivePrefix = "arXiv",
   eprint = {1304.7538},
     year = 2013,
    month = jun,
   volume = 770,
      eid = {L4},
    pages = {L4},
      doi = {10.1088/2041-8205/770/1/L4},
   adsurl = {http://adsabs.harvard.edu/abs/2013ApJ...770L...4M}
}

@ARTICLE{DiT+18,
       author = {{Di Teodoro}, Enrico M. and {McClure-Griffiths}, N.~M. and
         {Lockman}, Felix J. and {Denbo}, Sara R. and {Endsley}, Ryan and
         {Ford}, H. Alyson and {Harrington}, Kevin},
        title = "{Blowing in the Milky Way Wind: Neutral Hydrogen Clouds Tracing the Galactic Nuclear Outflow}",
      journal = {\apj},
         year = "2018",
        month = "Mar",
       volume = {855},
       number = {1},
          eid = {33},
        pages = {33},
          doi = {10.3847/1538-4357/aaad6a},
archivePrefix = {arXiv},
       eprint = {1802.02152},
 primaryClass = {astro-ph.GA},
       adsurl = {https://ui.adsabs.harvard.edu/abs/2018ApJ...855...33D}
}

@ARTICLE{DiTeodoro+2020,
       author = {{Di Teodoro}, Enrico M. and {McClure-Griffiths}, N.~M. and {Lockman}, Felix J. and {Armillotta}, Lucia},
        title = "{Cold gas in the Milky Way's nuclear wind}",
      journal = {\nat},
         year = 2020,
        month = aug,
       volume = {584},
       number = {7821},
        pages = {364-367},
          doi = {10.1038/s41586-020-2595-z},
archivePrefix = {arXiv},
       eprint = {2008.09121},
 primaryClass = {astro-ph.GA},
       adsurl = {https://ui.adsabs.harvard.edu/abs/2020Natur.584..364D}
}

@ARTICLE{Heyer+2025,
       author = {{Heyer}, M. and {Di Teodoro}, E. and {Loinard}, L. and {Lockman}, F.~J. and {McClure-Griffiths}, N.~M. and {Wang}, Q.~D.},
        title = "{Cold molecular gas in the hot nuclear wind of the Milky Way}",
      journal = {A\&A},
         year = 2025,
        month = mar,
       volume = {695},
          eid = {A60},
        pages = {A60},
          doi = {10.1051/0004-6361/202452875},
archivePrefix = {arXiv},
       eprint = {2502.08771},
 primaryClass = {astro-ph.GA},
       adsurl = {https://ui.adsabs.harvard.edu/abs/2025A&A...695A..60H}
}

@ARTICLE{Su+10,
   author = {{Su}, M. and {Slatyer}, T.~R. and {Finkbeiner}, D.~P.},
    title = "{Giant Gamma-ray Bubbles from Fermi-LAT: Active Galactic Nucleus Activity or Bipolar Galactic Wind?}",
  journal = {\apj},
archivePrefix = "arXiv",
   eprint = {1005.5480},
 primaryClass = "astro-ph.HE",
     year = 2010,
    month = dec,
   volume = 724,
    pages = {1044-1082},
      doi = {10.1088/0004-637X/724/2/1044},
   adsurl = {http://adsabs.harvard.edu/abs/2010ApJ...724.1044S}
}

@ARTICLE{2026A&A...714A..85Y,
       author = {{Yeung}, Michael C.~H. and {Mayer}, Martin G.~F. and {Strong}, Andy and {Freyberg}, Michael J. and {Ponti}, Gabriele and {Dennerl}, Konrad and {Mao}, Junjie and {Sasaki}, Manami and {Zheng}, Xueying and {Sanders}, Jeremy S. and {Zhang}, Yi and {Liu}, Jiejia and {Gu}, Liyi and {Becker}, Werner and {Haberl}, Frank and {Liu}, Teng and {Merloni}, Andrea and {Predehl}, Peter},
        title = "{The SRG/eROSITA diffuse soft X-ray background: II. Spectra and morphology of the eROSITA bubbles in the western Galactic hemisphere}",
      journal = {\aap},
         year = 2026,
        month = oct,
       volume = {714},
          eid = {A85},
        pages = {A85},
          doi = {10.1051/0004-6361/202558206},
       adsurl = {https://ui.adsabs.harvard.edu/abs/2026A&A...714A..85Y}
}

@misc{lin2026illuminatingm82simulatingxray,
      title={Illuminating M82: Simulating X-ray Emission from Galactic Winds in a Starburst Galaxy}, 
      author={Allison Lin and Lachlan Lancaster and Viraj Pandya and Drummond B. Fielding and Greg L. Bryan and John A. ZuHone and Sebastian Lopez and Laura A. Lopez and Christopher Carr},
      year={2026},
      eprint={2607.12087},
      archivePrefix={arXiv},
      primaryClass={astro-ph.GA},
      url={https://arxiv.org/abs/2607.12087}, 
}

@ARTICLE{2018ApJ...860..135S,
       author = {{Schneider}, Evan E. and {Robertson}, Brant E.},
        title = "{Introducing CGOLS: The Cholla Galactic Outflow Simulation Suite}",
      journal = {\apj},
         year = 2018,
        month = jun,
       volume = {860},
       number = {2},
          eid = {135},
        pages = {135},
          doi = {10.3847/1538-4357/aac329},
archivePrefix = {arXiv},
       eprint = {1803.01008},
 primaryClass = {astro-ph.GA},
       adsurl = {https://ui.adsabs.harvard.edu/abs/2018ApJ...860..135S}
}

@ARTICLE{Predehl+2020,
       author = {{Predehl}, P. and {Sunyaev}, R.~A. and {Becker}, W. and {Brunner}, H. and {Burenin}, R. and {Bykov}, A. and {Cherepashchuk}, A. and {Chugai}, N. and {Churazov}, E. and {Doroshenko}, V. and {Eismont}, N. and {Freyberg}, M. and {Gilfanov}, M. and {Haberl}, F. and {Khabibullin}, I. and {Krivonos}, R. and {Maitra}, C. and {Medvedev}, P. and {Merloni}, A. and {Nandra}, K. and {Nazarov}, V. and {Pavlinsky}, M. and {Ponti}, G. and {Sanders}, J.~S. and {Sasaki}, M. and {Sazonov}, S. and {Strong}, A.~W. and {Wilms}, J.},
        title = "{Detection of large-scale X-ray bubbles in the Milky Way halo}",
      journal = {\nat},
         year = 2020,
        month = dec,
       volume = {588},
       number = {7837},
        pages = {227-231},
          doi = {10.1038/s41586-020-2979-0},
archivePrefix = {arXiv},
       eprint = {2012.05840},
 primaryClass = {astro-ph.GA},
       adsurl = {https://ui.adsabs.harvard.edu/abs/2020Natur.588..227P}
}

@ARTICLE{Fox+15,
   author = {{Fox}, A.~J. and {Bordoloi}, R. and {Savage}, B.~D. and {Lockman}, F.~J. and 
	{Jenkins}, E.~B. and {Wakker}, B.~P. and {Bland-Hawthorn}, J. and 
	{Hernandez}, S. and {Kim}, T.-S. and {Benjamin}, R.~A. and {Bowen}, D.~V. and 
	{Tumlinson}, J.},
    title = "{Probing the Fermi Bubbles in Ultraviolet Absorption: A Spectroscopic Signature of the Milky Way's Biconical Nuclear Outflow}",
  journal = {\apjl},
archivePrefix = "arXiv",
   eprint = {1412.1480},
     year = 2015,
    month = jan,
   volume = 799,
      eid = {L7},
    pages = {L7},
      doi = {10.1088/2041-8205/799/1/L7},
   adsurl = {http://adsabs.harvard.edu/abs/2015ApJ...799L...7F}
}

@ARTICLE{Cashman+2023,
       author = {{Cashman}, Frances H. and {Fox}, Andrew J. and {Wakker}, Bart P. and {Ashley}, Trisha and {Massa}, Derck and {Jenkins}, Edward B. and {Krishnarao}, Dhanesh and {Benjamin}, Robert A. and {Bordoloi}, Rongmon and {Kim}, Tae-Sun},
        title = "{Caught in the Act: A Metal-rich High-velocity Cloud in the Inner Galaxy}",
      journal = {ApJ},
         year = 2023,
        month = feb,
       volume = {944},
       number = {1},
          eid = {65},
        pages = {65},
          doi = {10.3847/1538-4357/acaaa5},
archivePrefix = {arXiv},
       eprint = {2210.11583},
 primaryClass = {astro-ph.GA},
       adsurl = {https://ui.adsabs.harvard.edu/abs/2023ApJ...944...65C}
}

@ARTICLE{Bordoloi+17,
   author = {{Bordoloi}, R. and {Fox}, A.~J. and {Lockman}, F.~J. and {Wakker}, B.~P. and 
	{Jenkins}, E.~B. and {Savage}, B.~D. and {Hernandez}, S. and 
	{Tumlinson}, J. and {Bland-Hawthorn}, J. and {Kim}, T.-S.},
    title = "{Mapping the Nuclear Outflow of the Milky Way: Studying the Kinematics and Spatial Extent of the Northern Fermi Bubble}",
  journal = {\apj},
archivePrefix = "arXiv",
   eprint = {1612.01578},
     year = 2017,
    month = jan,
   volume = 834,
      eid = {191},
    pages = {191},
      doi = {10.3847/1538-4357/834/2/191},
   adsurl = {http://adsabs.harvard.edu/abs/2017ApJ...834..191B}
}

@ARTICLE{Ashley+2022,
       author = {{Ashley}, Trisha and {Fox}, Andrew J. and {Cashman}, Frances H. and {Lockman}, Felix J. and {Bordoloi}, Rongmon and {Jenkins}, Edward B. and {Wakker}, Bart P. and {Karim}, Tanveer},
        title = "{Diverse metallicities of Fermi bubble clouds indicate dual origins in the disk and halo}",
      journal = {NatAs},
         year = 2022,
        month = jul,
       volume = {6},
        pages = {968-975},
          doi = {10.1038/s41550-022-01720-0},
archivePrefix = {arXiv},
       eprint = {2207.08838},
 primaryClass = {astro-ph.GA},
       adsurl = {https://ui.adsabs.harvard.edu/abs/2022NatAs...6..968A}
}

@ARTICLE{2012ApJ...746..125H,
       author = {{Haardt}, Francesco and {Madau}, Piero},
        title = "{Radiative Transfer in a Clumpy Universe. IV. New Synthesis Models of the Cosmic UV/X-Ray Background}",
      journal = {\apj},
         year = 2012,
        month = feb,
       volume = {746},
       number = {2},
          eid = {125},
        pages = {125},
          doi = {10.1088/0004-637X/746/2/125},
       adsurl = {https://ui.adsabs.harvard.edu/abs/2012ApJ...746..125H}
}

@ARTICLE{Cunha+2007,
       author = {{Cunha}, Katia and {Sellgren}, Kris and {Smith}, Verne V. and {Ramirez}, Solange V. and {Blum}, Robert D. and {Terndrup}, Donald M.},
        title = "{Chemical Abundances of Luminous Cool Stars in the Galactic Center from High-Resolution Infrared Spectroscopy}",
      journal = {ApJ},
         year = 2007,
        month = nov,
       volume = {669},
       number = {2},
        pages = {1011-1023},
          doi = {10.1086/521813},
archivePrefix = {arXiv},
       eprint = {0707.2610},
 primaryClass = {astro-ph},
       adsurl = {https://ui.adsabs.harvard.edu/abs/2007ApJ...669.1011C}
}

@ARTICLE{Zubovas+2012,
       author = {{Zubovas}, Kastytis and {Nayakshin}, Sergei},
        title = "{Fermi bubbles in the Milky Way: the closest AGN feedback laboratory courtesy of Sgr A*?}",
      journal = {MNRAS},
         year = 2012,
        month = jul,
       volume = {424},
       number = {1},
        pages = {666-683},
          doi = {10.1111/j.1365-2966.2012.21250.x},
archivePrefix = {arXiv},
       eprint = {1203.3060},
 primaryClass = {astro-ph.GA},
       adsurl = {https://ui.adsabs.harvard.edu/abs/2012MNRAS.424..666Z}
}

@ARTICLE{Bland-Hawthorn+2019,
       author = {{Bland-Hawthorn}, Joss and {Maloney}, Philip R. and {Sutherland}, Ralph and {Groves}, Brent and {Guglielmo}, Magda and {Li}, Wenhao and {Curzons}, Andrew and {Cecil}, Gerald and {Fox}, Andrew J.},
        title = "{The Large-scale Ionization Cones in the Galaxy}",
      journal = {ApJ},
         year = 2019,
        month = nov,
       volume = {886},
       number = {1},
          eid = {45},
        pages = {45},
          doi = {10.3847/1538-4357/ab44c8},
archivePrefix = {arXiv},
       eprint = {1910.02225},
 primaryClass = {astro-ph.GA},
       adsurl = {https://ui.adsabs.harvard.edu/abs/2019ApJ...886...45B}
}

@ARTICLE{Crocker+2015,
       author = {{Crocker}, Roland M. and {Bicknell}, Geoffrey V. and {Taylor}, Andrew M. and {Carretti}, Ettore},
        title = "{A Unified Model of the Fermi Bubbles, Microwave Haze, and Polarized Radio Lobes: Reverse Shocks in the Galactic Center{\textquoteright}s Giant Outflows}",
      journal = {ApJ},
         year = 2015,
        month = aug,
       volume = {808},
       number = {2},
          eid = {107},
        pages = {107},
          doi = {10.1088/0004-637X/808/2/107},
archivePrefix = {arXiv},
       eprint = {1412.7510},
 primaryClass = {astro-ph.HE},
       adsurl = {https://ui.adsabs.harvard.edu/abs/2015ApJ...808..107C}
}

@ARTICLE{Smith+2018,
       author = {{Smith}, Matthew C. and {Sijacki}, Debora and {Shen}, Sijing},
        title = "{Supernova feedback in numerical simulations of galaxy formation: separating physics from numerics}",
      journal = {MNRAS},
         year = 2018,
        month = jul,
       volume = {478},
       number = {1},
        pages = {302-331},
          doi = {10.1093/mnras/sty994},
archivePrefix = {arXiv},
       eprint = {1709.03515},
 primaryClass = {astro-ph.GA},
       adsurl = {https://ui.adsabs.harvard.edu/abs/2018MNRAS.478..302S}
}

@ARTICLE{Smith+2021,
       author = {{Smith}, Matthew C. and {Bryan}, Greg L. and {Somerville}, Rachel S. and {Hu}, Chia-Yu and {Teyssier}, Romain and {Burkhart}, Blakesley and {Hernquist}, Lars},
        title = "{Efficient early stellar feedback can suppress galactic outflows by reducing supernova clustering}",
      journal = {MNRAS},
         year = 2021,
        month = sep,
       volume = {506},
       number = {3},
        pages = {3882-3915},
          doi = {10.1093/mnras/stab1896},
archivePrefix = {arXiv},
       eprint = {2009.11309},
 primaryClass = {astro-ph.GA},
       adsurl = {https://ui.adsabs.harvard.edu/abs/2021MNRAS.506.3882S}
}

@ARTICLE{Fielding+2017,
       author = {{Fielding}, Drummond and {Quataert}, Eliot and {Martizzi}, Davide and {Faucher-Gigu{\`e}re}, Claude-Andr{\'e}},
        title = "{How supernovae launch galactic winds?}",
      journal = {MNRAS},
         year = 2017,
        month = sep,
       volume = {470},
       number = {1},
        pages = {L39-L43},
          doi = {10.1093/mnrasl/slx072},
archivePrefix = {arXiv},
       eprint = {1704.01579},
 primaryClass = {astro-ph.GA},
       adsurl = {https://ui.adsabs.harvard.edu/abs/2017MNRAS.470L..39F}
}

@ARTICLE{Hu+2019,
       author = {{Hu}, Chia-Yu},
        title = "{Supernova-driven winds in simulated dwarf galaxies}",
      journal = {MNRAS},
         year = 2019,
        month = mar,
       volume = {483},
       number = {3},
        pages = {3363-3381},
          doi = {10.1093/mnras/sty3252},
archivePrefix = {arXiv},
       eprint = {1805.06614},
 primaryClass = {astro-ph.GA},
       adsurl = {https://ui.adsabs.harvard.edu/abs/2019MNRAS.483.3363H}
}

@ARTICLE{Rathjen+2021,
       author = {{Rathjen}, Tim-Eric and {Naab}, Thorsten and {Girichidis}, Philipp and {Walch}, Stefanie and {W{\"u}nsch}, Richard and {Dinnbier}, Frantis̆ek and {Seifried}, Daniel and {Klessen}, Ralf S. and {Glover}, Simon C.~O.},
        title = "{SILCC VI - Multiphase ISM structure, stellar clustering, and outflows with supernovae, stellar winds, ionizing radiation, and cosmic rays}",
      journal = {MNRAS},
         year = 2021,
        month = jun,
       volume = {504},
       number = {1},
        pages = {1039-1061},
          doi = {10.1093/mnras/stab900},
archivePrefix = {arXiv},
       eprint = {2103.14128},
 primaryClass = {astro-ph.GA},
       adsurl = {https://ui.adsabs.harvard.edu/abs/2021MNRAS.504.1039R}
}

@ARTICLE{Li+2017,
       author = {{Li}, Miao and {Bryan}, Greg L. and {Ostriker}, Jeremiah P.},
        title = "{Quantifying Supernovae-driven Multiphase Galactic Outflows}",
      journal = {ApJ},
         year = 2017,
        month = jun,
       volume = {841},
       number = {2},
          eid = {101},
        pages = {101},
          doi = {10.3847/1538-4357/aa7263},
archivePrefix = {arXiv},
       eprint = {1610.08971},
 primaryClass = {astro-ph.GA},
       adsurl = {https://ui.adsabs.harvard.edu/abs/2017ApJ...841..101L}
}

@ARTICLE{Gentry+2017,
       author = {{Gentry}, Eric S. and {Krumholz}, Mark R. and {Dekel}, Avishai and {Madau}, Piero},
        title = "{Enhanced momentum feedback from clustered supernovae}",
      journal = {MNRAS},
         year = 2017,
        month = feb,
       volume = {465},
       number = {2},
        pages = {2471-2488},
          doi = {10.1093/mnras/stw2746},
archivePrefix = {arXiv},
       eprint = {1606.01242},
 primaryClass = {astro-ph.GA},
       adsurl = {https://ui.adsabs.harvard.edu/abs/2017MNRAS.465.2471G}
}

@ARTICLE{Sarkar+2015,
       author = {{Sarkar}, Kartick Chandra and {Nath}, Biman B. and {Sharma}, Prateek},
        title = "{Multiwavelength features of Fermi bubbles as signatures of a Galactic wind}",
      journal = {MNRAS},
         year = 2015,
        month = nov,
       volume = {453},
       number = {4},
        pages = {3827-3838},
          doi = {10.1093/mnras/stv1806},
archivePrefix = {arXiv},
       eprint = {1505.03634},
 primaryClass = {astro-ph.GA},
       adsurl = {https://ui.adsabs.harvard.edu/abs/2015MNRAS.453.3827S}
}

@ARTICLE{Morris1996,
       author = {{Morris}, Mark and {Serabyn}, Eugene},
        title = "{The Galactic Center Environment}",
      journal = {\araa},
         year = 1996,
        month = jan,
       volume = {34},
        pages = {645-702},
          doi = {10.1146/annurev.astro.34.1.645},
       adsurl = {https://ui.adsabs.harvard.edu/abs/1996ARA&A..34..645M}
}

@INPROCEEDINGS{Henshaw2023,
       author = {{Henshaw}, J.~D. and {Barnes}, A.~T. and {Battersby}, C. and {Ginsburg}, A. and {Sormani}, M.~C. and {Walker}, D.~L.},
        title = "{Star Formation in the Central Molecular Zone of the Milky Way}",
    booktitle = {Protostars and Planets VII},
         year = 2023,
       editor = {{Inutsuka}, S. and {Aikawa}, Y. and {Muto}, T. and {Tomida}, K. and {Tamura}, M.},
       series = {Astronomical Society of the Pacific Conference Series},
       volume = {534},
        month = jul,
        pages = {83},
          doi = {10.48550/arXiv.2203.11223},
archivePrefix = {arXiv},
       eprint = {2203.11223},
 primaryClass = {astro-ph.GA},
       adsurl = {https://ui.adsabs.harvard.edu/abs/2023ASPC..534...83H}
}

@ARTICLE{Cashman+2021,
       author = {{Cashman}, Frances H. and {Fox}, Andrew J. and {Savage}, Blair D. and {Wakker}, Bart P. and {Krishnarao}, Dhanesh and {Benjamin}, Robert A. and {Richter}, Philipp and {Ashley}, Trisha and {Jenkins}, Edward B. and {Lockman}, Felix J. and {Bordoloi}, Rongmon and {Kim}, Tae-Sun},
        title = "{Molecular Gas within the Milky Way's Nuclear Wind}",
      journal = {ApJL},
         year = 2021,
        month = dec,
       volume = {923},
       number = {1},
          eid = {L11},
        pages = {L11},
          doi = {10.3847/2041-8213/ac3cbc},
archivePrefix = {arXiv},
       eprint = {2112.03335},
 primaryClass = {astro-ph.GA},
       adsurl = {https://ui.adsabs.harvard.edu/abs/2021ApJ...923L..11C}
}

@ARTICLE{Scannapieco&Bruggen15,
   author = {{Scannapieco}, E. and {Br{\"u}ggen}, M.},
    title = "{The Launching of Cold Clouds by Galaxy Outflows. I. Hydrodynamic Interactions with Radiative Cooling}",
  journal = {\apj},
archivePrefix = "arXiv",
   eprint = {1503.06800},
     year = 2015,
    month = jun,
   volume = 805,
      eid = {158},
    pages = {158},
      doi = {10.1088/0004-637X/805/2/158},
   adsurl = {http://adsabs.harvard.edu/abs/2015ApJ...805..158S}
}

@ARTICLE{Schneider+17,
       author = {{Schneider}, Evan E. and {Robertson}, Brant E.},
        title = "{Hydrodynamical Coupling of Mass and Momentum in Multiphase Galactic Winds}",
      journal = {\apj},
         year = "2017",
        month = "Jan",
       volume = {834},
       number = {2},
          eid = {144},
        pages = {144},
          doi = {10.3847/1538-4357/834/2/144},
archivePrefix = {arXiv},
       eprint = {1607.01788},
 primaryClass = {astro-ph.GA},
       adsurl = {https://ui.adsabs.harvard.edu/abs/2017ApJ...834..144S}
}

@ARTICLE{Gronke+20,
       author = {{Gronke}, Max and {Oh}, S. Peng},
        title = "{How cold gas continuously entrains mass and momentum from a hot wind}",
      journal = {\mnras},
         year = 2020,
        month = feb,
       volume = {492},
       number = {2},
        pages = {1970-1990},
          doi = {10.1093/mnras/stz3332},
archivePrefix = {arXiv},
       eprint = {1907.04771},
 primaryClass = {astro-ph.GA},
       adsurl = {https://ui.adsabs.harvard.edu/abs/2020MNRAS.492.1970G}
}

@article{Sormani+15,
    author = {Sormani, Mattia C. and Binney, James and Magorrian, John},
    title = {Gas flow in barred potentials},
    journal = {Monthly Notices of the Royal Astronomical Society},
    volume = {449},
    number = {3},
    pages = {2421-2435},
    year = {2015},
    month = {04},
    issn = {0035-8711},
    doi = {10.1093/mnras/stv441},
    url = {https://doi.org/10.1093/mnras/stv441},
    eprint = {https://academic.oup.com/mnras/article-pdf/449/3/2421/9383390/stv441.pdf},
}

@ARTICLE{2015MNRAS.451.3021D,
       author = {{Di Teodoro}, E.~M. and {Fraternali}, F.},
        title = "{$^{3D}$ BAROLO: a new 3D algorithm to derive rotation curves of galaxies}",
      journal = {\mnras},
         year = 2015,
        month = aug,
       volume = {451},
       number = {3},
        pages = {3021-3033},
          doi = {10.1093/mnras/stv1213},
archivePrefix = {arXiv},
       eprint = {1505.07834},
 primaryClass = {astro-ph.GA},
       adsurl = {https://ui.adsabs.harvard.edu/abs/2015MNRAS.451.3021D}
}

@ARTICLE{2020RAA....20..159S,
       author = {{Shen}, Juntai and {Zheng}, Xing-Wu},
        title = "{The bar and spiral arms in the Milky Way: structure and kinematics}",
      journal = {Research in Astronomy and Astrophysics},
         year = 2020,
        month = oct,
       volume = {20},
       number = {10},
          eid = {159},
        pages = {159},
          doi = {10.1088/1674-4527/20/10/159},
archivePrefix = {arXiv},
       eprint = {2012.10130},
 primaryClass = {astro-ph.GA},
       adsurl = {https://ui.adsabs.harvard.edu/abs/2020RAA....20..159S}
}

@ARTICLE{2019MNRAS.488.4663S,
       author = {{Sormani}, Mattia C. and {Tre{\ss}}, Robin G. and {Glover}, Simon C.~O. and {Klessen}, Ralf S. and {Barnes}, Ashley T. and {Battersby}, Cara D. and {Clark}, Paul C. and {Hatchfield}, H. Perry and {Smith}, Rowan J.},
        title = "{The geometry of the gas surrounding the Central Molecular Zone: on the origin of localized molecular clouds with extreme velocity dispersions}",
      journal = {\mnras},
         year = 2019,
        month = oct,
       volume = {488},
       number = {4},
        pages = {4663-4673},
          doi = {10.1093/mnras/stz2054},
archivePrefix = {arXiv},
       eprint = {1906.10129},
 primaryClass = {astro-ph.GA},
       adsurl = {https://ui.adsabs.harvard.edu/abs/2019MNRAS.488.4663S}
}

@ARTICLE{2025ApJ...981..154Z,
       author = {{Zhang}, Mengfei and {Li}, Miao and {Zhu}, Peixin},
        title = "{The Starburst Acceleration of High-velocity Clouds in the Galactic Center}",
      journal = {\apj},
         year = 2025,
        month = mar,
       volume = {981},
       number = {2},
          eid = {154},
        pages = {154},
          doi = {10.3847/1538-4357/adb619},
archivePrefix = {arXiv},
       eprint = {2502.17141},
 primaryClass = {astro-ph.GA},
       adsurl = {https://ui.adsabs.harvard.edu/abs/2025ApJ...981..154Z}
}

@ARTICLE{2017ApJS..232...25S,
       author = {{Savage}, Blair D. and {Kim}, Tae-Sun and {Fox}, Andrew J. and {Massa}, Derck and {Bordoloi}, Rongmon and {Jenkins}, Edward. B. and {Lehner}, Nicolas and {Bland-Hawthorn}, Joss and {Lockman}, Felix J. and {Hernandez}, Svea and {Wakker}, Bart P.},
        title = "{Probing the Outflowing Multiphase Gas {\ensuremath{\sim}}1 kpc below the Galactic Center}",
      journal = {\apjs},
         year = 2017,
        month = oct,
       volume = {232},
       number = {2},
          eid = {25},
        pages = {25},
          doi = {10.3847/1538-4365/aa8f4c},
archivePrefix = {arXiv},
       eprint = {1707.06942},
 primaryClass = {astro-ph.GA},
       adsurl = {https://ui.adsabs.harvard.edu/abs/2017ApJS..232...25S}
}

@ARTICLE{2012ApJ...756..181G,
       author = {{Guo}, Fulai and {Mathews}, William G.},
        title = "{The Fermi Bubbles. I. Possible Evidence for Recent AGN Jet Activity in the Galaxy}",
      journal = {\apj},
         year = 2012,
        month = sep,
       volume = {756},
       number = {2},
          eid = {181},
        pages = {181},
          doi = {10.1088/0004-637X/756/2/181},
archivePrefix = {arXiv},
       eprint = {1103.0055},
 primaryClass = {astro-ph.HE},
       adsurl = {https://ui.adsabs.harvard.edu/abs/2012ApJ...756..181G}
}

@ARTICLE{2009A&A...504...15P,
       author = {{Popping}, A. and {Dav{\'e}}, R. and {Braun}, R. and {Oppenheimer}, B.~D.},
        title = "{The simulated H I sky at low redshift}",
      journal = {\aap},
         year = 2009,
        month = sep,
       volume = {504},
       number = {1},
        pages = {15-32},
          doi = {10.1051/0004-6361/200911811},
archivePrefix = {arXiv},
       eprint = {0906.3067},
 primaryClass = {astro-ph.CO},
       adsurl = {https://ui.adsabs.harvard.edu/abs/2009A&A...504...15P}
}

@ARTICLE{Lockman+2020,
       author = {{Lockman}, Felix J. and {Di Teodoro}, Enrico M. and {McClure-Griffiths}, N.~M.},
        title = "{Observation of Acceleration of H I Clouds within the Fermi Bubbles}",
      journal = {ApJ},
         year = 2020,
        month = jan,
       volume = {888},
       number = {1},
          eid = {51},
        pages = {51},
          doi = {10.3847/1538-4357/ab55d8},
archivePrefix = {arXiv},
       eprint = {1911.06864},
 primaryClass = {astro-ph.GA},
       adsurl = {https://ui.adsabs.harvard.edu/abs/2020ApJ...888...51L}
}

@ARTICLE{Miller+2013,
       author = {{Miller}, Matthew J. and {Bregman}, Joel N.},
        title = "{The Structure of the Milky Way's Hot Gas Halo}",
      journal = {ApJ},
         year = 2013,
        month = jun,
       volume = {770},
       number = {2},
          eid = {118},
        pages = {118},
          doi = {10.1088/0004-637X/770/2/118},
archivePrefix = {arXiv},
       eprint = {1305.2430},
 primaryClass = {astro-ph.GA},
       adsurl = {https://ui.adsabs.harvard.edu/abs/2013ApJ...770..118M}
}

@ARTICLE{Noon+2023,
       author = {{Noon}, Karlie A. and {Krumholz}, Mark R. and {Di Teodoro}, Enrico M. and {McClure-Griffiths}, Naomi M. and {Lockman}, Felix J. and {Armillotta}, Lucia},
        title = "{Direct observations of the atomic-molecular phase transition in the Milky Way's nuclear wind}",
      journal = {MNRAS},
         year = 2023,
        month = sep,
       volume = {524},
       number = {1},
        pages = {1258-1268},
          doi = {10.1093/mnras/stad1890},
archivePrefix = {arXiv},
       eprint = {2304.06356},
 primaryClass = {astro-ph.GA},
       adsurl = {https://ui.adsabs.harvard.edu/abs/2023MNRAS.524.1258N}
}

@BOOK{Draine+2011,
       author = {{Draine}, Bruce T.},
        title = "{Physics of the Interstellar and Intergalactic Medium}",
         year = 2011,
       adsurl = {https://ui.adsabs.harvard.edu/abs/2011piim.book.....D}
}

@INCOLLECTION{Roberts+1975,
       author = {{Roberts}, Morton S.},
        title = "{Radio Observations of Neutral Hydrogen in Galaxies}",
    booktitle = {Galaxies and the Universe},
         year = 1975,
       editor = {{Sandage}, Allan and {Sandage}, Mary and {Kristian}, Jerome},
        pages = {309},
       adsurl = {https://ui.adsabs.harvard.edu/abs/1975gaun.book..309R}
}

@ARTICLE{Kataoka+2015,
       author = {{Kataoka}, J. and {Tahara}, M. and {Totani}, T. and {Sofue}, Y. and {Inoue}, Y. and {Nakashima}, S. and {Cheung}, C.~C.},
        title = "{Global Structure of Isothermal Diffuse X-Ray Emission along the Fermi Bubbles}",
      journal = {ApJ},
         year = 2015,
        month = jul,
       volume = {807},
       number = {1},
          eid = {77},
        pages = {77},
          doi = {10.1088/0004-637X/807/1/77},
archivePrefix = {arXiv},
       eprint = {1505.05936},
 primaryClass = {astro-ph.HE},
       adsurl = {https://ui.adsabs.harvard.edu/abs/2015ApJ...807...77K}
}

@ARTICLE{Dobler+2008,
       author = {{Dobler}, Gregory and {Finkbeiner}, Douglas P.},
        title = "{Extended Anomalous Foreground Emission in the WMAP Three-Year Data}",
      journal = {ApJ},
         year = 2008,
        month = jun,
       volume = {680},
       number = {2},
        pages = {1222-1234},
          doi = {10.1086/587862},
archivePrefix = {arXiv},
       eprint = {0712.1038},
 primaryClass = {astro-ph},
       adsurl = {https://ui.adsabs.harvard.edu/abs/2008ApJ...680.1222D}
}

@ARTICLE{Carretti+2013,
       author = {{Carretti}, Ettore and {Crocker}, Roland M. and {Staveley-Smith}, Lister and {Haverkorn}, Marijke and {Purcell}, Cormac and {Gaensler}, B.~M. and {Bernardi}, Gianni and {Kesteven}, Michael J. and {Poppi}, Sergio},
        title = "{Giant magnetized outflows from the centre of the Milky Way}",
      journal = {Nature},
         year = 2013,
        month = jan,
       volume = {493},
       number = {7430},
        pages = {66-69},
          doi = {10.1038/nature11734},
archivePrefix = {arXiv},
       eprint = {1301.0512},
 primaryClass = {astro-ph.GA},
       adsurl = {https://ui.adsabs.harvard.edu/abs/2013Natur.493...66C}
}

@ARTICLE{Bland-Hawthorn+2003,
       author = {{Bland-Hawthorn}, Joss and {Cohen}, Martin},
        title = "{The Large-Scale Bipolar Wind in the Galactic Center}",
      journal = {ApJ},
         year = 2003,
        month = jan,
       volume = {582},
       number = {1},
        pages = {246-256},
          doi = {10.1086/344573},
archivePrefix = {arXiv},
       eprint = {astro-ph/0208553},
 primaryClass = {astro-ph},
       adsurl = {https://ui.adsabs.harvard.edu/abs/2003ApJ...582..246B}
}

@ARTICLE{Bland-Hawthorn+2016,
       author = {{Bland-Hawthorn}, Joss and {Gerhard}, Ortwin},
        title = "{The Galaxy in Context: Structural, Kinematic, and Integrated Properties}",
      journal = {ARA\&A},
         year = 2016,
        month = sep,
       volume = {54},
        pages = {529-596},
          doi = {10.1146/annurev-astro-081915-023441},
archivePrefix = {arXiv},
       eprint = {1602.07702},
 primaryClass = {astro-ph.GA},
       adsurl = {https://ui.adsabs.harvard.edu/abs/2016ARA&A..54..529B}
}

@ARTICLE{2020MNRAS.498.2440Y,
       author = {{Yuan}, Yuxuan and {Krumholz}, Mark R. and {Burkhart}, Blakesley},
        title = "{Understanding biases in measurements of molecular cloud kinematics using line emission}",
      journal = {\mnras},
         year = 2020,
        month = oct,
       volume = {498},
       number = {2},
        pages = {2440-2455},
          doi = {10.1093/mnras/staa2432},
archivePrefix = {arXiv},
       eprint = {2007.13488},
 primaryClass = {astro-ph.GA},
       adsurl = {https://ui.adsabs.harvard.edu/abs/2020MNRAS.498.2440Y}
}

@ARTICLE{2021ApJ...910..109Y,
       author = {{Yan}, Qing-Zeng and {Yang}, Ji and {Yang}, Su and {Sun}, Yan and {Wang}, Chen},
        title = "{On the Beam Filling Factors of Molecular Clouds}",
      journal = {\apj},
         year = 2021,
        month = apr,
       volume = {910},
       number = {2},
          eid = {109},
        pages = {109},
          doi = {10.3847/1538-4357/abe628},
archivePrefix = {arXiv},
       eprint = {2102.06917},
 primaryClass = {astro-ph.GA},
       adsurl = {https://ui.adsabs.harvard.edu/abs/2021ApJ...910..109Y}
}

@ARTICLE{2018ApJ...858...16G,
       author = {{Gong}, Munan and {Ostriker}, Eve C. and {Kim}, Chang-Goo},
        title = "{The X $_{CO}$ Conversion Factor from Galactic Multiphase ISM Simulations}",
      journal = {\apj},
         year = 2018,
        month = may,
       volume = {858},
       number = {1},
          eid = {16},
        pages = {16},
          doi = {10.3847/1538-4357/aab9af},
archivePrefix = {arXiv},
       eprint = {1803.09822},
 primaryClass = {astro-ph.GA},
       adsurl = {https://ui.adsabs.harvard.edu/abs/2018ApJ...858...16G}
}

@ARTICLE{2019MNRAS.486.4463Y,
       author = {{Yu}, Xiaoling and {Shi}, Yong and {Chen}, Yanmei and {Law}, David R. and {Bizyaev}, Dmitry and {Bing}, Longji and {Li}, Songlin and {Zhou}, Luwenjia and {Chen}, Jianhang and {Riffel}, Rogemar A. and {Riffel}, Rog{\'e}rio and {Zhang}, Kai and {Chen}, Yongyun and {Pan}, Kaike},
        title = "{What drives the velocity dispersion of ionized gas in star-forming galaxies?}",
      journal = {\mnras},
         year = 2019,
        month = jul,
       volume = {486},
       number = {4},
        pages = {4463-4472},
          doi = {10.1093/mnras/stz1146},
archivePrefix = {arXiv},
       eprint = {1904.11431},
 primaryClass = {astro-ph.GA},
       adsurl = {https://ui.adsabs.harvard.edu/abs/2019MNRAS.486.4463Y}
}

@ARTICLE{2006PASP..118..590R,
       author = {{Rosolowsky}, Erik and {Leroy}, Adam},
        title = "{Bias-free Measurement of Giant Molecular Cloud Properties}",
      journal = {\pasp},
         year = 2006,
        month = apr,
       volume = {118},
       number = {842},
        pages = {590-610},
          doi = {10.1086/502982},
archivePrefix = {arXiv},
       eprint = {astro-ph/0601706},
 primaryClass = {astro-ph},
       adsurl = {https://ui.adsabs.harvard.edu/abs/2006PASP..118..590R}
}

@ARTICLE{1993A&A...269...54B,
       author = {{Breitschwerdt}, D. and {McKenzie}, J.~F. and {Voelk}, H.~J.},
        title = "{Galactic winds. II. Role of the disk-halo interface in cosmic ray driven galactic winds.}",
      journal = {\aap},
         year = 1993,
        month = mar,
       volume = {269},
        pages = {54-66},
       adsurl = {https://ui.adsabs.harvard.edu/abs/1993A&A...269...54B}
}

@ARTICLE{2022NatAs...6..584Y,
       author = {{Yang}, H.-Y. Karen and {Ruszkowski}, Mateusz and {Zweibel}, Ellen G.},
        title = "{Fermi and eROSITA bubbles as relics of the past activity of the Galaxy's central black hole}",
      journal = {Nature Astronomy},
         year = 2022,
        month = mar,
       volume = {6},
        pages = {584-591},
          doi = {10.1038/s41550-022-01618-x},
archivePrefix = {arXiv},
       eprint = {2203.02526},
 primaryClass = {astro-ph.HE},
       adsurl = {https://ui.adsabs.harvard.edu/abs/2022NatAs...6..584Y}
}

@ARTICLE{2020ApJ...894..117Z,
       author = {{Zhang}, Ruiyu and {Guo}, Fulai},
        title = "{Simulating the Fermi Bubbles as Forward Shocks Driven by AGN Jets}",
      journal = {\apj},
         year = 2020,
        month = may,
       volume = {894},
       number = {2},
          eid = {117},
        pages = {117},
          doi = {10.3847/1538-4357/ab8bd0},
archivePrefix = {arXiv},
       eprint = {2003.03625},
 primaryClass = {astro-ph.HE},
       adsurl = {https://ui.adsabs.harvard.edu/abs/2020ApJ...894..117Z}
}

@ARTICLE{2018MNRAS.479.3042G,
       author = {{Girichidis}, Philipp and {Naab}, Thorsten and {Hanasz}, Micha{\l} and
         {Walch}, Stefanie},
        title = "{Cooler and smoother - the impact of cosmic rays on the phase structure of galactic outflows}",
      journal = {\mnras},
         year = 2018,
        month = sep,
       volume = {479},
       number = {3},
        pages = {3042-3067},
          doi = {10.1093/mnras/sty1653},
archivePrefix = {arXiv},
       eprint = {1805.09333},
 primaryClass = {astro-ph.GA},
       adsurl = {https://ui.adsabs.harvard.edu/abs/2018MNRAS.479.3042G}
}

@ARTICLE{2018MNRAS.474.5076J,
       author = {{Johnson}, H.~L. and {Harrison}, C.~M. and {Swinbank}, A.~M. and {Tiley}, A.~L. and {Stott}, J.~P. and {Bower}, R.~G. and {Smail}, Ian and {Bunker}, A.~J. and {Sobral}, D. and {Turner}, O.~J. and {Best}, P. and {Bureau}, M. and {Cirasuolo}, M. and {Jarvis}, M.~J. and {Magdis}, G. and {Sharples}, R.~M. and {Bland-Hawthorn}, J. and {Catinella}, B. and {Cortese}, L. and {Croom}, S.~M. and {Federrath}, C. and {Glazebrook}, K. and {Sweet}, S.~M. and {Bryant}, J.~J. and {Goodwin}, M. and {Konstantopoulos}, I.~S. and {Lawrence}, J.~S. and {Medling}, A.~M. and {Owers}, M.~S. and {Richards}, S.},
        title = "{The KMOS Redshift One Spectroscopic Survey (KROSS): the origin of disc turbulence in z {\ensuremath{\approx}} 1 star-forming galaxies}",
      journal = {\mnras},
         year = 2018,
        month = mar,
       volume = {474},
       number = {4},
        pages = {5076-5104},
          doi = {10.1093/mnras/stx3016},
archivePrefix = {arXiv},
       eprint = {1707.02302},
 primaryClass = {astro-ph.GA},
       adsurl = {https://ui.adsabs.harvard.edu/abs/2018MNRAS.474.5076J}
}

@ARTICLE{2011ApJ...741...69D,
       author = {{Davies}, R. and {F{\"o}rster Schreiber}, N.~M. and {Cresci}, G. and {Genzel}, R. and {Bouch{\'e}}, N. and {Burkert}, A. and {Buschkamp}, P. and {Genel}, S. and {Hicks}, E. and {Kurk}, J. and {Lutz}, D. and {Newman}, S. and {Shapiro}, K. and {Sternberg}, A. and {Tacconi}, L.~J. and {Wuyts}, S.},
        title = "{How Well Can We Measure the Intrinsic Velocity Dispersion of Distant Disk Galaxies?}",
      journal = {\apj},
         year = 2011,
        month = nov,
       volume = {741},
       number = {2},
          eid = {69},
        pages = {69},
          doi = {10.1088/0004-637X/741/2/69},
archivePrefix = {arXiv},
       eprint = {1108.0285},
 primaryClass = {astro-ph.CO},
       adsurl = {https://ui.adsabs.harvard.edu/abs/2011ApJ...741...69D}
}

@ARTICLE{2020ApJ...898...80Y,
       author = {{Yan}, Qing-Zeng and {Yang}, Ji and {Su}, Yang and {Sun}, Yan and {Wang}, Chen},
        title = "{Distances and Statistics of Local Molecular Clouds in the First Galactic Quadrant}",
      journal = {\apj},
         year = 2020,
        month = jul,
       volume = {898},
       number = {1},
          eid = {80},
        pages = {80},
          doi = {10.3847/1538-4357/ab9f9c},
archivePrefix = {arXiv},
       eprint = {2006.13654},
 primaryClass = {astro-ph.GA},
       adsurl = {https://ui.adsabs.harvard.edu/abs/2020ApJ...898...80Y}
}

@ARTICLE{2022MNRAS.512.4765S,
       author = {{Seifried}, D. and {Beuther}, H. and {Walch}, S. and {Syed}, J. and {Soler}, J.~D. and {Girichidis}, P. and {W{\"u}nsch}, R.},
        title = "{On the accuracy of H I observations in molecular clouds - More cold H I than thought?}",
      journal = {\mnras},
         year = 2022,
        month = jun,
       volume = {512},
       number = {4},
        pages = {4765-4784},
          doi = {10.1093/mnras/stac607},
archivePrefix = {arXiv},
       eprint = {2109.10917},
 primaryClass = {astro-ph.GA},
       adsurl = {https://ui.adsabs.harvard.edu/abs/2022MNRAS.512.4765S}
}

@ARTICLE{DiTeodoro+2026,
       author = {{Di Teodoro}, Enrico M. and {Heyer}, Mark and {Krumholz}, Mark R. and {Armillotta}, Lucia and {Lockman}, Felix J. and {Afruni}, Andrea and {Busch}, Michael P. and {McClure-Griffiths}, Naomi M. and {Noon}, Karlie A. and {Peschken}, Nicolas and {Yu}, Qingzheng},
        title = "{A survey of molecular clouds in the Galactic center's outflow}",
      journal = {A\&A},
         year = 2026,
        month = feb,
       volume = {706},
          eid = {A330},
        pages = {A330},
          doi = {10.1051/0004-6361/202557732},
       adsurl = {https://ui.adsabs.harvard.edu/abs/2026A&A...706A.330D}
}

@ARTICLE{2017ApJ...847...59H,
    author = {{Hummels}, C.~B. and {Smith}, B.~D. and {Silvia}, D.~W.},
    title = "{Trident: A Universal Tool for Generating Synthetic Absorption Spectra from Astrophysical Simulations}",
    journal = {\apj},
    archivePrefix = "arXiv",
    eprint = {1612.03935},
    primaryClass = "astro-ph.IM",
    year = 2017,
    month = sep,
    volume = 847,
    eid = {59},
    pages = {59},
    doi = {10.3847/1538-4357/aa7e2d},
    adsurl = {http://adsabs.harvard.edu/abs/2017ApJ...847...59H}
}

@ARTICLE{Leary+2026,
       author = {{Leary}, Hannah and {Richie}, Helena M. and {Schneider}, Evan},
        title = "{Effects of Numerical Resolution on Simulated Cloud─Wind Interactions}",
      journal = {ApJ},
         year = 2026,
        month = feb,
       volume = {998},
       number = {2},
          eid = {281},
        pages = {281},
          doi = {10.3847/1538-4357/ae356c},
archivePrefix = {arXiv},
       eprint = {2602.10082},
 primaryClass = {astro-ph.GA},
       adsurl = {https://ui.adsabs.harvard.edu/abs/2026ApJ...998..281L}
}

@ARTICLE{1959ApJ...129..243S,
       author = {{Schmidt}, Maarten},
        title = "{The Rate of Star Formation.}",
      journal = {\apj},
         year = 1959,
        month = mar,
       volume = {129},
        pages = {243},
          doi = {10.1086/146614},
       adsurl = {https://ui.adsabs.harvard.edu/abs/1959ApJ...129..243S}
}

@ARTICLE{2020NatAs...4..377N,
       author = {{Nogueras-Lara}, Francisco and {Sch{\"o}del}, Rainer and {Gallego-Calvente}, Aurelia Teresa and {Gallego-Cano}, Eulalia and {Shahzamanian}, Banafsheh and {Dong}, Hui and {Neumayer}, Nadine and {Hilker}, Michael and {Najarro}, Francisco and {Nishiyama}, Shogo and {Feldmeier-Krause}, Anja and {Girard}, Julien H.~V. and {Cassisi}, Santi},
        title = "{Early formation and recent starburst activity in the nuclear disk of the Milky Way}",
      journal = {Nature Astronomy},
         year = 2020,
        month = jan,
       volume = {4},
        pages = {377-381},
          doi = {10.1038/s41550-019-0967-9},
archivePrefix = {arXiv},
       eprint = {1910.06968},
 primaryClass = {astro-ph.GA},
       adsurl = {https://ui.adsabs.harvard.edu/abs/2020NatAs...4..377N}
}

@ARTICLE{1991MNRAS.252..210B,
       author = {{Binney}, James and {Gerhard}, Ortwin E. and {Stark}, Antony A. and {Bally}, John and {Uchida}, Keven I.},
        title = "{Understanding the kinematics of Galactic Centre gas.}",
      journal = {\mnras},
         year = 1991,
        month = sep,
       volume = {252},
        pages = {210},
          doi = {10.1093/mnras/252.2.210},
       adsurl = {https://ui.adsabs.harvard.edu/abs/1991MNRAS.252..210B}
}

@ARTICLE{2015MNRAS.449.2421S,
       author = {{Sormani}, Mattia C. and {Binney}, James and {Magorrian}, John},
        title = "{Gas flow in barred potentials}",
      journal = {\mnras},
         year = 2015,
        month = may,
       volume = {449},
       number = {3},
        pages = {2421-2435},
          doi = {10.1093/mnras/stv441},
archivePrefix = {arXiv},
       eprint = {1502.02740},
 primaryClass = {astro-ph.GA},
       adsurl = {https://ui.adsabs.harvard.edu/abs/2015MNRAS.449.2421S}
}

@ARTICLE{2022ApJ...924...26S,
       author = {{Steinwandel}, Ulrich P. and {Dolag}, Klaus and {Lesch}, Harald and {Burkert}, Andreas},
        title = "{Driving Galactic Outflows with Magnetic Fields at Low and High Redshift}",
      journal = {\apj},
         year = 2022,
        month = jan,
       volume = {924},
       number = {1},
          eid = {26},
        pages = {26},
          doi = {10.3847/1538-4357/ac2ffd},
archivePrefix = {arXiv},
       eprint = {2012.12905},
 primaryClass = {astro-ph.GA},
       adsurl = {https://ui.adsabs.harvard.edu/abs/2022ApJ...924...26S}
}

@ARTICLE{Koyama+02,
       author = {{Koyama}, Hiroshi and {Inutsuka}, Shu-ichiro},
        title = "{An Origin of Supersonic Motions in Interstellar Clouds}",
      journal = {\apjl},
         year = 2002,
        month = jan,
       volume = {564},
       number = {2},
        pages = {L97-L100},
          doi = {10.1086/338978},
archivePrefix = {arXiv},
       eprint = {astro-ph/0112420},
 primaryClass = {astro-ph},
       adsurl = {https://ui.adsabs.harvard.edu/abs/2002ApJ...564L..97K}
}

@ARTICLE{2026A&A...712A.148P,
       author = {{Peschken}, Nicolas and {Di Teodoro}, Enrico M. and {Hanasz}, Micha{\l} and {Armillotta}, Lucia and {Tress}, Robin and {Naab}, Thorsten and {W{\'o}lta{\'n}ski}, Dominik and {Gawryszczak}, Artur},
        title = "{Simulating winds in the Galactic centre: I. Supernova-driven multiphase outflows and HI cloud acceleration}",
      journal = {\aap},
         year = 2026,
        month = aug,
       volume = {712},
          eid = {A148},
        pages = {A148},
          doi = {10.1051/0004-6361/202659639},
archivePrefix = {arXiv},
       eprint = {2606.16553},
 primaryClass = {astro-ph.GA},
       adsurl = {https://ui.adsabs.harvard.edu/abs/2026A&A...712A.148P}
}

\begin{appendix}

\section{Extended quiescent epoch}
\label{app:A}
\begin{figure}[!h]
\includegraphics[width=0.5\textwidth, trim = 0.0cm 0 0 0, clip]{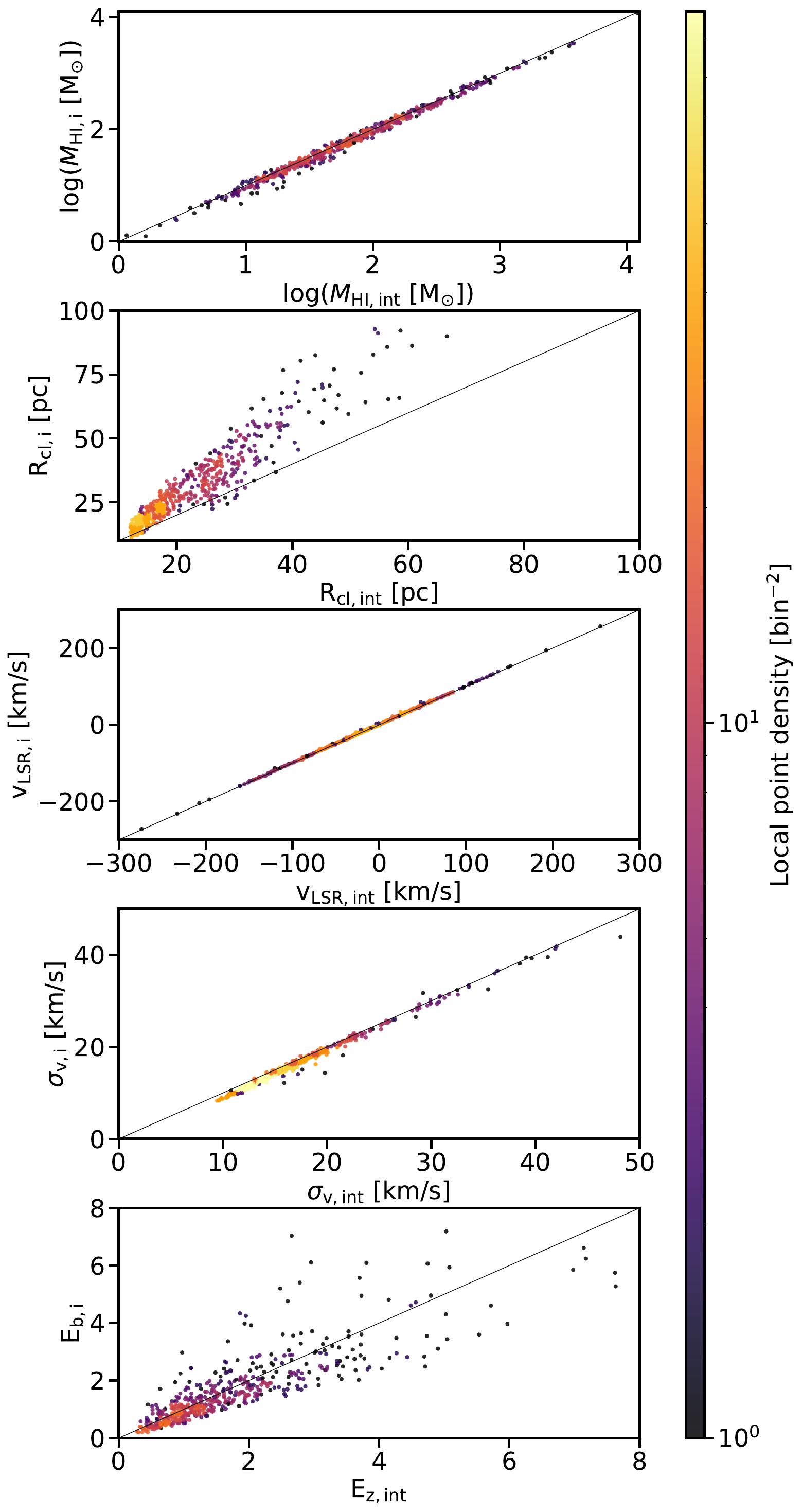}
\caption{Same as Fig. \ref{fig:corr_3d_ideal}, but for the extended quiescent time.}
\label{fig:corr_ideal_smooth_quiet}
\end{figure}

\begin{figure*}
\centering
\includegraphics[width=1.05\textwidth, trim = 2.0cm 0cm 0cm 4cm, clip]{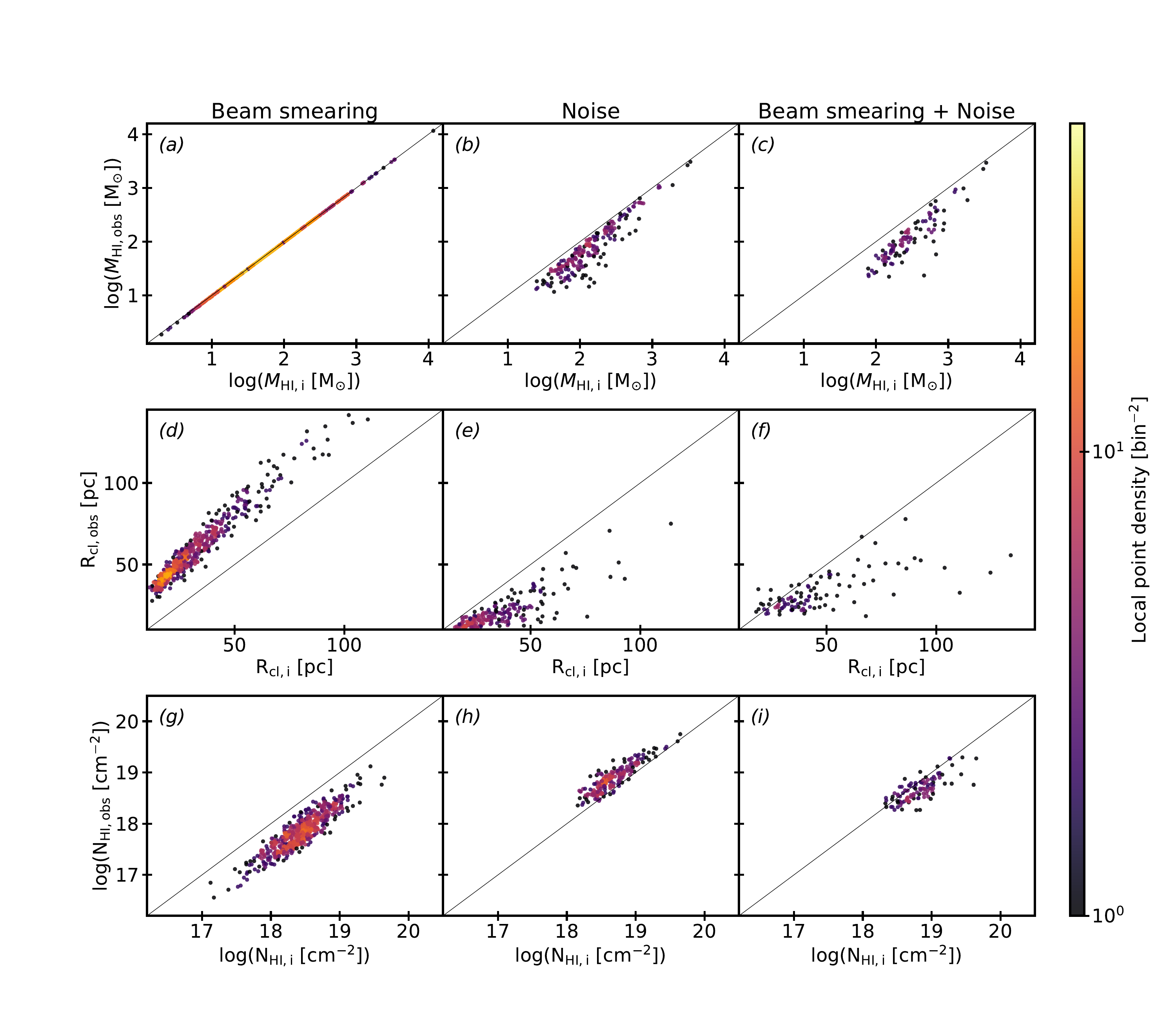}
\caption{Same as Fig.~\ref{fig:all_correlations1}, but for the quiescent period.}
\label{fig:all_correlations_quiet1}
\end{figure*}

\begin{figure*}
\centering
\includegraphics[width=1.05\textwidth, trim = 2.0cm 0cm 0cm 0cm, clip]{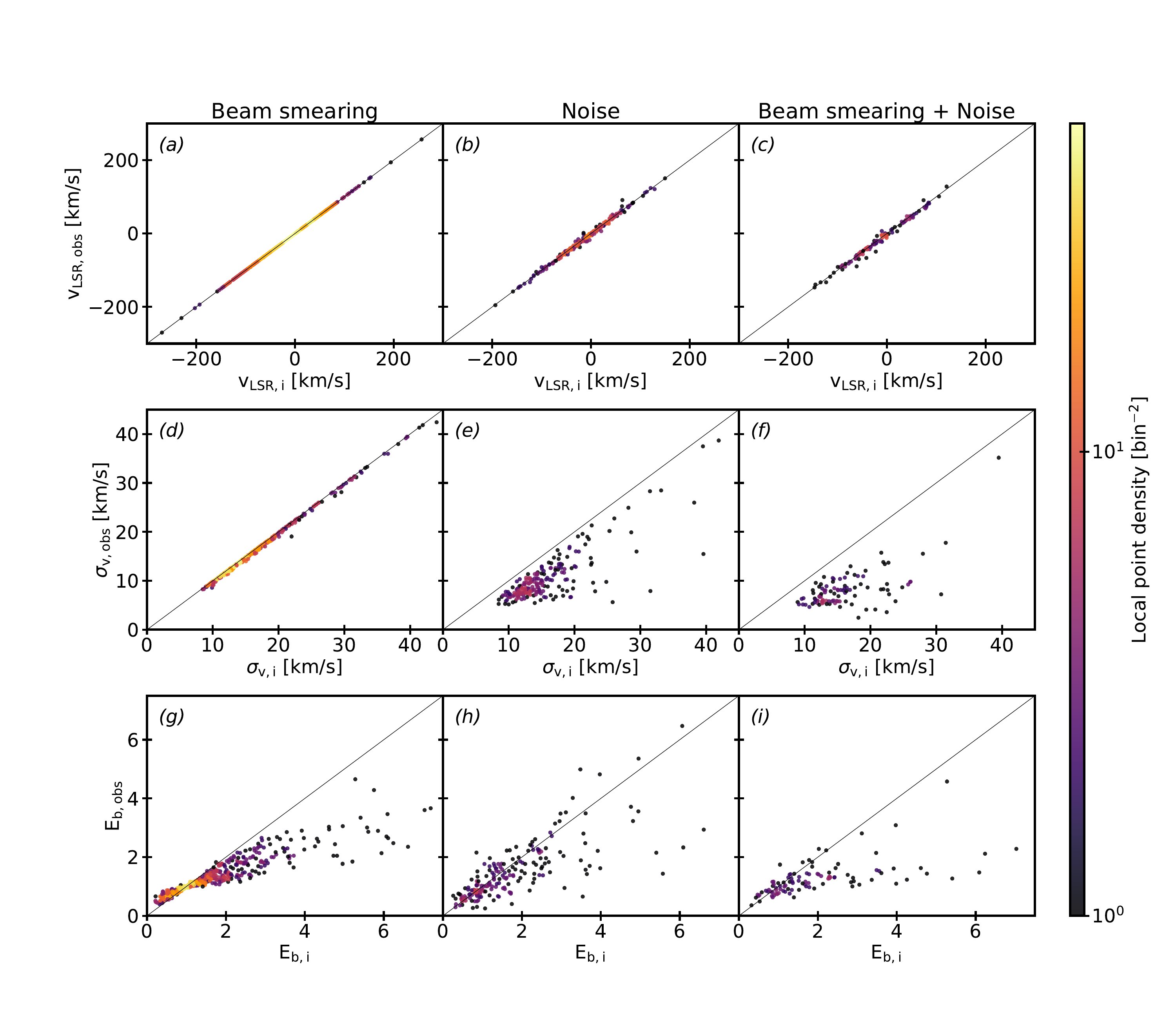}
\caption{Same as Fig.~\ref{fig:all_correlations2}, but for the quiescent period.}
\label{fig:all_correlations_quiet2}
\end{figure*}

In this Appendix, we show the figures for the cloud population at the extended quiescent time. Figures~\ref{fig:corr_ideal_smooth_quiet}, \ref{fig:all_correlations_quiet1} and \ref{fig:all_correlations_quiet2} are analogous to Figures~\ref{fig:corr_3d_ideal}, \ref{fig:all_correlations1} and \ref{fig:all_correlations2}, respectively.

\section{Observational effects and cloud radii}
\label{app:B}

\begin{figure}
\includegraphics[width=0.5\textwidth, trim = 0.5cm 0 0 0, clip]{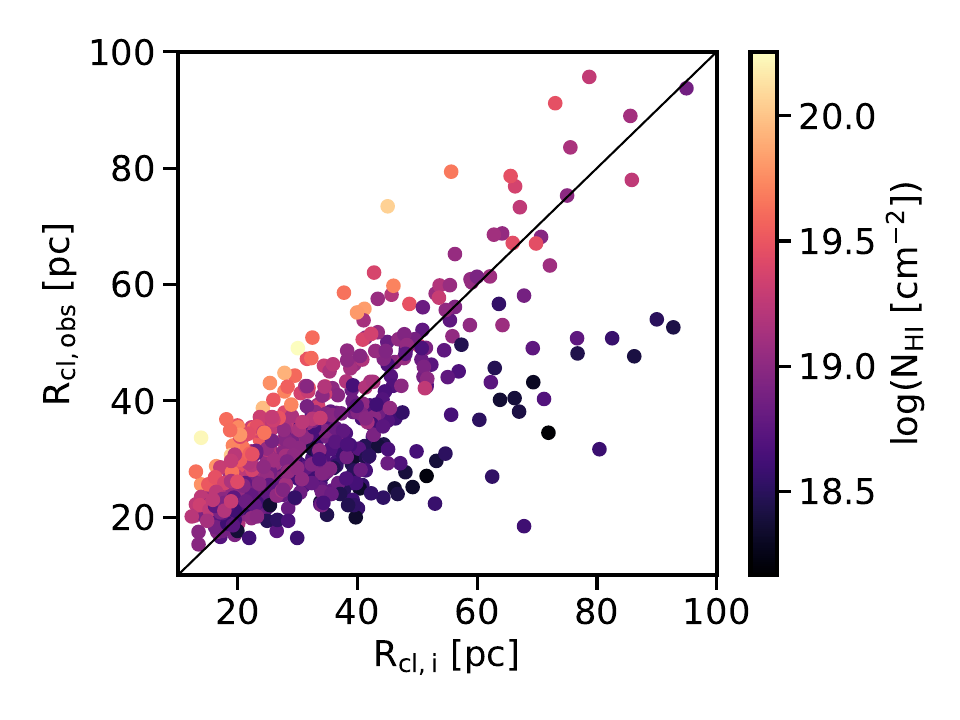}
\caption{Cloud radii when both beam smearing and noise are taken into account, as a function of their radii in the ideal observational case (no beam smearing, minimal noise), for the combined populations of clouds of the post-starburst and quiescent time. Each point is colour-coded by its column density measured in the ideal case. Low density clouds tend have a radius reduced by observational biases, while in contrast high density clouds appear larger.}
\label{fig:radii_wholepop}
\end{figure}

Here we show how the change in the measured radius of a cloud under realistic observational conditions depends on its column density. Fig.~\ref{fig:radii_wholepop} shows the recovered radii under both observational biases as a function of their values in the ideal case for the combined post-starburst and quiescent cloud populations. The points are colour-coded by the column density measured in the ideal case. The figure shows a clear transition across the one-to-one relation, with clouds below $\sim10^{19}$~cm$^{-2}$ tending to have underestimated radii because noise removes part of their emission, while denser clouds tend to have overestimated radii because beam smearing dominates.

\end{appendix}

\end{document}